\documentclass[aps,prb,twocolumn,superscriptaddress,10pt,article,nofootinbib,longbibliography]{revtex4-2}
\usepackage{graphicx,amsfonts,amssymb,amsmath,hyperref,hypcap,enumerate,bm,color,float,multirow,tabularray,placeins,adjustbox}

\AddToHook{env/align/begin}{\openup-9pt}

\newcommand{\beq}{\begin{equation}}
\newcommand{\eneq}{\end{equation}}

\def\qq{\mathbf{q}}
\def\kk{\mathbf{k}}

\def\pp{\mathbf{p}}

\def\LL{\mathbf{L}}

\def\rr{\mathbf{r}}

\def\QQ{\mathbf{Q}}

\def\jj{\mathbf{j}}

\def\qq{\mathbf{q}}
\def\pp{\mathbf{p}}
\def\pp{\mathbf{p}}

\def\QQ{\mathbf{Q}}

\def\nn{\mathbf{n}}
\def\dd{\mathbf{d}}

\def\JJ{\mathbf{J}}
\def\vv{\mathbf{v}}

\begin{document}
\setlength{\abovedisplayskip}{10pt}
\setlength{\belowdisplayskip}{10pt}
\setlength{\abovedisplayshortskip}{2pt}
\setlength{\belowdisplayshortskip}{2pt}

\title{Pair-mixing induced Time-reversal-breaking superconductivity}
\author{Saswata Mandal}
\affiliation{Department of Physics, The Pennsylvania State University, University Park,  Pennsylvania 16802, USA}
\author{Chao-Xing Liu$^*$}
\affiliation{Department of Physics, The Pennsylvania State University, University Park,  Pennsylvania 16802, USA}
\begin{abstract}
Experimental evidences of spontaneous time-reversal (TR) symmetry breaking have been reported for the superconducting ground state in the transition metal dichalcogenide (TMD) superconductor 4H$_b$-TaS$_2$ or chiral molecule intercalated TaS$_2$ hybrid superlattices, and is regarded as the evidence of the emergent chiral superconductivity. However, the $T_c$ of these TMD superconductors is at the same order as pristine 1H or 2H-TaS$_2$ that does not show any signature of TR breaking and is believed to be conventional Bardeen–Cooper–Schrieffer superconductors. To resolve this puzzle, we proposed a new type of pair-mixing states that mix the dominant conventional s-wave pairing channel with the subdominant chiral p-wave pairing channel via the finite Cooper pair momentum based on the symmetry analysis of the Ginzburg-Landau theory. Our analysis shows the fourth order terms in the chiral p-wave channel can lead to a variety of pair-mixing states with spontaneous TR breaking. These TR-breaking superconducting states also reveal zero-field junction-free superconducting diode effect that is observed in chiral molecule intercalated TaS$_2$ superlattices. 
\end{abstract}
\date{\today}

\maketitle

{\it Introduction---} In the Bardeen–Cooper–Schrieffer (BCS) theory, the conventional s-wave spin-singlet pairing state preserves time-reversal (TR) symmetry. Spontaneous TR breaking in superconductors (SCs) when observed, is widely considered as a hallmark of unconventional superconductivity. Experimental evidence of TR breaking has been reported in a variety of superconducting (SC) materials, including Sr$_2$RuO$_4$\cite{Sr2RuO4_2,Sr2RuO4_1}, LaNiC$_2$\cite{LaNiC2, csire2018nonunitary}, LaNiGa$_2$\cite{LaNiGa2, ghosh2020quantitative}, UTe$_2$\cite{ran2019nearly,ishihara2023chiral,jiao2020chiral,metz2019point}, other heavy-fermion SCs \cite{schemm2014observation,avers2020broken,heffner1990new,schemm2015evidence,PhysRevLett.91.067003,maisuradze2010evidence}, cuprates \cite{zhao2023time,kaur2003broken,weber1990evidence,simon2002detection}, iron based SCs \cite{matsuura2023two,PhysRevB.95.214511, hu2020pairing, zaki2021time} and others\cite{TRSrevGhosh,ghosh2020recent, maccari2025tuning, csire2022magnetically}. More recently, experimental evidence of TR breaking, including non-zero $\mu$SR signal below $T_c$\cite{ribak2020}, the existence of magnetic memory below $T_c$ \cite{persky2022},  $\pi$-phase shift in Little-Parks effect and zero-field superconducting diode effect (SDE)\cite{wan2024,almoalem2024}, has been identified in a class of transition metal dichalcogenide (TMD) SCs, notably 4Hb-TaS$_2$\cite{ribak2020,meyer1975properties,liu2025crystal,huang2025pressure,wang2025evidence,liu2014coexistence,dentelski2021robust,zhou2025nodeless} and chiral molecule intercalated TaS$_2$ hybrid superlattices\cite{wan2024}. These experimental observations motivate theoretical speculation regarding the potential existence of chiral superconductivity in TaS$_2$ compounds\cite{ribak2020, wan2024,silber2024b,MengF2024,PhysRevB.102.075138,wu2025unveiling,kumar2023first,persky2022,almoalem2024,ramires2025pure}. However, spontaneous TR breaking has {\it not} been observed in the pristine 1H or 2H-TaS$_2$ films, which are generally considered conventional s-wave BCS SCs\cite{persky2022,ising2018,anisoS,HcS2017} with Ising-type of spin-orbit coupling, similar to other TMD SCs \cite{saito2016,xi2016,lu2015evidence}. 4Hb-TaS$_2$ consists of alternating 1H and 1T TaS$_2$ layers and exhibits the transition temperature $T_c$ around 2.7 K \cite{ribak2020,silber2024b,MengF2024}.
This value is comparable to $T_c \sim$ 2.2 K observed in 2H-TaS$_2$ \cite{atomic2016,yang2018}, implying that superconductivity in 4Hb-TaS$_2$ is likely to originate from the 1H TaS$_2$ layers. Similarly, superconductivity in chiral molecule intercalated TaS$_2$ hybrid superlattices should also originate from 1H-TaS$_2$ layers\cite{wan2024}. This raises a fundamental question: whether the pairing state in 4Hb-TaS$_2$ and chiral molecule intercalated TaS$_2$ belongs to the conventional Ising type of BCS state, or the unconventional TR-breaking chiral SC state. 

We note that, compared to the pristine 1H or 2H-TaS$_2$ films, 1T-TaS$_2$ layers in 4Hb-TaS$_2$ and chiral molecule intercalation layers lower the local symmetry of 1H-TaS$_2$ layer. In 4Hb-TaS$_2$, a chiral charge density wave exists in the 1T-TaS$_2$ layers\cite{zhao}, accompanied with lattice distortion that cause the distances between the 1H-TaS$_2$ layer and its two adjacent 1T-TaS$_2$ layers to become unequal\cite{yang2024}, which breaks the local mirror symmetry along the z-axis. Similarly, in Ref.~\cite{wan2024}, the intercalation of chiral molecules between 1H-TaS$_2$ layers also breaks any local mirror or inversion symmetry. As different pairing channels in SCs can be classified by irreducible representations (irrep) of the crystal symmetry group \cite{sigristuedaPhenomenologicalTheoryUnconventional1991a}, lowering crystal symmetry will allow the mixing of different pairing channels \cite{gor2001superconducting,yoshida2014parity,bauer2012non, shaffer2023weak, mockli2009s_if, shaffer2020nodal, hamill2021two, haim2022mecha}. This motivates us to explore the possibility of the pair-mixing between BCS state and chiral SC state in these TaS$_2$ systems. 

In this work, based on the analysis of Ginzburg-Landau (GL) theory, we theoretically propose a finite-momentum-assisted pair-mixing state with dominant conventional s-wave pairing channel belonging to the $A$ irrep and a secondary multi-component chiral pairing channel belonging to the $E$ irrep of the magnetic point group (MPG) $31'$ \cite{bilbao1,bilbao2,bilbao3} in 4Hb-TaS$_2$ and chiral molecule intercalated TaS$_2$. We demonstrate that the sub-dominant multi-component $E$ pairing can induce TR breaking via the finite orbital angular momentum of the Cooper pairs. As an example, we discuss the TR-breaking vortex-antivortex phase and TR-preserving rotation-breaking stripe phase, and identify the SC phase diagram for these two SC phases as a function of GL parameters. We also demonstrate the existence of the SDE in the TR-breaking vortex-antivortex phase. Our theory illustrates the complex interplay between chiral structures that lower local crystal symmetry and spontaneous TR breaking in such SC compounds.

\begin{figure}[!htbp]
\includegraphics[width=\columnwidth]{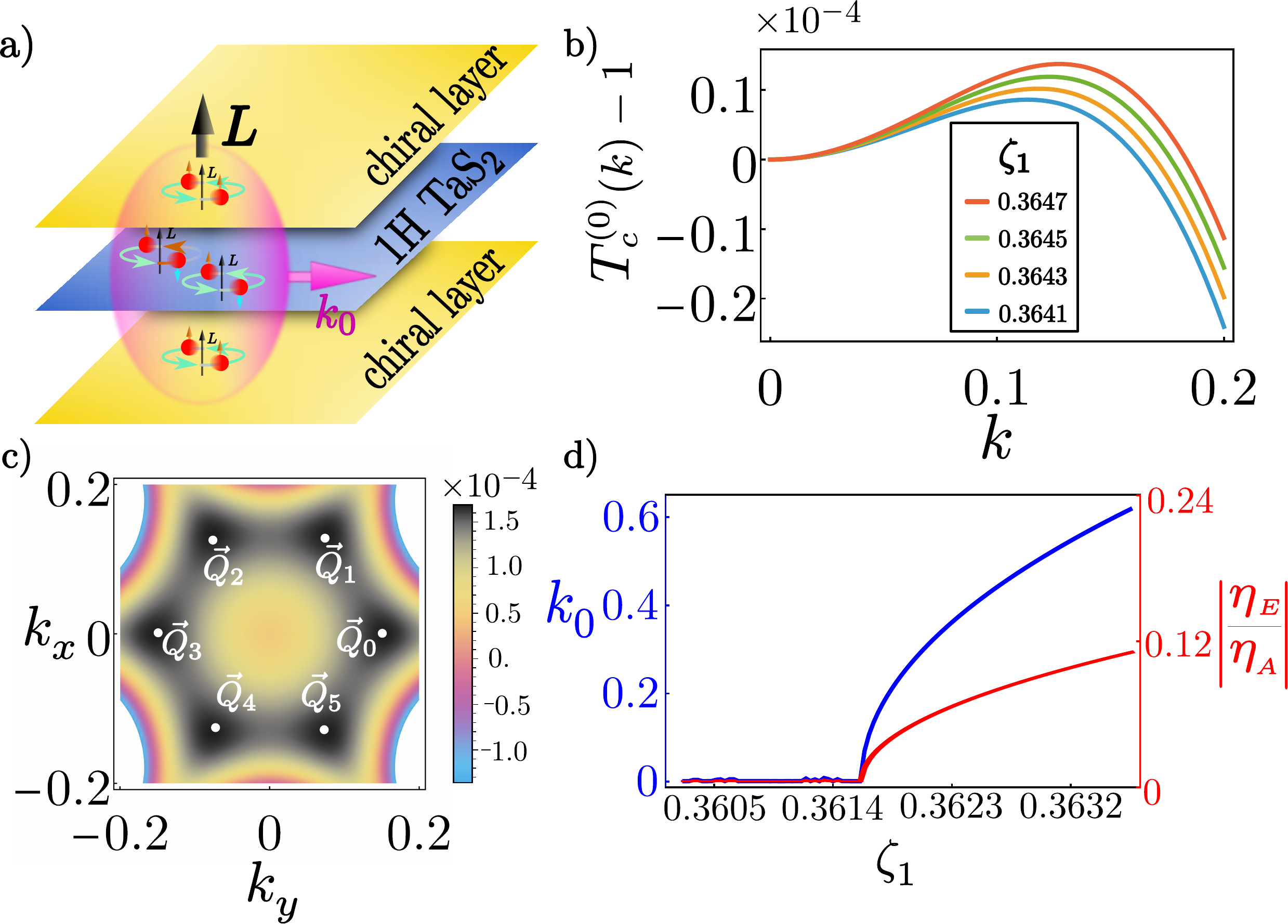}
\caption{(a) Schematics of 1H-TaS$_2$ layer sandwiched between chiral layers formed by either 1T-TaS$_2$ layer or chiral molecule layers. The s-wave singlet pairing in 1H-TaS$_2$ layer are mixed with p-wave triplet pairing in chiral layers at a finite momentum $\kk_0$. 
(b) $T_c^{(0)}(k)-1$ as a function of $k$ with $T_{c,A}=2.7K, T_{c,E}=2K, \gamma_{A}=0.076, \gamma_{E}=0.01$ at different values of $\zeta_1$. (c) Color denotes $T_c(\kk)-1$ as a function of $\kk$ with $\zeta_2=\zeta_4=0.13$, $\zeta_1=0.3643$, and all other parameters set to be same as (b). $\QQ_n$ with $n\in \mathbb{Z}_6$ labels six momenta for the maximal $T_c(\kk)$. (d) The blue and red curves depict the $k_0$ and the ratio $\left|\frac{\eta_E}{\eta_A}\right|$ as a function of $\zeta_{1}$, respectively, with the parameters to be the same as (b). All quantities above have been regularized to be dimensionless. ~\label{fig:pairmixing}
}
\end{figure}

{\it Pair-mixing states with a finite Cooper pair momentum -} 
Instead of 3D TaS$_2$ systems, we consider a minimal model of a 1H-TaS$_2$ SC monolayer sandwiched by two insulating layers in this work, as depicted in Fig.\ref{fig:pairmixing}(a). Here the insulating layers correspond to 1T-TaS$_2$ layer with the chiral CDW in 4Hb-TaS$_2$ \cite{ribak2020,persky2022} or the chiral molecule intercalation layer\cite{wan2024}. The insulating layers break all mirror symmetries of 1H TaS$_2$ monolayer and thus are dubbed ``chiral layers" below. 

We first consider the symmetry classification of pairing channels for this sandwich structure. The 1H-TaS$_2$ monolayer is described by MPG $\bar{6}m2.1'$, including $D_{3h}$ point group and TR symmetry \cite{ribak2020}. The chiral layers are expected to break all mirror symmetries, and thus reduce the symmetry group to MPG $31'$ with only $C_{3z}$ and TR \cite{bilbao1,bilbao2,bilbao3}. Table \ref{tab:irrepMPG31'} lists the irrep for different pairing channels of MPG $31'$. Similar to other TMD SCs, experiments on pristine TaS$_2$ have revealed s-wave pairings with TR symmetry \cite{wan2024,ribak2020,anisoS}, naturally belonging to the $A$ irrep in Table \ref{tab:irrepMPG31'}. In 4H$_b$ TaS$_2$ and chiral molecule intercalated TaS$_2$, a variety of experiments have revealed the evidence of spontaneous TR breaking \cite{ribak2020,persky2022,wan2024,almoalem2024} or nematic phases \cite{silber2024b}, necessitating the presence of multi-component order parameter \cite{sigristTimeReversalSymmetryBreaking1998,sigrist2009,shaffer2024superconducting,csire2022magnetically}. In Table \ref{tab:irrepMPG31'}, only the pairing channel with the $^1E^2E$ irrep  (called $E$-irrep below) is two-dimensional and can host both the chiral and nematic phases. As a comparison, we also discuss the pairing symmetry classification for MPG $\bar{6}m2.1'$, as well as the compatibility relation, in SM \cite{SM_Mandal2025} Table II and III. 

We search for possible pair-mixing between BCS ($A$) and chiral pairing ($E$), as schematically shown in Fig.\ref{fig:pairmixing}(a). At zero Cooper pair momentum ($\kk=0$), crystal symmetry forbids the mixing between the pairing channels that belong to different irreps. However, this obstacle can be overcome by considering a finite center-of-mass Cooper pair momentum $\kk$. Table \ref{tab:irrepMPG31'} also lists the irreps for $k$-polynomials, in which we notice $i \{k^{3n+1}_+,k^{3n+1}_-\}$ and $\{k_+^{3n+2},k_-^{3n+2}\}$ for any integer $n$ also belong to the $E$ irrep. Since $E\otimes E = 2A\oplus E$, we can construct pair-mixing terms invariant under MPG $31'$ as
\begin{equation}\label{eq:freenpairmixing}
\begin{aligned}\mathcal{F}_{A-E}&=\sum\limits_{\kk}\eta_{A}^*(i\zeta_1^* k_-+\zeta^*_2 k_+^2+\zeta^*_4k_-^4)\eta_{E,+}\\&+\eta_{A}^*(i\zeta_1 k_++\zeta_2 k_-^2+\zeta_4k_+^4)\eta_{E,-}+h.c.\end{aligned}
\end{equation}
up to $k^4$ orders, where $\eta_{A}$ and $\eta_{E,\pm}$ denote the SC order parameters for the $A$-irrep channel and two components of $E$-irrep channel, respectively\cite{sigristuedaPhenomenologicalTheoryUnconventional1991a}, $\zeta_n$ is the parameter of the $k_{\pm}^n$ term for $n=1,2,4$. 
To be more concrete, we consider s-wave spin-singlet pairing ($is_y$) for the $A$ channel $\eta_A$ and p-wave spin-triplet pairing ($p_\pm(s_z\pm s_0)$) for the $E$ channel $\eta_{E,\pm}$ in MPG $31'$. For these pairing channels, we note that the pair-mixing terms $\mathcal{F}_{A-E}$ in Eq.(\ref{eq:freenpairmixing}) cannot exist for MPG $\bar{6}m2.1'$ due to the mirror symmetry $M_z$ along the z axis. As discussed in SM\cite{SM_Mandal2025} Sec.A, the p-wave triplet pairing, order parameter $\eta_{E,\pm}$ is odd
under $M_z$, whereas the spin-singlet pairing $\eta_{A}$ and any momentum polynomial of $k_{\pm}$
are even under $M_z$. Therefore, the existence of $\mathcal{F}_{A-E}$ in Eq.(\ref{eq:freenpairmixing}) requires to break $M_z$ symmetry in MPG $\bar{6}m2.1'$. 
Up to the quadratic terms, the GL free energy is written as $\mathcal{F}^{(2)}=\mathcal{F}^{(2)}_{A}+\mathcal{F}^{(2)}_{E}+\mathcal{F}_{A-E}$, where $\mathcal{F}^{(2)}_{A}$ and $\mathcal{F}^{(2)}_{E}$ are the standard quadratic terms for the $A$ and $E$ channels, respectively (See SM\cite{SM_Mandal2025} Sec.A for details). From $\frac{\partial\mathcal{F}^{(2)}}{\partial \eta_{\Gamma}(\kk)^*}=0$ with $\Gamma = A, E+, E-$, the linearized GL-equation can be derived as
\begin{eqnarray}\label{eq:linGL}
 &&(\tilde{T}-M(\kk) ) \Psi(\kk) = 0, \nonumber \\
 &&M(\kk)=\scalebox{1.2}{$\left(\begin{smallmatrix}t_E&-\mu_1^*(\kk)&0\\-\mu_1(\kk)&t_{A}&-\mu_2(\kk)\\0&-\mu_2^*(\kk)&t_E\end{smallmatrix}\right)$}
\end{eqnarray}
where $t_{\Gamma=E,A}=\tilde{T}_{c,\Gamma}-\tilde{\gamma}_{\Gamma}\tilde{k}^2$, $\mu_1(\kk)=i\tilde{\zeta}_1^{*}\tilde{k}_-+\tilde{\zeta}_2^{*}\tilde{k}_+^2+\tilde{\zeta}_4^{*}\tilde{k}_-^4$,
$\mu_2(\kk)=i\tilde{\zeta}_1 \tilde{k}_++\tilde{\zeta}_2 \tilde{k}_-^2+\tilde{\zeta}_4 \tilde{k}_+^4$ and $\Psi(\kk) = \{\eta_{E,+}(\kk),\eta_{A}(\kk),\eta_{E,-}(\kk)\}^T$ is the eigen-vector. We have rescaled all the parameters with tilde to be dimensionless by $\tilde{T}=T/T_{c,A}$, $\tilde{\gamma}_{\Gamma}=\frac{\gamma_{\Gamma}}{\tilde{\alpha}_{\Gamma}}, \tilde{\kk} = \frac{k}{\sqrt{T_{c,A}/\tilde{\gamma}_{A}}} (\cos\theta,\sin\theta)$, 
$\tilde{\zeta}_{1}=\frac{\zeta_{1}}{\sqrt{\tilde{\alpha}_{A}\tilde{\alpha}_{E}\tilde{\gamma}_AT_{c,A}}}$, $\tilde{\zeta}_2=\frac{\zeta_{2}}{\tilde{\gamma}_A\sqrt{\tilde{\alpha}_{A}\tilde{\alpha}_{E}}}$, $\tilde{\zeta}_4=\frac{\zeta_{4}T_{c,A}}{\tilde{\gamma}_A^2\sqrt{\tilde{\alpha}_{A}\tilde{\alpha}_{E}}}$. The rescaling quantities $\tilde{\alpha}_{\Gamma}$ and $T_{c,\Gamma}$ are defined from the coefficient of the quadratic term in $\mathcal{F}^{(2)}$ before rescaling, namely $\alpha_{\Gamma}=\tilde{\alpha}_{\Gamma}(T-T_{c,\Gamma})$ (See Eq.(A28) in SM Sec.A). 
Below we always choose the dimensionless quantities after rescaling and have dropped the notation of tilde for all the parameters in Eq.(\ref{eq:linGL}). 
The nonzero solution of linearized GL-equation requires Det$(T\mathbb{I}-M(\kk))=0$, which gives rise to (See SM\cite{SM_Mandal2025} Sec.A for details)
\begin{eqnarray}~\label{eq:Tsoln}
&&T_c(\kk)= T_c^{(0)}(k)+\Delta T^{(1)}(k,\theta),
\end{eqnarray}
where $T_c^{(0)}(k)$ is the isotropic part and $\Delta T^{(1)}(k,\theta)$ depends on the momentum angle $\theta$. We have already chosen the largest solution when solving $T(\kk)$. The critical temperature $T_c$ is determined from the maximal value of $T(\kk)$ with respect to $\kk$ in Eq.\eqref{eq:Tsoln}, i.e. $T_c=\max\limits_{\kk}T_c(\kk)$. For the isotropic part ($\zeta_2=\zeta_4=0$), Fig.\ref{fig:pairmixing}(b) shows $T_c^{(0)}(k)$ as a function of $k$ for different values of $\zeta_1$. The maximum of $T_c^{(0)}(k)$ appears at $k=0$ for a small $\zeta_1$, while with increasing $\zeta_1$ above the critical value $\zeta_{1c}$, the maximum of $T(\kk)$ is shifted to a finite momentum, denoted as $k_0$. The dependence of $k_0$ with respect to $\zeta_1$ is shown by the blue curve in Fig.\ref{fig:pairmixing}(d). In the isotropic case ($\zeta_2=\zeta_4=0$), the eigenvector solution $\Psi^{(0)}(\kk)$ can be analytically solved as
\begin{eqnarray}\label{eq:GL2eigstate}
    \Psi^{(0)}(\kk) = \scalebox{0.9}{$\begin{pmatrix}
\eta^{(0)}_{E,+}(\kk) \\
\eta^{(0)}_{A}(\kk) \\
\eta^{(0)}_{E,-}(\kk)
\end{pmatrix}$} = \frac{1}{\sqrt{N(\kk)}}\scalebox{0.9}{$\begin{pmatrix}
i\zeta_1k_+ \\
T_c^{(0)}(k)-T_{c,E}-\gamma_{E}k^2 \\
i\zeta_1^{*}k_-
\end{pmatrix}$},\nonumber \\
\end{eqnarray}
where $N(\kk)$ is the normalization factor. We next consider the first order correction of the anisotropic $\zeta_2$ and $\zeta_4$ terms, which acts on $\Psi^{(0)}(\kk)$ and gives rise to $\Delta T^{(1)}(\kk)$. $\Delta T^{(1)}(\kk)$ splits the $T_c$ for different $\theta$ values and results in six momenta, namely $\{\QQ_n=k_0\cos(2n\pi/3)\hat{e}_x+k_0\sin(2n\pi/3)\hat{e}_y;  \forall n \in \mathbb{Z}_6\}$ related by $C_{3z}$ and $\mathcal{T}$ of MPG31$'$, for the critical temperature $T_c(\QQ_n)$. A typical $\kk$ dependence of $T_c(\kk)$ is shown in Fig.\ref{fig:pairmixing}(c) with six $\QQ_n$ labelled by the white dots. Strikingly, for $\zeta>\zeta_{1c}$, the SC ground state is a mixture of the $A$ and $E$ pairing channels due to non-zero $k_0$, as shown by the blue curve in Fig.\ref{fig:pairmixing}(d), and the ratio $|\eta_E/\eta_A|$ with \scalebox{0.95}{$|\eta_E| = \sqrt{|\eta_{E,+}^{(0)}(\QQ_n)|^2+|\eta_{E,-}^{(0)}(\QQ_n)|^2}$} also becomes non-zero, as exhibited by the red curve in Fig.\ref{fig:pairmixing}(d). The pairing states at six $\QQ_n$ are degenerate for $\mathcal{F}^{(2)}$and thus we need to consider the fourth-order terms to identify the possible SC ground states below. 

\begin{table}[h]
\centering
\caption{Representation ($\mathcal{D}_\mathcal{S}^{\Gamma}$) table and basis functions for the symmetry operator $\mathcal{S}$ in the little group at $\Gamma$ of MPG$31'$. Here $s_0$ and $\tau_0$ are 2-by-2 identity matrices while $s_{x,y,z}$ and $\tau_{x,y,z}$ are Pauli matrices. $s$ is for Cooper pair spin while $\tau$ is for the basis of the 2D $E$ irrep. $p_\pm=p_x\pm ip_y$ is the relative momentum of Cooper pair, whereas $k_{\pm}=k_x\pm ik_y$ is the center-of-mass momentum. 
\label{tab:irrepMPG31'}}
\begin{adjustbox}{max width=\columnwidth}
\begin{tabular}{c c c c c} 
\hline
\hline
\noalign{\vspace{2pt}}
Irrep($\Gamma$) &Basis functions& $\mathcal{D}^{\Gamma}_E$& $\mathcal{D}^{\Gamma}_{C_{3z}}$& $\mathcal{D}^{\Gamma}_{\mathcal{T}}$\\[2pt]
\hline
\noalign{\vspace{4pt}}
$A$&$is_y,is_x,i^{\frac{1\pm1}{2}}\{p_+(s_0-s_z)\pm p_-(s_0+s_z)\}$&1&1&1\\[4pt]
\multirow{3}{*}{$^1E^2E$}&$\{p_+(s_z-s_0),p_-(s_z+s_0)\}$&\multirow{3}{*}{$\tau_0$}&\multirow{3}{*}{$e^{i\frac{2\pi}{3}\tau_z}$}&\multirow{3}{*}{$\tau_x$}\\&\footnotesize{$i\{k_+^{3n+1},k_-^{3n+1}\},\{k_-^{3n+2},k_+^{3n+2}\}$}&&&\\&$p_+^{3l+1}(s_z+s_0),p_-^{3l+1}(s_z-s_0)$&&&\\[4pt]
\hline
\hline
\end{tabular}
\end{adjustbox}
\end{table}



{\it Time-Reversal-Breaking Pair-Mixing States -}
For $\zeta>\zeta_{1c}$, the eigen-states $\Psi^{(0)}(\QQ_n)$ at six $\QQ_n$ span the $U(6)$ subspace and the general form of this projected SC state should be a linear superposition of the eigen-vectors within this subspace 
\begin{eqnarray}~\label{eq:momgstate}\Psi_{\alpha}(\rr)=\sum\limits_{n\in \mathbb{Z}_6} P_{\QQ_n}\Psi_{\alpha}^{(0)}(\QQ_n)e^{i\QQ_n\cdot\rr},\end{eqnarray}
where $P_{\QQ_n}$ is the expansion coefficient of the eigenstate $\Psi^{(0)}(\QQ_n)$ and $\alpha=2$ for the $A$ component and $\alpha=1,3$ for the $E$ components of the pair function. Based on the MPG $31'$ and $U(1)$ gauge invariance, the fourth-order terms within the $U(6)$ subspace can be constructed as (See SM\cite{SM_Mandal2025} Sec.B1) \cite{tsunetsugu2008}
\begin{equation}\label{eq:fourthGLen}\begin{aligned}&\mathcal{F}^{(4)}=c_0\left[(\sum\limits_{i\in\mathbb{Z}_6}|P_{\QQ_i}|^2)^2\right]+c_1\left[\sum\limits_{i\in\mathbb{Z}_6}|P_{\QQ_i}|^2|P_{\QQ_{i+1}}|^2\right]\\&+c_2\left[\sum\limits_{i\in\mathbb{Z}_6}|P_{\QQ_i}|^2|P_{\QQ_{i+2}}|^2\right]+c_3\left[\sum\limits_{i\in\mathbb{Z}_6}|P_{\QQ_i}|^2|P_{-\QQ_i}|^2\right]\\&+c_4\sum\limits_{i\in\mathbb{Z}_6}P_{\QQ_i}P_{-\QQ_i}(P^*_{\QQ_{i+1}}P^*_{-\QQ_{i+1}}+P^*_{\QQ_{i+2}}P^*_{-\QQ_{i+2}})\end{aligned}\end{equation}
with five independent parameters $c_i$ ($i=0,...,4$). Here $c_0$ is the isotropic term that does not split the $U(6)$ subspace and we choose $c_0\gg |c_i|$ ($i=1,...,4$) to ensure the positive definiteness of the GL free energy. To simplify the problem, we below consider $c_4=0$, and all the other terms ($c_{0,1,2,3}$) in the free energy only rely on the amplitude of $P_{\QQ_n}$. 

We first search for all possible local minima of the free energy $\mathcal{F}=\sum_{n}[T-T_c(\QQ_n)]|P_{\QQ_n}|^2 +\mathcal{F}^{(4)}$ by solving the GL equation $\frac{\partial \mathcal{F}}{\partial |P_{\QQ_n}|}=0$ with $n\in\mathbb{Z}_6$. The solution ans\"atze can be labelled by the vector $\mathcal{P}_{\lambda}=(P^{\lambda}_{\QQ_0},\cdots,P^{\lambda}_{\QQ_5})^T$ with $\lambda$ denoting different local minima. After obtaining all possible solutions, we can compare their minimal free energy $\mathcal{F}_{min,\lambda}$ and the lowest energy solution determines the SC ground state. Depending on different values of the $c_{i}$ parameters ($i=0, 1, ..., 4$), we find in total 9 possible inequivalent solutions for the SC ground state, denoted by 
$\lambda \in \{{\bf 1,2a,2c,3a,3d,4a,4b,5,6}\}$. In SM\cite{SM_Mandal2025} Sec.B2, we provide a complete discussion on the vector form, physical meaning and symmetry properties of all these possible SC ground states. 

\begin{figure}[!htbp]
\includegraphics[width=\linewidth]{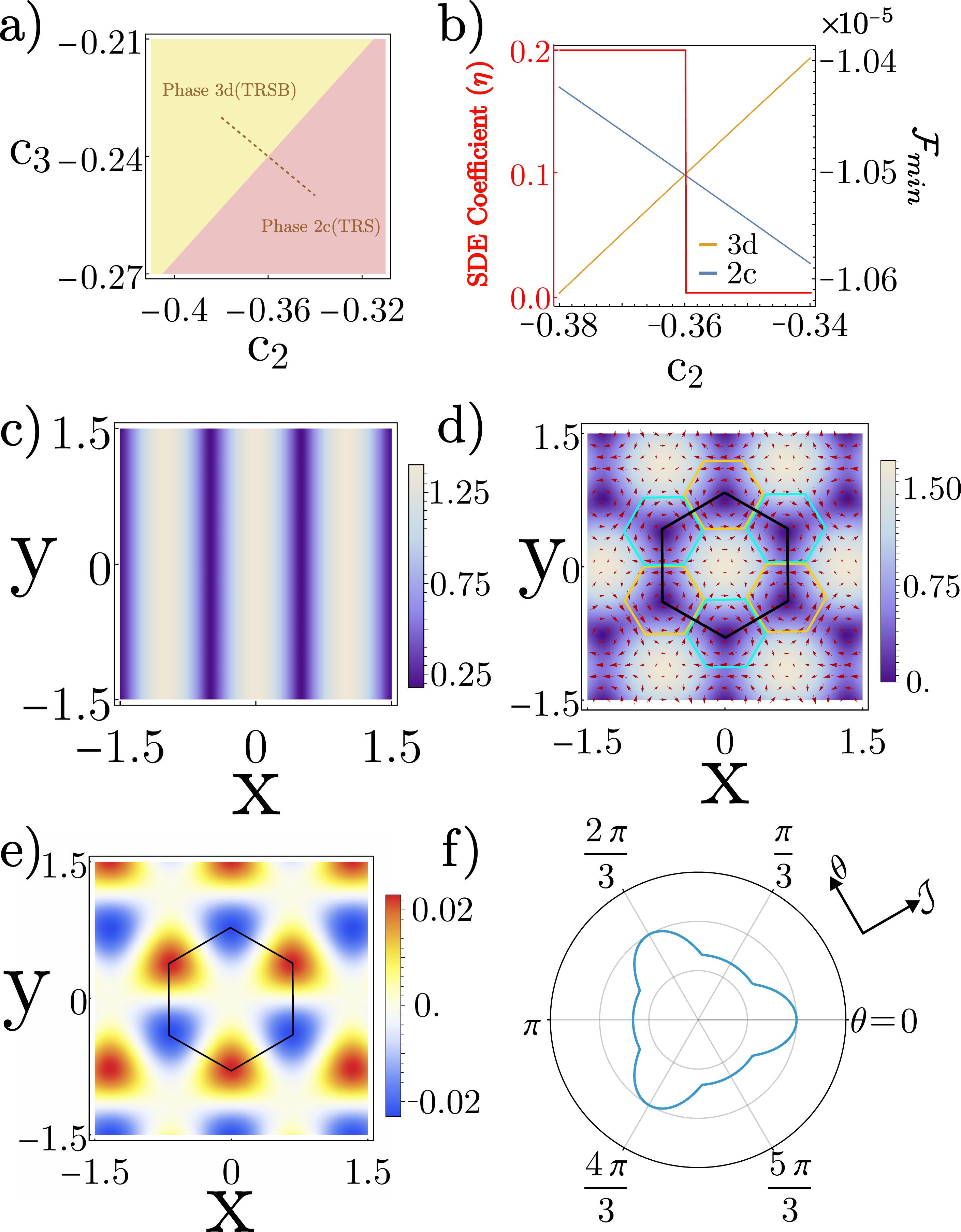}
\caption{ (a) Phase diagram showing the stripe and vortex-antivortex phases for $c_1=c_2<0,c_0=1,c_4=0$. (b) Minimal free energy $\mathcal{F}_{min}$ for {\bf phase 2c} (blue curve for $\mathcal{F}_{min,2c}$) and {\bf phase 3d} (yellow curve for $\mathcal{F}_{min,3d}$) across the dotted path in (a). Here $\mathcal{F}_{min}$ is in units of $T_{c,A}$. The red curve represents the SDE coefficient $\eta$ across the dotted path in (a). (c) and (d) represent the real space distribution $|\Psi(\rr)|$ of pair function for the {\bf Phase 2c} and {\bf 3d}, respectively. In (d), the black hexagon denotes the unit cell of vortex-antivortex lattice for {\bf Phase 3d}, the red arrows depict the direction and magnitude of local supercurents, and the cyan and orange hexagons show the regions for vortcies and antivortices, respectively. (e) Distribution of $L_z(\rr)$ for {\bf Phase 3d}. (f) Distribution of normalized critical current $\mathcal{J}(\theta)$ as a function of angle $\theta$. Here we choose $\zeta_1=0.3647$ and all other parameters are same as Fig.\ref{fig:pairmixing}(b). All in-plane coordinates $\{x,y\}$ are in units of $\pi/k_0$.
\label{fig:phasefig2}}
\end{figure}

Next we consider {\bf phase 2c} and {\bf 3d} as an example. With the choice of the parameters $c_0=1,c_1=c_2<0,c_4=0$, we find either {\bf phase 2c} or {\bf 3d} to be the SC ground state, separated by the phase boundary at $c_3=\frac{2c_2}{3}$  (See SM\cite{SM_Mandal2025} Sec.B3 for detailed derivation), as shown in Fig.\ref{fig:phasefig2}(a). We also calculate the minimal free energies $\mathcal{F}_{min,2c}$ and $\mathcal{F}_{min,3d}$ for {\bf phase 2c} and {\bf 3d}, as depicted by the blue and yellow lines in Fig.\ref{fig:phasefig2}(b), respectively, which are plotted along the dotted path shown in Fig.\ref{fig:phasefig2}(a). The energy crossing between these two phases suggests a first-order phase transition. {\bf Phase 2c} is described by the vector $\mathcal{P}_{2c}=\frac{e^{i\theta}\Delta_0}{\sqrt{2}}(e^{i\phi_1},0,0,e^{-i\phi_1},0,0)^T$, where $\Delta_0$ is the pairing strength, while $\theta$ and $\phi_1$ are overall and relative phase factors, respectively. {\bf Phase 2c} preserves $\mathcal{T}$ but breaks $C_{3z}$ and is a stripe phase with an oscillating order parameter amplitude shown in Fig.\ref{fig:phasefig2}(c). In contrast, {\bf Phase 3d} is characterized by the vector $\mathcal{P}_{3d}=\frac{e^{i\theta}\Delta_0}{\sqrt{3}}(e^{i\phi_1},0,e^{i\phi_2},0,e^{-i(\phi_1+\phi_2)},0)^T$, which preserves $C_{3z}$ but breaks $\mathcal{T}$. The order parameter of {\bf Phase 3d} forms a hexagonal lattice of vortex-antivortex pairs, thus dubbed {\em vortex-antivortex} phase\cite{agterberg2011,liangfuval}, as shown in Fig.\ref{fig:phasefig2}(d). The black hexagon in Fig.\ref{fig:phasefig2}(d) shows one unit cell of the hexagonal lattice with the order parameter amplitude dropping to a minimal value at six corners of the hexagon. The red arrows in Fig.\ref{fig:phasefig2}(d) reveal the local supercurrent flows that wind clockwise and counter-clockwise around each black hexagon corner (see the cyan and yellow small hexagons), thus leading to the formation of vortices and antivortices, respectively. The spontaneous TR breaking of {\bf Phase 3d} can be exhibited by the z-directional orbital angular momenta for the p-wave components (E-irrep) of the order parameter \cite{furusakiSpontaneousHallEffect2001,mineevbook1999,venderbos2016},
$L_z(\rr)=|\Psi_{1}(\rr)|^2-|\Psi_{3}(\rr)|^2$, 
where $|\Psi_{1,3}(\rr)|$ denotes the real space $E$ component of the order parameters, as defined in Eqs.\eqref{eq:GL2eigstate} and \eqref{eq:momgstate} (See SM\cite{SM_Mandal2025} Sec.B4). A non-zero $L_z(\rr)$ implying $\Psi_{1}(\rr)\neq e^{i\alpha}\Psi_{3}(\rr)$ breaks TR since $\Psi_{1}(\rr)$ is changed to $\Psi_{3}(\rr)$ under $\mathcal{T}$. In Fig.\ref{fig:phasefig2}(e), the peak (red color) and dip (blue color) of $L_z(\rr)$ appear at six corners of the unit cell (black hexagon), suggesting that the p-wave components (E-irrep) of order parameter emerges when the s-wave component (A-irrep) is suppressed in the vortex cores. 

Since our system contains chiral layers and thus is non-centrosymmetric, spontaneous TR breaking can potentially lead to zero-field junction-free SDE\cite{PhysRevB.98.054510,wan2024, FePtzeroB, linZerofieldSuperconductingDiode2022, revzerofield}. Thus, we evaluate critical current $J_c(\theta)$ as a function of the angle $\theta$ of the current direction. 
Fig.\ref{fig:phasefig2}(f) depicts the normalized critical current $\mathcal{J}(\theta)=J_c(\theta)/J_c(0)$ for {\bf Phase 3d}, which reveals a strong anisotropy with three-fold rotation symmetry. 
Consequently, the SDE coefficient $\eta$, defined by $\eta=\frac{1-\mathcal{J}(\pi)}{1+\mathcal{J}(\pi)}$, 
jumps from zero in {\bf Phase 2c} due to TR symmetry, to a large non-zero value in {\bf Phase 3d} across the phase transition, as depicted by the red line in Fig.\ref{fig:phasefig2}(b). Non-zero $\eta$ is also found in other TR-breaking phases, including {\bf Phase 2a, 3a, 4a \text{and} 4b}, as discussed in SM\cite{SM_Mandal2025} Sec.C. We note that the zero-field junction-free SDE has been reported in  chiral molecule intercalated TaS$_2$\cite{wan2024}, consistent with our TR-breaking phases. We further propose to distinguish different TR-breaking phases via examining the angular dependence of normalized critical current $\mathcal{J}(\theta)$ in SM\cite{SM_Mandal2025} Sec.C.

{\it Conclusion and Discussion -}
In this work, we proposed a pair-mixing mechanism for unconventional superconductivity with spontaneous TR breaking. This TR breaking SC phase is featured by a finite Cooper pair momentum and can be probed through the zero-field junction-free SDE. Although our theory is formulated for the 2D TaS$_2$ monolayer sandwiched between two chiral layers in Fig.\ref{fig:pairmixing}(a), we would expect similar physics occurs in bulk 4Hb-TaS$_2$ \cite{ribak2020} and chiral molecule intercalated TaS$_2$ hybrid superlattices \cite{wan2024} once different 1H TaS$_2$ layers in the bulk are decoupled, revealing quasi-2D SC behaviors. We also noticed that $\pi$-phase shift in Little-Parks experiments has also been observed in both TaS$_2$ systems\cite{almoalem2024,wan2024}, providing another evidence of TR breaking, similar to that in superconductor/ferromagnet hetero-junctions \cite{ryazanovr2004,PhysRevB.70.144505,buzdin2005proximity}. The TR-breaking vortex-antivortex phase has also recently been proposed in rhombohedral graphene \cite{liangfuval,han2025signatures}. However, there is one substantial difference. TR-breaking in rhombohedral graphene is from orbital magnetism and already exists above $T_c$, while TR-breaking only occurs spontaneously below $T_c$ in our pair-mixing mechanism. This pair-mixing mechanism also provides a new path for realizing TR breaking superconductivity via engineering the local chemical environment of conventional SC layers.

{\it Acknowledgment -}
We would like to acknowledge D. F. Agterberg and H. Miao for the helpful discussions. SM and CXL acknowledge the support from the NSF-MERSEC (Grant No. MERSEC DMR 2011750).

\bibliography{ref}
\onecolumngrid
\appendix

\section{Quadratic terms in Ginzburg-Landau free energy}\label{app:1}

In this section, we will discuss the pair-mixing terms between the $A$ and $E$ pairing due to non-zero momentum up to the quadratic terms of the order parameters in the Ginzburg-Landau (GL) theory. To consider the pair-mixing, we consider the form of the gap function in the momentum space as \cite{sigristuedaPhenomenologicalTheoryUnconventional1991a} 
\begin{equation}~\label{eq:gapnorder}
\Delta(\kk,\pp)=\sum\limits_{\Gamma,m}\eta_{\Gamma,m}(\kk)\hat{\Delta}_{\Gamma,m}(\pp),
\end{equation} 
where $\kk$ represents the total momentum of the Cooper pair, which describes the modulation of the gap function in the real space, while $\pp$ represents the relative momentum of the Cooper pair, which represents the orbital part of the Cooper pair wavefunction \cite{sigrist2009}. In Eq.(\ref{eq:gapnorder}), $\hat{\Delta}_{\Gamma,m}(\pp)$ represents the basis function of the gap function expansion for the pairing channel belonging to the irreducible representation (irrep) $\Gamma$ of the crystal symmetry group \cite{agterberg2009a,agterberg2020b,barzykin2002a,wang2018a}, denoted as $\mathcal{G}$, at certain high-symmetry momentum with $m$ denoting different components of the basis function. The expansion coefficient $\eta_{\Gamma,m}(\kk)$ in Eq.(\ref{eq:gapnorder}) can be viewed as the order parameter corresponding to the pairing channel of the basis function $\hat{\Delta}_{\Gamma,m}(\pp)$. One should note that $\hat{\Delta}_{\Gamma,m}(\pp)$ has a matrix form in the spin space of Cooper pairs, while $\eta_{\Gamma,m}(\kk)$ for $\Gamma=A,E$ is component of the order parameter vector
\begin{eqnarray}\label{eq_SM:def_Psi}
\Psi(\kk)=\left(\begin{matrix}\eta_{E+}(\kk)\\\eta_{A}(\kk)\\\eta_{E-}(\kk)\end{matrix}\right). 
\end{eqnarray} 
where we explicitly considered $\eta_A$ as the order parameter for the dominant s-wave pairing and $\eta_{E\pm}$ as the order parameter for the chiral p-wave pairing. As the basis wavefunction $\hat{\Delta}_{\Gamma,m}(\pp)$ belongs to the irrep $\Gamma$ of the crystal symmetry group $\mathcal{G}$, it transforms as
\begin{eqnarray}\label{eq:defrepmats}
D_\mathcal{S}\hat{\Delta}_{\Gamma,m}(\mathcal{S}^{-1}\pp)D_\mathcal{S}^T=
\sum_{n}\hat{\Delta}_{\Gamma,n}(\pp) (\mathcal{D}^{\Gamma}_\mathcal{S})_{nm}
\end{eqnarray}
where $D_\mathcal{S}$ is the transformation matrix of electron wavefunction basis for the symmetry operator $\mathcal{S} \in \mathcal{G}$, while $\mathcal{D}^{\Gamma}_{\mathcal{S}}$ represents the representation matrix of the symmetry $\mathcal{S}$ for the irrep $\Gamma$. The basis function $\hat{\Delta}_{\Gamma,m}(\pp)$ is composed of the orbital wavefunction, which is written as a polynomial of the relative momentum $\pp$ (e.g. s-wave, p-wave, $...$), and spin wavefunction, which can be represented by Pauli matrices $s$ (e.g. spin-singlet and spin-triplet). Table \ref{tab:irrepMPG6m21'} lists the possible basis function $\hat{\Delta}_{\Gamma,m}(\pp)$ for the irrep $\Gamma$ and its transformation matrices. 


\begin{table}[h]
\begin{center}
\begin{tabular}{| c | c | c | c | c | c | c |} 
\hline
Irrep($\Gamma$)&Basis functions $\hat{\Delta}_{\Gamma,m}(\pp)$ & $\mathcal{D}^{\Gamma}_E$& $\mathcal{D}^{\Gamma}_{C_{3z}}$&$\mathcal{D}^{\Gamma}_{\mathcal{M}_x}$&$\mathcal{D}^{\Gamma}_{\mathcal{I}C_{6z}}$&$\mathcal{D}^{\Gamma}_{\mathcal{T}}$\\[3pt]
\hline
\hline
$A_1'$&$is_y$&1&1&1&1&1\\[3mm]
$A_1''$&$is_x$&1&1&-1&1&1\\[3mm]
$A_2'$&$p_+(s_0-s_z)-p_-(s_0+s_z)$&1&1&-1&-1&1\\[3mm]
$A_2''$&$ip_+(s_0-s_z)+ip_-(s_0+s_z)$&1&1&1&-1&1\\[3mm]
\multirow{2}{*}{$E'$}&\footnotesize{$i\{k_+^{3n+1},k_-^{3n+1}\},\{k_-^{3n+2},k_+^{3n+2}\}$}&\multirow{2}{*}{$\tau_0$}&\multirow{2}{*}{$e^{i2\pi\tau_z/3}$}&\multirow{2}{*}{$-\tau_x$}&\multirow{2}{*}{$e^{-i2\pi\tau_z/3}$}&\multirow{2}{*}{$\tau_x$}\\
&\footnotesize{$\{p_-^{3l+2},p_+^{3l+2}\}is_y$}&&&&&\\[3mm]
$E''$&$\{p_{+}^{3l+1}(s_z+s_0),p_{-}^{3l+1}(s_z-s_0)\}$&$\tau_0$&$e^{i2\pi \tau_z/3}$&$-\tau_x$&$e^{i\pi \tau_z/3}$&$\tau_x$\\[3mm]
\hline
\end{tabular}
\end{center}
\caption{Possible basis functions for irreps ($\Gamma$) of MPG $\bar{6}m21'$ of the 1H-TaS$_2$ monolayers without chiral layer induced mirror symemtry breaking. $s,\tau$ labels the spin of Cooper pairs and the basis of the 2D $E$ irreps respectively. $p_{\pm}=p_x\pm ip_y$ is for the relative momenta of each Cooper pair, whereas $k_{\pm}=k_x\pm ik_y$ is the center-of-mass momenta. The third to last columns give the representation matrix ($\mathcal{D}_\mathcal{S}^{\Gamma}$) for a typical symmetry element ($\mathcal{S}$) from each conjugacy class of the local symmetry group. \label{tab:irrepMPG6m21'}
} 

\end{table}

\begin{table}[h]
\begin{center}
\begin{tabular}{| c | c | c | c | c | c | c | } 
\hline
MPG$\bar{6}m21'$&MPG$31'$\\
\hline\hline
$A_1'$&\\
$A_1''$&\multirow{2}{*}{$A$}\\
$A_2'$&\\
$A_2''$&\\
\hline
$E'$&\multirow{2}{*}{$^1E^2E$}\\
$E''$&\\
\hline
\end{tabular}
\end{center}
\caption{Compatibility relations between the MPG $\bar{6}m21'$ and its subgroup $31'$. 
\label{tab:compatibility}}
\end{table}

We consider two magnetic point groups (MPGs), $\bar{6}m2.1'$ and $31'$, which describe for a pure 1H-TaS$_2$ layer and the sandwich hetero-structure discussed in Fig.1(a) of the main text, respectively. The MPG $31'$ can be generated by three-fold rotation $C_{3z}$ and time reversal (TR) $\mathcal{T}$, while the MPG $\bar{6}m21'$ can be generated by $\mathcal{T}$, x-directional mirror symmetry $M_x$ and improper six-fold rotation $\mathcal{I}C_{6z}$ that combines inversion $\mathcal{I}$ and six-fold rotation $C_{6z}$. For these symmetry operations, the transformation matrices are given by 
\begin{eqnarray}\label{eq_SM:symops1_31}
&D_{C_{3z}}=e^{-i\frac{\pi}{3}s_z}; \quad D_\mathcal{T}=-is_y\mathcal{K}
\end{eqnarray}
for MPG $31'$. There are additional symmetry generators
\begin{eqnarray}\label{eq_SM:symops1_6m21}
&D_{\mathcal{I}C_{6z}}=e^{-i\frac{\pi}{6}s_z};  \quad D_{M_x}=is_x,
\end{eqnarray}
for MPG $\bar{6}m2.1'$. Here $\mathcal{K}$ is complex conjugation for the anti-unitary time reversal operations and $s$ are the Pauli matrices acting on individual spin $1/2$ basis. The transformation of the momentum $\kk$ is define as
\begin{eqnarray}\label{eq:symops1}
&\hat{C}_{3z}k_{\pm}=e^{\pm i\frac{2\pi}{3}}k_{\pm};  \quad \hat{\mathcal{I}}\hat{C}_{6z}k_{\pm}=-e^{\pm i\frac{\pi}{3}}k_{\pm}; \quad \hat{\mathcal{T}}k_{\pm}=- k_{\mp}; \quad \hat{M}_xk_{\pm}=-k_{\mp}.
\end{eqnarray}

With the above transformation rules, we can construct the representation matrix $\mathcal{D}^{\Gamma}_\mathcal{S}$ for the basis wavefunction $\hat{\Delta}_{\Gamma,m}(\pp)$ using Eq.\eqref{eq:defrepmats}, which is presented in Table\ref{tab:irrepMPG6m21'} for MPG $\bar{6}m21'$. For the MPG $31'$, the representation matrix can be constructed from the compatibility relation in Table\ref{tab:compatibility} using the MPG $\bar{6}m21'$, shown in Table I of main text. With these transformation matrices, we can classify all the basis functions for the gap functions according to the irreps of $\mathcal{G} = \bar{6}m2.1'$ or $31'$ in Table\ref{tab:irrepMPG6m21'} and \ref{tab:compatibility}. The gap function can be expanded on the basis functions in Eq.\eqref{eq:gapnorder}, from which we can derive the representation matrices for the order parameters $\eta_{\Gamma,m}(\kk)$ that belongs to the irrep $\Gamma$. 
For the unitary symmetry transformation $\mathcal{S}$ acting on the expansion \eqref{eq:gapnorder} for the general form of gap function $\Delta(\kk,\pp)$, we obtain
\begin{eqnarray}\label{eq:ordtrans}
D_\mathcal{S}\Delta(\kk,\pp)D_\mathcal{S}^T&=&
\sum\limits_{\Gamma,m}\eta_{\Gamma,m}(\kk) D_\mathcal{S}\hat{\Delta}_{\Gamma,m}(\pp)D_\mathcal{S}^T \nonumber \\
&=&\sum\limits_{\Gamma,n}\Big\{\sum\limits_{m} (\mathcal{D}_{\mathcal{S}}^{\Gamma})_{nm} \eta_{\Gamma,m}(\kk)\Big\}\hat{\Delta}_{\Gamma,n}(\mathcal{S}\pp).
\end{eqnarray}

After the symmetry transformation, we can expand the gap function $\tilde{\Delta}(\mathcal{S}\kk, \mathcal{S}\pp)=D_\mathcal{S}\Delta(\kk,\pp)D^T_\mathcal{S}$ with the basis function $\hat{\Delta}_{\Gamma,n}(\mathcal{S}\pp)$ at the total momentum $\mathcal{S}\kk$ and the expansion coefficient, denoted as $\tilde{\eta}_{\Gamma,n}(\mathcal{S}\kk)$, should be defined as
\begin{eqnarray}\label{eq:symcomparison}
\tilde{\Delta}(\mathcal{S}\kk,\mathcal{S}\pp)=\sum\limits_{\Gamma,n}\tilde{\eta}_{\Gamma,n}(\mathcal{S}\kk)\hat{\Delta}_{\Gamma,n}(\mathcal{S}\pp),
\end{eqnarray}
leading to the transformation form of $\tilde{\eta}_{\Gamma,n}$ as
\begin{eqnarray}\label{eq:orderptrans}
   \tilde{\eta}_{\Gamma,n}(\mathcal{S}\kk) = \sum\limits_{m}(\mathcal{D}_{\mathcal{S}}^{\Gamma})_{nm} \eta_{\Gamma,m}(\kk)\quad \rightarrow \quad 
   \tilde{\eta}_{\Gamma,n}(\kk) = \sum\limits_{m}(\mathcal{D}_{\mathcal{S}}^{\Gamma})_{nm} \eta_{\Gamma,m}(\mathcal{S}^{-1}\kk)
\end{eqnarray}
The representation matrices $\mathcal{D}_{\mathcal{S}}^{\Gamma}$ for all the generators in the MPG $31'$ are shown in Table I of the main text,
and those for the MPG $\bar{6}m2.1'$ are shown in Table \ref{tab:irrepMPG6m21'}. 

For any anti-unitary symmetry transformation, we can rewrite it as a conjugation followed by a unitary operation, leading to
\begin{eqnarray}D_{\mathcal{S}}\eta_{\Gamma,m}(\kk)D_{\mathcal{S}}^{-1}=\eta^*_{\Gamma,m}(\kk),\end{eqnarray}
from which we have
\begin{eqnarray}D_\mathcal{S}\Delta(\kk,\pp)D_\mathcal{S}^T&=&
\sum\limits_{\Gamma,m}\eta_{\Gamma,m}^*(\kk) D_\mathcal{S}\hat{\Delta}_{\Gamma,m}(\pp)D_\mathcal{S}^T \nonumber \\
&=&\sum\limits_{\Gamma,n}\Big\{\sum\limits_{m} (\mathcal{D}_{\mathcal{S}}^{\Gamma})_{nm} \eta_{\Gamma,m}^*(\kk)\Big\}\hat{\Delta}_{\Gamma,n}(\mathcal{S}\pp)
\end{eqnarray}
and
\begin{eqnarray}\label{eq:orderpantitrans}
   \tilde{\eta}_{\Gamma,n}(\mathcal{S}\kk) = \sum\limits_{m}(\mathcal{D}_{\mathcal{S}}^{\Gamma})_{nm} \eta^*_{\Gamma,m}(\kk)\quad \rightarrow \quad 
   \tilde{\eta}_{\Gamma,n}(\kk) = \sum\limits_{m}(\mathcal{D}_{\mathcal{S}}^{\Gamma})_{nm} \eta^*_{\Gamma,m}(\mathcal{S}^{-1}\kk).
\end{eqnarray}

From the transformations of the order parameter under symmetry $\mathcal{S}$ in Eqs.\eqref{eq:orderptrans} and \eqref{eq:orderpantitrans}, we will construct the Ginzburg-Landau free energy for the $E$ and $A$ pairing channels of the MPG $31'$. The transformation of the order parameters in Eq.(\ref{eq:orderptrans}) can be rewritten in a matrix form as
\begin{eqnarray}\label{eq:GL2symtrans}
&\tilde{\Psi}_{\alpha}(\mathcal{S}\kk)=\sum_{\beta}\mathcal{R}^{\mathcal{S}}_{\alpha\beta}\Psi_{\beta}(\kk)\nonumber\\
&\mathcal{R}^{\mathcal{S}}=\mathcal{J}\left(\begin{matrix}\mathcal{D}^{A}_{\mathcal{S}}&{\bf 0}_{1\times2}\\{\bf 0}_{2\times1}&\mathcal{D}^{E}_{\mathcal{S}}\end{matrix}\right)\mathcal{J}^T,\quad\mathcal{J}=\left(\begin{matrix}0&1&0\\1&0&0\\0&0&1\end{matrix}\right)
\end{eqnarray}
where $\mathcal{R}^{\mathcal{S}}$ is the representation matrix of the symmetry transformation $\mathcal{S}$ on the vector space of $\Psi(\kk)$ and ${\bf 0}_{a\times b}$ is a zero matrix with dimensions ($a,b$). The permutation matrix $\mathcal{J}$ re-arranges the order parameters from different channels in the form of $\Psi(\kk)$ as 
\begin{eqnarray}\label{eq:psidef2}
    \Psi(\kk)=\left(\begin{matrix}\eta_{E+}(\kk)\\\eta_{A}(\kk)\\\eta_{E-}(\kk)\end{matrix}\right)=\mathcal{J}\left(\begin{matrix}\eta_{A}(\kk)\\\eta_{E+}(\kk)\\\eta_{E-}(\kk)\end{matrix}\right).
\end{eqnarray}
For anti-unitary transformations such as $\mathcal{T}$ symmetry, the extra conjugation operation leads to  
\begin{eqnarray}\tilde{\Psi}_{\alpha}(\mathcal{S}\kk)=\sum_{\beta}\mathcal{R}^{\mathcal{S}}_{\alpha\beta}\Psi^*_{\beta}(\kk). \end{eqnarray}The representation matrices for MPG $31'$ are given by 
\begin{eqnarray}\mathcal{R}^{C_{3z}}=\left(\begin{matrix}e^{i\frac{2\pi}{3}}&0&0\\0&1&0\\0&0&e^{-i\frac{2\pi}{3}}\end{matrix}\right);\quad\mathcal{R}^{\mathcal{T}}=\left(\begin{matrix}0&0&1\\0&1&0\\1&0&0\end{matrix}\right)\end{eqnarray}
for $C_{3z}$ and $\mathcal{T}$ symmetry, respectively. The representation matrices for the remaining generators of MPG $\bar{6}m21'$ are defined as
\begin{eqnarray}\mathcal{R}^{M_x}=\left(\begin{matrix}0&0&-1\\0&1&0\\-1&0&0\end{matrix}\right);\quad\mathcal{R}^{\mathcal{I}C_{6z}}=\left(\begin{matrix}e^{i\frac{\pi}{3}}&0&0\\0&1&0\\0&0&e^{-i\frac{\pi}{3}}\end{matrix}\right)\end{eqnarray}
for $M_{x}$ and $\mathcal{I}C_{6z}$ symmetry, respectively. We note that the out-of-plane $\mathcal{M}_z$ symmetry can be generated from $\mathcal{I}C_{6z}$ by \begin{eqnarray}\mathcal{M}_z=(\mathcal{I}C_{6z})^3,\end{eqnarray}
and the corresponding representation matrix of $\mathcal{M}_z$ is
\begin{eqnarray}\mathcal{R}^{M_z}=(\mathcal{R}^{\mathcal{I}C_{6z}})^3=\left(\begin{matrix}-1&0&0\\0&1&0\\0&0&-1\end{matrix}\right). 
\end{eqnarray}From the symmetry property of the order parameter $\Psi(\kk)$, we next construct the phenomenological Ginzburg-Landau (GL) free energy given by
\begin{eqnarray}\label{eq_SM:GLfreeenergy_1}
    &\mathcal{F}=\mathcal{F}^{(2)}+\mathcal{F}^{(4)}.
\end{eqnarray}
where $\mathcal{F}^{(2)}$ is the term in the second order of order parameters $\eta_{\Gamma,m}(\kk)$ and $\mathcal{F}^{(4)}$ is the fourth-order term. We first consider the second-order term, written as
\begin{eqnarray}\label{eq:compactGL2_1}
&\mathcal{F}^{(2)}=\sum\limits_{\kk,\alpha.\beta}\Psi_{\alpha}^*(\kk)\mathcal{H}_{\alpha\beta}(\kk)\Psi_{\beta}(\kk),
\end{eqnarray}
where $\alpha, \beta$ labels the different components of the order parameter $\Psi(\kk)$ in Eq.\eqref{eq_SM:def_Psi}, $T$ is temperature, $\mathbb{I}$ is a 3-by-3 identity matrix, $\mathcal{H}$ is the momentum dependent matrix of the generalized Hamiltonian. 
To derive the form of $\mathcal{H}_{\alpha\beta}(\kk)$, we first consider the unitary symmetry transformation $\mathcal{S}$, under which the free energy that should be invariant. We may consider the free energy after the transformation of the basis $\Psi(\kk)$ in Eq.\eqref{eq:GL2symtrans}, given by
\begin{eqnarray}\label{eq:GL2unitrans}
\mathcal{F}^{(2)}&&=\sum\limits_{\kk,\alpha.\beta}\tilde{\Psi}_{\alpha}^*(\kk)\mathcal{H}_{\alpha\beta}(\kk)\tilde{\Psi}_{\beta}(\kk)\nonumber\\&&=\sum\limits_{\substack{\kk,\alpha,\beta\\\alpha',\beta'}}\Psi_{\alpha'}^*(\mathcal{S}^{-1}\kk)(\mathcal{R}^{\mathcal{S}}_{\alpha\alpha'})^*\mathcal{H}_{\alpha\beta}(\kk)\mathcal{R}^{\mathcal{S}}_{\beta\beta'}\Psi_{\beta}(\mathcal{S}^{-1}\kk)\nonumber\\&&=\sum\limits_{\substack{\kk,\alpha,\beta\\\alpha',\beta'}}\Psi_{\alpha'}^*(\kk)(\mathcal{R}^{\mathcal{S}})^{\dagger}_{\alpha'\alpha}\mathcal{H}_{\alpha\beta}(\mathcal{S}\kk)\mathcal{R}^{\mathcal{S}}_{\beta\beta'}\Psi_{\beta}(\kk),
\end{eqnarray}
where we have used Eq.\eqref{eq:GL2symtrans} in the second step. The invariance of the free energy under symmetry transformation requires 
\begin{eqnarray}\label{eq:unihamtrans}
\mathcal{H}_{\alpha'\beta'}(\kk)=\sum\limits_{\alpha,\beta}(\mathcal{R}^{\mathcal{S}})^{\dagger}_{\alpha'\alpha}\mathcal{H}_{\alpha\beta}(\mathcal{S}\kk)\mathcal{R}^{\mathcal{S}}_{\beta\beta'},
\end{eqnarray}
which provides the symmetry constraint on the form of the matrix $\mathcal{H}(\kk)$. Similarly, for anti-unitary TR symmetry, we have
\begin{eqnarray}\label{eq:GL2antiunitrans}
\mathcal{F}^{(2)}&&=\sum\limits_{k,\alpha.\beta}\tilde{\Psi}_{\alpha}^*(\kk)\mathcal{H}_{\alpha\beta}(\kk)\tilde{\Psi}_{\beta}(\kk)\nonumber\\&&=\sum\limits_{\substack{k,\alpha,\beta\\\alpha',\beta'}}\Psi_{\alpha'}(-\kk)(\mathcal{R}^{\mathcal{T}}_{\alpha\alpha'})^*\mathcal{H}_{\alpha\beta}(\kk)\mathcal{R}^{\mathcal{T}}_{\beta\beta'}\Psi^*_{\beta'}(-\kk)\nonumber\\&&=\sum\limits_{\substack{k,\alpha,\beta\\\alpha',\beta'}}\Psi^*_{\beta'}(\kk)(\mathcal{R}^{\mathcal{T}})^{\dagger}_{\alpha'\alpha}\mathcal{H}_{\alpha\beta}(-\kk)\mathcal{R}^{\mathcal{T}}_{\beta\beta'}\Psi_{\alpha'}(\kk),
\end{eqnarray}
which leads to 
\begin{eqnarray}\label{eq:aunihamtrans}
\mathcal{H}_{\beta'\alpha'}(\kk)=\sum\limits_{\alpha,\beta}(\mathcal{R}^{\mathcal{T}})^{\dagger}_{\alpha'\alpha}\mathcal{H}_{\alpha\beta}(-\kk)\mathcal{R}^{\mathcal{T}}_{\beta\beta'}.
\end{eqnarray}In addition to the symmetry constraints in Eqs.\eqref{eq:unihamtrans} and \eqref{eq:aunihamtrans}, the matrix $\mathcal{H}(\kk)$ also demands hermicity,
\begin{eqnarray}\label{eq_SM:hermicity_M}
    \mathcal{H}(\kk) = \mathcal{H}^\dagger(\kk),
\end{eqnarray}
as the free energy $\mathcal{F}$ is real. Based on the above symmetry constraints from MPG 31$'$, the GL free energy up to the second-order terms in order parameter is given by
\begin{equation}\begin{aligned}~\label{eq:freen2app}
&\mathcal{F}^{(2)}=\mathcal{F}^{(2)}_{A}+\mathcal{F}^{(2)}_{E}+\mathcal{F}_{A-E}\\
&\mathcal{F}^{(2)}_{A}=\sum\limits_{k}(\alpha_{A}+\gamma_{A}k^2)|\eta_{A}(\kk)|^2\\
&\mathcal{F}_{E}^{(2)}=\sum\limits_{\kk}(\alpha_{E}+\gamma_{E}k^2)(|\eta_{E,+}(\kk)|^2+|\eta_{E,-}(\kk)|^2)\\
&\mathcal{F}_{A-E}=\sum\limits_{\kk}\eta_{A}^*(i\zeta_1^* k_-+\zeta^*_2 k_+^2+\zeta^*_4k_+^4)\eta_{E,+} +\eta_{A}^*(i\zeta_1 k_++\zeta_2 k_-^2+\zeta_4k_-^4)\eta_{E,-}+h.c.
\end{aligned}\end{equation}
where $T_{c,\Gamma}$ are the critical temperature of the $\Gamma=A,E$ irrep channels without any pair-mixing. The GL coefficients 
\begin{eqnarray}
    \alpha_{\Gamma}=\tilde{\alpha}_{\Gamma}(T-T_{c,\Gamma})
\end{eqnarray}
and $\gamma_{\Gamma}$ represent constant and quadratic in center of mass momentum $\mathcal{O}(k^2)$ terms, respectively ($\Gamma=A,E$). The parameters $\zeta_n$ represent the strength of the pair mixing between $A$ and $E$ irrep channels, coupled with $n^{th}$ power of the center of mass momentum $\kk$. All these parameters are material dependent. To rewrite the free energy $\mathcal{F}^{(2)}$ in a compact form, we rescale the coefficients as $\gamma'_{\Gamma}=\gamma_{\Gamma}/\tilde{\alpha}_{\Gamma}, \zeta'_{1,2,4}=\zeta_{1,2,4}/\sqrt{\tilde{\alpha}_{A}\tilde{\alpha}_{E}}$, where $\Gamma=A,E$ are the different irrep channels. Furthermore, we rescale the order parameter vector components as $\eta'_{\Gamma}(\kk)=\frac{\eta_{\Gamma}(\kk)}{\sqrt{\tilde{\alpha}_{\Gamma}}}$ . After the regularization, we rewrite the free energy in Eq.\eqref{eq:compactGL2_1} as
\begin{eqnarray}\label{eq:compactGL2_2}
&\mathcal{F}^{(2)}=\sum\limits_{\kk,\alpha.\beta}\Psi_{\alpha}^{'*}(\kk)\{T \mathbb{I}-M_{\alpha\beta}(\kk)\}\Psi'_{\beta}(\kk)
\end{eqnarray}
with
\begin{eqnarray}\label{eq:compactGL2}
&M(\kk)=\begin{pmatrix}T_{c,E}-\gamma'_{E}k^2&i\zeta_1^{\prime}k_+-\zeta_2^{\prime}k_-^2-\zeta_4^{\prime}k_+^4&0\\-i\zeta_1^{\prime*}k_--\zeta_2^{\prime*}k_+^2-\zeta_4^{\prime*}k_-^4&T_{c,A}-\gamma'_{A}k^2&-i\zeta_1' k_+-\zeta'_2 k_-^2-\zeta'_4 k_+^4\\0&i\zeta_1^{\prime*} k_--\zeta^{\prime*}_2 k_+^2-\zeta^{\prime*}_4 k_-^4&T_{c,E}-\gamma'_{E}k^2\end{pmatrix}. 
\end{eqnarray}
Below we always renamed the rescaled order parameter vector $\Psi'(\kk)$ in \eqref{eq:compactGL2_2} as $\Psi(\kk)$. The matrix $M(\kk)$ also respects all the symmetry constraints of MPG 31$'$ in Eqs.\eqref{eq:unihamtrans}, \eqref{eq:aunihamtrans} and \eqref{eq_SM:hermicity_M}.
For the MPG $\bar{6}m21'$ with extra symmetry operations, including x-directional mirror symmetry $\mathcal{M}_x$ and the combined symmetry $\mathcal{I}C_{6z}$ symmetry with $\mathcal{I}$ to be inversion and $C_{6z}$ to be six-fold rotation and z-directional mirror symmetry  $\mathcal{M}_z$, we can implement a similar symmetry construction of the GL free energy as
\begin{eqnarray}\label{eq:compactGL2freen6m21}
&\mathcal{F}^{(2)}_{\bar{6}m21'}=\sum\limits_{\kk,\alpha.\beta}\Psi_{\alpha}^*(\kk)\{T \mathbb{I}-M^{\bar{6}m21'}_{\alpha\beta}(\kk)\}\Psi_{\beta}(\kk).
\end{eqnarray}
Since we focus on pair mixing between s-wave ($A_1'$) and p-wave($E''$), the order parameter basis $\Psi(\kk)$ is written as
\begin{eqnarray}
    \Psi(\kk)=\left(\begin{matrix}\eta_{E''+}(\kk)\\\eta_{A_1}(\kk)\\\eta_{E''-}(\kk)\end{matrix}\right).
\end{eqnarray}
In this basis, with all the regularization as before, we find the $M^{\bar{6}m21'}(\kk)$ matrix that respects all the symmetry constraints Eq. \eqref{eq:unihamtrans} and \eqref{eq:aunihamtrans} for MPG $\bar{6}$m21' is given by 
\begin{eqnarray}\label{eq:compactGL26m21}
&M^{\bar{6}m21'}(\kk)=\begin{pmatrix}T_{c,E}-\gamma'_{E}k^2&0&0\\0&T_{c,A}-\gamma'_{A}k^2&0\\0&0&T_{c,E}-\gamma'_{E}k^2\end{pmatrix}.
\end{eqnarray} Comparing Eq.(\ref{eq:compactGL2}) with (\ref{eq:compactGL26m21}), we find that the off-diagonal pair mixing terms disappear in MPG $\bar{6}$m21'. Therefore the s-wave and p-wave pair mixing terms are limited to the MPG 31$'$.

For the derivation below, we will only focus on the GL free energy (\ref{eq:compactGL2}) and (\ref{eq:freen2app}) for the MPG $31'$.  We next consider the linearized GL equation and thus drop the fourth-order term $\mathcal{F}^{(4)}$ in this section. The linearized GL equation can be derived as
\begin{eqnarray}\label{eq_SM:Tc_eigenequation1}
    \frac{\partial\mathcal{F}^{(2)}}{\partial \Psi^{\dagger}(\kk)}=0\Rightarrow M(\kk)\Psi(\kk)=T\Psi(\kk),
\end{eqnarray}
where $M(\kk)$ is given in Eq.(\ref{eq:compactGL2}). 
We can numerically solve this eigen-equation problem of $M(\kk)$ matrix and obtain three eigen-values, denoted as $T_\lambda(\kk)$ with $\lambda=1,2,3$. We may arrange these eigen-values in descending orders, $T_1(\kk) > T_2(\kk) > T_3(\kk)$, and the critical temperature can be obtained by maximizing the function $T_1(\kk)$ with respect to $\kk$, i.e.
\begin{eqnarray}
    T_c = \max\limits_{\kk}(T_1(\kk)).
\end{eqnarray}

To gain more analytical understanding, we can consider solving the eigen-equation problem perturbatively. We separate the $M(\kk)$ matrix into two parts, 
\begin{eqnarray}M(\kk)=M^{(0)}(\kk)+M^{(1)}(\kk)\end{eqnarray}
where $M^{(0)}(\kk)$ has the full-rotation symmetry and is given by
\begin{eqnarray}\label{eq_SM:M0kk}
&&M^{(0)}(\kk)=\left(\begin{matrix}T_{c,E}-\gamma'_{E}k^2&i\zeta_1^{\prime}k_+&0\\-i\zeta_1^{\prime*}k_-&T_{c,A}-\gamma'_{A}k^2&-i\zeta_1' k_+\\0&i\zeta_1^{\prime*} k_-&T_{c,E}-\gamma'_{E}k^2\end{matrix}\right)\end{eqnarray}
while $M^{(1)}(\kk)$ is the anisotropic term, 
\begin{eqnarray}\label{eq_SM:M1kk}
&&M^{(1)}(\kk)=\left(\begin{matrix}0&-\zeta_2^{\prime}k_-^2-\zeta_4'k_+^4&0\\-\zeta_2^{\prime*}k_+^2-\zeta_4^{\prime*}k_-^4&0&-\zeta'_2 k_-^2-\zeta_4^{\prime}k_+^4\\0&-\zeta^{\prime*}_2 k_+^2-\zeta_4^{\prime*}k_-^4&0\end{matrix}\right).
\end{eqnarray}We first consider the eigen-equation problem for $M^{(0)}(\kk)$, 
\begin{eqnarray}\label{eq:linGLeqSM}
    M^{(0)}(\kk) \Psi^{(0)}(\kk) = T^{(0)}(\kk) \Psi^{(0)}(\kk),
\end{eqnarray}
which can be solved analytically. It has three eigenvalues,
\begin{eqnarray}\label{eq:GL2eigvals}
    &&T^{(0)}_{\nu=\pm 1}(\kk)=\frac{1}{2}\Bigg[T_{c,A}+T_{c,E}-k^2(\gamma'_A+\gamma'_E)+\nu \sqrt{\Big\{ (T_{c,A} - T_{c,E}) + (\gamma'_{E} - \gamma'_{A}) k^2 \Big\}^2 + 8k^2 |\zeta_1'|^2}\Bigg]\nonumber\\
    &&T^{(0)}_{\nu=0}(\kk)=T_{c,E}-\gamma'_Ek^2.
\end{eqnarray}
The corresponding wavefunctions are given by
\begin{eqnarray}\label{eq_SM:GL2eigstate}
    \Psi_{\nu=0,\pm 1}^{(0)}(\kk) = \frac{1}{\sqrt{N_{i=0,\pm 1}(\kk)}} \begin{pmatrix}
i\zeta'_1k_+\\
T_{\nu=0,\pm 1}^{(0)}(k)-T_{c,E}+\gamma_{E}k^2 \\
i\zeta_1^{\prime*}k_-
\end{pmatrix},\nonumber \\
\end{eqnarray} where $N_{\nu=0,\pm1}$ are the normalization coefficient for the eigenstates. We assume $T_{c,\Gamma},\gamma'_{\Gamma}$ for $\Gamma=A,E$ to be positive numbers and $T_{c,A}-T_{c,E}>(\gamma'_{A}-\gamma'_{E})k^2$ for all relevant and smaller scales of magnitude of momenta, i.e.-- when $k\in[0,1]$.

From these three eigen-values in Eq.\eqref{eq:GL2eigvals}, we pick the largest one ($\lambda=+1$), denoted as $T_c^{(0)}(k)$,  
\begin{eqnarray}\label{eq_SM:Tc0kk}
    T_c^{(0)}(k)=T_1^{(0)}(\kk)
\end{eqnarray}
and the corresponding eigen-state wavefunction is denoted as 
\begin{eqnarray}\label{eq:critwavefn}\Psi^{(0)}_c(\kk)=\Psi_{1}^{(0)}(\kk)\end{eqnarray} 
It should be noted that $T_c^{(0)}(k)$ only depend on $k=|\kk|$ since $M^{(0)}(\kk)$ is isotropic. Consequently, the maximum of $T_c^{(0)}(k)$ only depend on the momentum amplitude $k$, but not momentum angle $\theta$ ($\kk = k (\cos\theta, \sin\theta)$), due to the full rotation symmetry. 

To break the full rotation symmetry, we consider $M^{(1)}(\kk)$ in Eq.(\ref{eq_SM:M1kk}) perturbatively, and the first order correction to the eigen-value is given by 
\begin{eqnarray}
&&\Delta T_c^{(1)}(\kk) = \{\Psi^{(0)}_{c}(\kk)\}^{\dagger}M^{(1)}(\kk)\Psi^{(0)}_{c}(\kk)
\end{eqnarray}
with the eigen-state $\Psi^{(0)}_{c}(\kk)$ for the largest eigen-value $T_c^{(0)}(k)$ of $M^{(0)}(\kk)$. 
With the expression of $\Psi^{(0)}_{c}(\kk)$ in Eq. (4) in main text or Eq.\eqref{eq:critwavefn}, we obtain the correction to be
\begin{eqnarray}\label{eq_SM:Tc1kk}\Delta T_c^{(1)}(\kk)=&&0.\end{eqnarray} Thus, we calculate the second order correction,
\begin{eqnarray}
&&\Delta T_c^{(2)}(\kk) = \frac{|(\Psi^{(0)}_{0})^{\dagger}(\kk)M^{(1)}(\kk)\Psi^{(0)}_{c}(\kk)|^2}{T^{(0)}_c(\kk)-T^{(0)}_0(\kk)}+\frac{|(\Psi^{(0)}_{-1})^{\dagger}(\kk)M^{(1)}(\kk)\Psi^{(0)}_{c}(\kk)|^2}{T^{(0)}_c(\kk)-T^{(0)}_{-1}(\kk)}.
\end{eqnarray}
Using the eigenstates in \eqref{eq_SM:GL2eigstate}, we find this perturbation to be
\begin{eqnarray}\label{eq_SM:Tc2kk}\Delta T_c^{(2)}(\kk)\approx\frac{2k^4\Big\{|\zeta'_2|^2+k^4|\zeta'_4|^2+2k^2\Re(e^{-i6\theta}\zeta'_2\zeta^{\prime *}_4)\Big\}}{\sqrt{\Big\{T_{c,A}-T_{c,E}+k^2(\gamma'_{E}-\gamma'_{A})\Big\}^2+8k^2|\zeta'_1|^2}}.\end{eqnarray} With the results in Eqs.(\ref{eq_SM:Tc0kk}), (\ref{eq_SM:Tc1kk}) and (\ref{eq_SM:Tc2kk}), the critical temperature is given by 
\begin{eqnarray}~\label{eq:Tsolnapp}
T_c = \text{max}(T_c(\kk)); \quad T_c(\kk) = T_c^{(0)}(k)+\Delta T_c^{(1)}(\kk)+\Delta T_c^{(2)}(\kk).
\end{eqnarray} The numerical and analytical solutions of $T_c(\kk)$ are discussed in Fig. 1(b) and 1(c) of the main text. To get $T_c$, we need to maximize $T_c(\kk)$ with respect to $\kk$, which can be achieved in two steps. First, we maximize the isotropic part of $T_c(\kk)$, i.e. $T_c^{(0}(\kk)$, with respect to magnitude of momenta $|\kk|=k_0$. As $T_c^{(0}(\kk)$ is isotropic, we obtain a degenerate subspace of momenta space with full rotational symmetry, characterized by the angle $\theta$ of the momentum, for the highest $T_c$. Second, the non-zero second order correction $\Delta T_c^{(2)}(\kk)$ remove this degeneracy and pick six values of the angle $\theta$ for the maximal $T_c$. Thus, the symmetry of the degenerate subspace is reduced from continuous rotational symmetry down to six momenta that are related by three fold-rotation and time reversal symmetry. 

The direction of these momenta will depend on the values of the pair-mixing coupling parameters $\zeta'_{2,4}$. For example, if we assume $\zeta'_{2,4}$ are real or pure imaginary while $\zeta'_2\zeta^{\prime *}_4>0$, these momenta can be denoted by the set
\begin{eqnarray}\label{eq:sixpointsset}\mathcal{Q}_6 = \{\QQ_j=k_0\cos(\theta_{\QQ_j})\hat{e}_x+k_0\sin(\theta_{\QQ_j})\hat{e}_y;  \theta_{\QQ_j}=\frac{j\pi}{3}; \quad \forall j \in \mathbb{Z}_6\},\end{eqnarray}
as depicted in Fig.1(c) in main text, and the corresponding critical temperature is denoted as 
\begin{eqnarray} \label{eq_SM:Tc_quadraticTerm}
    T_{c,0} = T_c(\kk = \QQ_j); \quad \QQ_j \in \mathcal{Q}_6 
\end{eqnarray}
where $T_c(\kk)$ is given in Eq.(\ref{eq:Tsolnapp}) and we can choose any $j$ value for the same $T_{c,0}$.

\section{Fourth Order terms and Time reversal breaking Phase}~\label{app:2}
\subsection{Symmetry construction of the fourth-order terms}
The GL free energy with the quadratic terms of order parameters can be solved by the eigen-problem of the linearized GL equation \eqref{eq:linGLeqSM} and the relevant eigenvalue for the maximal value of $T(\kk)$ and the corresponding eigenstates are given in Eq.(3)(expanded in Eq.\eqref{eq_SM:Tc0kk} and Eq.\eqref{eq_SM:Tc1kk}) and Eq.(4) in main text, respectively. We note that when the parameter $\zeta'$ is smaller than a critical value $\zeta'_{1c}$ ($\zeta'_1< \zeta'_{1c}$), the largest value of $T(\kk)$ occurs at $\kk = 0$, corresponding to the zero-momentum pairing in the $A$ irrep channel. However, when $\zeta' > \zeta'_{1c}$, the highest $T(\kk)$ occurs at non-zero $\kk$, given by the momenta set in Eq.\eqref{eq:sixpointsset}. These six momenta $\QQ_j$ are related to each other by $C_{3z}$ and TR symmetry, so the eigen-states $\Psi^{(0)}_c(\QQ_j)$ at these six momenta are degenerate and form a $U(6)$ subspace. The general form of the ground state should be a linear superposition of all $\Psi^{(0)}_c(\QQ_j)$ in Eq.(5) of the main text. At the quadratic order, any linear superposition should share the same energy, but this degeneracy is expected to be split by the fourth-order terms of the order parameters. Thus, we will construct all the fourth-order terms of the GL free energy within the $U(6)$ subspace in this section. 


We consider the order parameter $\Psi$ in the real space, which is related to $\Psi(\kk)$ by the Fourier transform
\begin{eqnarray} \label{eq_SM:Fourier_Psi}
&\Psi_\alpha(\rr)=\frac{1}{V}\sum\limits_{\kk}\Psi_\alpha(\kk)e^{i\kk\cdot\rr}, 
\end{eqnarray}
where $\alpha$ labels the order parameter components, i.e.-- $\Psi_{1}(\kk)=\eta_{E+}(\kk),\Psi_{2}(\kk)=\eta_{A}(\kk),\Psi_{3}(\kk)=\eta_{E-}(\kk)$. Under the unitary symmetry transformation $\mathcal{S}$, the real space order parameter $\Psi_{\alpha}(\rr)$ is transformed as
\begin{eqnarray}\label{eq:realeigtransuni}
\tilde{\Psi}_\alpha(\rr)&=&\frac{1}{V}\sum\limits_{\kk}\tilde{\Psi}_\alpha(\kk)e^{i\kk\cdot\rr} = \frac{1}{V}\sum\limits_{\kk,\beta}\mathcal{R}^{\mathcal{S}}_{\alpha\beta}\Psi_{\beta}(\mathcal{S}^{-1}\kk) e^{i\kk\cdot\rr} \nonumber \\
&=& \frac{1}{V} \sum\limits_{\kk',\beta}\mathcal{R}^{\mathcal{S}}_{\alpha\beta}\Psi_{\beta}(\kk') e^{i \mathcal{S}\kk'\cdot \rr} = \frac{1}{V} \sum\limits_{\kk',\beta}\mathcal{R}^{\mathcal{S}}_{\alpha\beta}\Psi_{\beta}(\kk') e^{i \kk'\cdot \mathcal{S}^{-1}\rr} \nonumber\\
&=& \sum_\beta \mathcal{R}^{\mathcal{S}}_{\alpha\beta} \Psi_\beta(\mathcal{S}^{-1}\rr)
\end{eqnarray}
where $V$ is the volumn of the system and we have used Eq.\eqref{eq:GL2symtrans} in the second step, $\kk' = \mathcal{S}^{-1} \kk$ in the third step, and the definition of Fourier transform in the last step. 
For anti-unitary symmetry $\mathcal{T}$, this becomes

\begin{eqnarray}\label{eq:realeigtransantiuni}
\tilde{\Psi}_\alpha(\rr)&=&\frac{1}{V}\sum\limits_{\kk}\tilde{\Psi}_\alpha(\kk)e^{i\kk\cdot\rr} = \frac{1}{V}\sum\limits_{\kk,\beta}\mathcal{R}^{\mathcal{T}}_{\alpha\beta}\Psi^*_{\beta}(-\kk) e^{i\kk\cdot\rr} \nonumber \\
&=& \frac{1}{V} \sum\limits_{\kk',\beta}\mathcal{R}^{\mathcal{T}}_{\alpha\beta}\Psi^*_{\beta}(\kk') e^{i (-\kk')\cdot \rr} = \frac{1}{V} \sum\limits_{\kk',\beta}\mathcal{R}^{\mathcal{T}}_{\alpha\beta}\Psi^*_{\beta}(\kk') e^{-i \kk'\cdot \rr} \nonumber\\
&=& \sum_\beta \mathcal{R}^{\mathcal{T}}_{\alpha\beta} \Psi^*_\beta(\rr)
\end{eqnarray} where  $\kk'=-\kk$ in the third equality. As we have described above, the quadratic terms in the GL free energy will give the maximal $T_c$ for degenerate eigen-states $\Psi^{(0)}_c(\QQ_j)$ located at these six momenta ($\QQ_j\in\mathcal{Q}_6$) and we hope to consider the solution within this $U(6)$ subspace, so we consider the wavefunction ansatz
\begin{eqnarray}\label{eq:fteigvec}
\Psi_\alpha(\kk)=\sum\limits_{\QQ_j\in\mathcal{Q}_6}\delta(\kk-\QQ_j) P_{\QQ_j}\Psi_{c,\alpha}^{(0)}(\QQ_j)
\end{eqnarray}
where $P_{\QQ_j}$ is a superposition coefficient that accounts for the contribution from $\Psi_{c,\alpha}^{(0)}(\QQ_j)$, given in Eq.\eqref{eq:critwavefn}, at different momenta $\QQ_j$ in the ground state. Substituting Eq.(\ref{eq:fteigvec}) into the Fourier transform in Eq.(\ref{eq_SM:Fourier_Psi}) leads to 
\begin{eqnarray}\label{eq:realeigvec}
\Psi_\alpha(\rr)=\frac{1}{V}\sum\limits_{\QQ_j\in\mathcal{Q}_6} P_{\QQ_j}\Psi_{c,\alpha}^{(0)}(\QQ_j)e^{i\QQ_j\cdot\rr}. 
\end{eqnarray}
Eq.(\ref{eq:realeigvec}) can be viewed as the mode expansion of $\Psi(\rr)$ on the eigen-state $\Psi^{(0)}(\QQ_j)$ with the expansion parameter $P_{\QQ_j}$. We can introduce a vector \begin{eqnarray}\label{eq:Pvector}
    \mathcal{P}=(P_{\QQ_0},\cdots,P_{\QQ_5})^T, 
\end{eqnarray}
to describe the order parameter in this $U(6)$ space, and hope to figure out the transformation form of the vector $\mathcal{P}$ under the symmetry operations in the MPG $31'$ that is generated by $C_{3z}$ and $\mathcal{T}$, namely
\begin{eqnarray}\hat{\mathcal{C}}_{3z}\QQ_j=\QQ_{j+2}; \quad \hat{\mathcal{T}}\QQ_j=-\QQ_j=\QQ_{j+3}\end{eqnarray}
where $\QQ_j=\QQ_{j+6}$ from Eq.\eqref{eq:sixpointsset}. We consider the symmetry transformation of $\Psi_\alpha(\rr)$ in Eq.\eqref{eq:realeigtransuni} and apply the expansion ansatz in Eq.(\ref{eq:realeigvec}) to derive
\begin{eqnarray}\label{eq:unirealsteps}
\tilde{\Psi}_\alpha(\rr) &&=\sum_\beta\mathcal{R}_{\alpha\beta}^{\mathcal{S}}\Psi_\beta(\mathcal{S}^{-1}\rr)\nonumber \\
&&=\frac{1}{V}\sum\limits_{\QQ_j\in\mathcal{Q}_6,\beta}\mathcal{R}_{\alpha\beta}^{\mathcal{S}}P_{\QQ_j}\Psi^{(0)}_{c,\beta}(\QQ_j)e^{i\QQ_j\cdot\mathcal{S}^{-1}\rr}\nonumber \\
&&=\frac{1}{V}\sum\limits_{\QQ_j\in\mathcal{Q}_6}P_{\QQ_j}\tilde{\Psi}_{c,\alpha}^{(0)}(\mathcal{S}\QQ_j)e^{i\QQ_j\cdot\mathcal{S}^{-1}\rr}\nonumber\\
&&=\frac{1}{V}\sum\limits_{\tilde{\QQ}_j=\mathcal{S}\QQ_j}P_{\mathcal{S}^{-1}\tilde{\QQ}_j}\tilde{\Psi}_{c,\alpha}^{(0)}(\tilde{\QQ}_j)e^{i(\mathcal{S}^{-1}\tilde{\QQ}_j)\cdot(\mathcal{S}^{-1}\rr)}\nonumber \\
&&=\frac{1}{V}\sum\limits_{\tilde{\QQ}_j\in\mathcal{Q}_6}P_{\mathcal{S}^{-1}\tilde{\QQ}_j}\tilde{\Psi}_{c,\alpha}^{(0)}(\tilde{\QQ}_j)e^{i\tilde{\QQ}_j\cdot\rr},
\end{eqnarray}
where we have used Eq.\eqref{eq:GL2symtrans} at the third equality. By definition, we should have
\begin{eqnarray}\label{eq:realtransideal}\tilde{\Psi}(\rr)=\frac{1}{V}\sum\limits_{\QQ_j\in\mathcal{Q}_6}\tilde{P}_{\QQ_j}\tilde{\Psi}_c^{(0)}(\QQ_j)e^{i\QQ_j\cdot\rr}\end{eqnarray}
after symmetry transformation. Thus, we find 
\begin{eqnarray}\label{eq_SM:symmetry_P_unitary}
    \tilde{P}_{\QQ_j} = P_{\mathcal{S}^{-1}\QQ_j} 
\end{eqnarray}
for the unitary symmetry transformation $\mathcal{S}$. For the anti-unitary time-reversal transformation 

\begin{eqnarray}\label{eq:antiunirealsteps}
\tilde{\Psi}_\alpha(\rr) &&=\sum_\beta\mathcal{R}_{\alpha\beta}^{\mathcal{T}}\Psi^*_\beta(\rr)\nonumber \\
&&=\frac{1}{V}\sum\limits_{\QQ_j\in\mathcal{Q}_6,\beta}\mathcal{R}_{\alpha\beta}^{\mathcal{T}}P^*_{\QQ_j}\Psi_{c,\beta}^{(0)*}(\QQ_j)e^{-i\QQ_j\cdot\rr}\nonumber \\
&&=\frac{1}{V}\sum\limits_{\QQ_j\in\mathcal{Q}_6}P^*_{\QQ_j}\tilde{\Psi}_{c,\alpha}^{(0)}(-\QQ_j)e^{-i\QQ_j\cdot\rr}\nonumber\\
&&=\frac{1}{V}\sum\limits_{\tilde{\QQ}_j=-\QQ_j}P^*_{-\tilde{\QQ}_j}\tilde{\Psi}_{c,\alpha}^{(0)}(\tilde{\QQ}_j)e^{i\tilde{\QQ}_j\cdot\rr}
\end{eqnarray}
where the second equality comes from anti-unitary symmetry transformation in Eq. \eqref{eq:GL2symtrans}. By comparing with the transformed form in Eq.\eqref{eq:realtransideal} and we find
\begin{eqnarray}\label{eq_SM:symmetry_P_antiunitary}
    \tilde{P}_{\QQ_j} = P^*_{-\QQ_j} 
\end{eqnarray}
for TR symmetry. Based on Eq.(\ref{eq_SM:symmetry_P_unitary}) and Eq.\eqref{eq_SM:symmetry_P_antiunitary}, the symmetry operations in MPG $31'$ should transform the order parameter vector $\mathcal{P}$ in the following way: 
\begin{eqnarray}\label{eq_SM:C3z_T_Pvector}
&C_{3z}:\tilde{P}_{\QQ_j}=P_{\QQ_{j-2}}\Rightarrow \tilde{\mathcal{P}}=U_{C_{3z}}\mathcal{P}\\
&\mathcal{T}:\tilde{P}_{\QQ_j}=P^*_{-\QQ_j}=P^*_{\QQ_{j+3}}\Rightarrow \tilde{\mathcal{P}}=U_{\mathcal{T}}\mathcal{K}\mathcal{P} \quad \forall \QQ_j \in \mathcal{Q}_6\\&
U_{C_{3z}}=\left(\begin{matrix}0&0&0&0&1&0\\0&0&0&0&0&1\\1&0&0&0&0&0\\0&1&0&0&0&0\\0&0&1&0&0&0\\0&0&0&1&0&0\end{matrix}\right);\quad U_{\mathcal{T}}=\left(\begin{matrix}0&0&0&1&0&0\\0&0&0&0&1&0\\0&0&0&0&0&1\\1&0&0&0&0&0\\0&1&0&0&0&0\\0&0&1&0&0&0\end{matrix}\right).
\end{eqnarray}Before the symmetry construction of the $\mathcal{F}^{(4)}$ term in Eq.(\ref{eq_SM:GLfreeenergy_1}), we first project the $\mathcal{F}^{(2)}$ term in Eq.(\ref{eq:compactGL2}) into the $U(6)$ subspace of the order parameter vector $\mathcal{P}$. We substitute the order parameter ansatz in Eq.(\ref{eq:fteigvec}) into Eq.(\ref{eq:compactGL2_1}) and obtain
\begin{eqnarray}
\mathcal{F}^{(2)}\approx &&\sum\limits_{\kk,\substack{\QQ_j,\QQ_k\in\mathcal{Q}_6},\alpha\beta}|P_{\kk}|^2\delta(\kk-\QQ_j)\delta(\kk-\QQ_k)(\Psi_{c,\alpha}^{(0)}(\kk))^{\dagger}(T-M_{\alpha\beta}(\kk)))\Psi_{c,\beta}^{(0)}(\kk). 
\end{eqnarray}
Since $\Psi^{(0)}$ can be viewed as an approximated eigen-state of $M(\QQ_j)$ matrix with its eigen-value to be $T_{c,0}$ given in Eq.(\ref{eq_SM:Tc_quadraticTerm}), the $\mathcal{F}^{(2)}$ term can be simplified as
\begin{eqnarray}\label{eq_SM:freen2simpl}\mathcal{F}^{(2)}&&=(T-T_{c,0})\sum\limits_{\QQ_j\in\mathcal{Q}_6}|P_{\QQ_j}|^2.
\end{eqnarray}
We next aim to construct the fourth-order term $\mathcal{F}^{(4)}$ from the symmetry consideration. The generic fourth-order terms in the GL free energy for the $A$ and $E$ irrep pairing channels take the form
\begin{eqnarray}\label{eq:momGL4_0}
    && \mathcal{F}^{(4)}= \mathcal{F}^{(4)}_{A} + \mathcal{F}^{(4)}_{E} \\
    && \mathcal{F}^{(4)}_{A} = \frac{1}{V}\int d^2\rr  b_{A} |\eta_{A}(\rr)|^2 |\eta_{A}(\rr)|^2\nonumber \\
    && \mathcal{F}^{(4)}_{E} = \frac{1}{V}\int d^2\rr\sum\limits_{m,n=\pm} b_{E,mn} |\eta_{E,m}(\rr)|^2 |\eta_{E,n}(\rr)|^2\nonumber 
\end{eqnarray}
where $b_{A}$ and $b_{E,mn}$ are the fourth-order coefficients and $\eta_A,\eta_{E,\pm}$ are the order parameter components of $\Psi(\rr)$ in Eq.\eqref{eq:psidef2}. Here we only consider the local terms for the fourth-order GL free energy. Next, we substitute Eq.(\ref{eq:realeigvec}) into $\mathcal{F}^{(4)}$ to project $\Psi_\alpha$ into the $U(6)$ subspace and find 
\begin{eqnarray}
&\mathcal{F}^{(4)}=\sum\limits_{i,j,k,l\in\mathbb{Z}_6}\delta_{\QQ_j+\QQ_k,\QQ_l+\QQ_l}c(\QQ_j,\QQ_k,\QQ_l,\QQ_l) P_{\QQ_j}P_{\QQ_k}P_{\QQ_l}^*P_{\QQ_l}^*\\
&c(\QQ_j,\QQ_k,\QQ_l,\QQ_l) = b_A\eta^{(0)}_{c,A}(\QQ_j)\eta^{(0)}_{c,A}(\QQ_k)\eta^{(0)*}_{c,A}(\QQ_l)\eta^{(0)*}_{c,A}(\QQ_l)\nonumber\\
&+\sum_{mn} b_{E,mn}\eta^{(0)}_{c,E,m}(\QQ_j)\eta^{(0)}_{c,E,n}(\QQ_k)\eta^{(0)*}_{c,E,m}(\QQ_l)\eta^{(0)*}_{c,E,n}(\QQ_l),\label{eq_SM:freeGL4mom_1}
\end{eqnarray}
where $\eta^{(0)}_{c,A}(\kk),\eta^{(0)}_{c,E,\pm}(\kk)$ are the order parameter components of the eigen-state $\Psi^{(0)}_c(\kk)$. 
Since six $\QQ_j$ points have same magnitude $|\QQ_j|=k_0$ for any $j=0, 1, ..., 5$, the momentum conservation in $\delta_{\QQ_j+\QQ_k,\QQ_l+\QQ_l}$ appears in only three ways, resulting in 
\begin{eqnarray}
&&\mathcal{F}^{(4)}=\sum\limits_{\QQ_{i,j,k}\in\mathcal{Q}_6} \Big(\mathcal{F}^{(4,1)}(\QQ_j,\QQ_k)+\mathcal{F}^{(4,2)}(\QQ_j,\QQ_k)+\mathcal{F}^{(4,3)}(\QQ_j,\QQ_l)\Big)\label{eq:genfreen4_1}\\
&&\mathcal{F}^{(4,1)}(\QQ_j,\QQ_k)= c(\QQ_j,\QQ_k,\QQ_j,\QQ_k) |P_{\QQ_j}|^2|P_{\QQ_k}|^2  \quad \text{if} \quad \QQ_j=\QQ_l\cap\QQ_k=\QQ_l\label{eq:momconsGL4_1}\\
&&\mathcal{F}^{(4,2)}(\QQ_j,\QQ_k)=c(\QQ_j,\QQ_k,\QQ_k,\QQ_j) |P_{\QQ_j}|^2|P_{\QQ_k}|^2  \quad \text{if}\quad \QQ_j=\QQ_l\cap\QQ_k=\QQ_l\label{eq:momconsGL4_2}\\
&&\mathcal{F}^{(4,3)}(\QQ_j,\QQ_l)=c(\QQ_j,-\QQ_j,\QQ_l,-\QQ_l) P_{\QQ_j}P_{-\QQ_j}(P_{\QQ_l}P_{-\QQ_l})^* \quad \text{if}\quad \QQ_j=-\QQ_k\cap\QQ_l=-\QQ_l. \label{eq:momconsGL4_3}
\end{eqnarray}

Using the definitions of order parameters in Eq. \eqref{eq_SM:GL2eigstate}, we can compute the coefficients as
\begin{eqnarray}
c(\QQ_j,\QQ_k,\QQ_j,\QQ_k)&=&\frac{b_A}{N_1^2(k_0)}(T_c^{(0)}(k_0)-T_{c,E}+\gamma'_Ek_0^2)^4+ \sum\limits_{mn}\frac{b_{E,mn}}{N_1^2(k_0)}|\zeta'_1|^4k_0^4\label{eq:cmnsum_1}\\
c(\QQ_j,\QQ_k,\QQ_k,\QQ_j)&=&\frac{b_A}{N_1^2(k_0)}(T_c^{(0)}(k_0)-T_{c,E}+\gamma'_Ek_0^2)^4\nonumber\\&+& \sum\limits_{m}\Big[\frac{b_{E,mm}}{N_1^2(k_0)}|\zeta'_1|^4k_0^4+\frac{b_{E,m,-m}}{N_1^2(k_0)}|\zeta'_1|^4k_0^4\cos{\{2m(\theta_{\QQ_j}-\theta_{\QQ_l})\}}\Big]\label{eq:cmnsum_2}\\
c(\QQ_j,-\QQ_j,\QQ_l,-\QQ_l)&=&\frac{b_A}{N_1^2(k_0)}(T_c^{(0)}(k_0)-T_{c,E}+\gamma'_Ek_0^2)^4\nonumber\\&+& \sum\limits_{m}\Big[\frac{b_{E,m,-m}}{N_1^2(k_0)}|\zeta'_1|^4k_0^4+\frac{b_{E,mm}}{N_1^2(k_0)}|\zeta'_1|^4k_0^4\cos{\{2m(\theta_{\QQ_j}-\theta_{\QQ_l})\}}\Big]\label{eq:cmnsum_3}. 
\end{eqnarray}
where $\theta_{\QQ_j}=\frac{j\pi}{3}$ is defined in Eq.\eqref{eq:sixpointsset}. The derivation of
Eq.\eqref{eq:cmnsum_3} has used the fact that $c_{mn}(...)=c_{nm}(...)$, since $b_{E,nm}=b_{E,mn}$ in Eq. \eqref{eq:momGL4_0}. 
We first consider the first two terms in Eq.\eqref{eq:genfreen4_1}, $\mathcal{F}^{(4,1)}(\QQ_j,\QQ_k)$ and $\mathcal{F}^{(4,2)}(\QQ_j,\QQ_k)$. From Eq.\eqref{eq:cmnsum_1}, the coefficients $c(\QQ_j,\QQ_k,\QQ_j,\QQ_k)$ is independent of $\QQ_j,\QQ_k$. From Eq. \eqref{eq:cmnsum_2}, $c(\QQ_j,\QQ_k,\QQ_j,\QQ_k)$ depends on the momentum angle $\theta_{\QQ_j}-\theta_{\QQ_l}$ and can take two different values, one for $\theta_{\QQ_j}-\theta_{\QQ_l}=\{0,\pi\}$, namely,
\begin{eqnarray}c(\QQ_j,\QQ_k,\QQ_k,\QQ_j)=\frac{b_A}{N_1^2(k_0)}(T_c^{(0)}(k_0)-T_{c,E}+\gamma'_Ek_0^2)^4+ \sum\limits_{m,n}\Big[\frac{b_{E,mn}}{N_1^2(k_0)}|\zeta'_1|^4k_0^4\Big],\end{eqnarray}
and the other for $\theta_{\QQ_j}-\theta_{\QQ_l}=\{\frac{\pi}{3},\frac{2\pi}{3},\frac{4\pi}{3},\frac{5\pi}{3}\}$, namely,
\begin{eqnarray}c(\QQ_j,\QQ_k,\QQ_k,\QQ_j)=\frac{b_A}{N_1^2(k_0)}(T_c^{(0)}(k_0)-T_{c,E}+\gamma'_Ek_0^2)^4+ \sum\limits_{m}\frac{|\zeta'_1|^4k_0^4}{N_1^2(k_0)}\Big[b_{E,mm}-\frac{b_{E,m,-m}}{2}\Big].\end{eqnarray}
These values are fixed for any $\QQ_{j,k,l}$. Combining these results, the first two terms in Eq.\eqref{eq:genfreen4_1} give rise to
\begin{eqnarray}\label{eq_SM:F41andF_42}
&&\sum\limits_{\QQ_{i,j}\in\mathcal{Q}_6}\mathcal{F}^{(4,1)}(\QQ_j,\QQ_k)+\mathcal{F}^{(4,2)}(\QQ_j,\QQ_k)=\sum\limits_{\substack{\QQ_{i}\in\mathcal{Q}_6\\q=\{0,1,2,3\}}}\mathcal{F}^{(4,1)}(\QQ_j,\QQ_{j+q})+\mathcal{F}^{(4,2)}(\QQ_j,\QQ_{j+q})\nonumber\\&&=\tilde{c}_0\left[\sum\limits_{\QQ_j\in\mathcal{Q}_6}|P_{\QQ_j}|^4\right]+\tilde{c}_1\left[\sum\limits_{\QQ_j\in\mathcal{Q}_6}|P_{\QQ_j}|^2|P_{\QQ_{j+1}}|^2\right]+\tilde{c}_2\left[\sum\limits_{\QQ_j\in\mathcal{Q}_6}|P_{\QQ_j}|^2|P_{\QQ_{j+2}}|^2\right]\nonumber\\&&+\tilde{c}_{3,1}\left[\sum\limits_{\QQ_j\in\mathcal{Q}_6}|P_{\QQ_j}|^2|P_{-\QQ_j}|^2\right]
\end{eqnarray}
where
\begin{eqnarray}
&&\tilde{c}_p=c(\QQ_j,\QQ_{j+p},\QQ_{i},\QQ_{j+p})+c(\QQ_j,\QQ_{j+p},\QQ_{j+p},\QQ_j) 
\end{eqnarray}
for $p=0,1,2$ and
\begin{eqnarray}
    &&\tilde{c}_{3,1}=c(\QQ_j,-\QQ_{i},\QQ_{i},-\QQ_j)+c(\QQ_j,-\QQ_{i},-\QQ_{i},\QQ_j)=2c(\QQ_j,-\QQ_{i},\QQ_{i},-\QQ_j).
\end{eqnarray}
It should be noted that the values of the parameters $\tilde{c}_p$ and $\tilde{c}_{3,1}$ are independent of $\QQ_j$. 


For the last term in Eq.\eqref{eq:genfreen4_1}, $c(\QQ_j,-\QQ_j,\QQ_l,-\QQ_l)$ also depends on the momentum angle $\theta_{\QQ_j}-\theta_{\QQ_l}$ and is given by 

\begin{eqnarray}c(\QQ_j,-\QQ_j,\pm\QQ_j,\mp\QQ_j)=\frac{b_A}{N_1^2(k_0)}(T_c^{(0)}(k_0)-T_{c,E}+\gamma'_Ek_0^2)^4+ \sum\limits_{m,n}\Big[\frac{b_{E,mn}}{N_1^2(k_0)}|\zeta'_1|^4k_0^4\Big],\label{eq:cmnsum3_1}\end{eqnarray}for $\theta_{\QQ_j}-\theta_{\QQ_l}\in\{0,\pi\}$, and
\begin{eqnarray}\label{eq:cmnsum3_2}c(\QQ_j,-\QQ_j,\QQ_l,-\QQ_l)=\frac{b_A}{N_1^2(k_0)}(T_c^{(0)}(k_0)-T_{c,E}+\gamma'_Ek_0^2)^4+ \sum\limits_{m}\frac{|\zeta'_1|^4k_0^4}{N_1^2(k_0)}\Big[b_{E,m,-m}-\frac{b_{E,mm}}{2}\Big]\end{eqnarray}
for $\theta_{\QQ_j}-\theta_{\QQ_l}=\{\frac{\pi}{3},\frac{2\pi}{3},\frac{4\pi}{3},\frac{5\pi}{3}\}$. 

With the expression Eq.\eqref{eq:cmnsum_3}, the last term in Eq.\eqref{eq:genfreen4_1} becomes
\begin{eqnarray}\label{eq_SM:F43}
&&\sum\limits_{\QQ_{i,k}\in\mathcal{Q}_6}\mathcal{F}^{(4,3)}(\QQ_j,\QQ_l)=\sum\limits_{\substack{\QQ_{i}\in\mathcal{Q}_6\\q=\{0,1,2,3\}}}\mathcal{F}^{(4,3)}(\QQ_j,\QQ_{j+q})\nonumber\\&=&\tilde{c}_{3,2}\left[\sum\limits_{\QQ_j\in\mathcal{Q}_6}|P_{\QQ_j}|^2|P_{-\QQ_j}|^2\right]+\tilde{c}_4\sum\limits_{\QQ_j\in\mathcal{Q}_6}\left[P_{\QQ_j}P_{-\QQ_j}(P^*_{\QQ_{j+1}}P^*_{-\QQ_{j+1}}+P^*_{\QQ_{j+2}}P^*_{-\QQ_{j+2}})\right]
\end{eqnarray}
where
\begin{eqnarray}
&&\tilde{c}_{3,2}=c(\QQ_j,-\QQ_j,\QQ_j,-\QQ_j)+c(\QQ_j,-\QQ_j,-\QQ_j,\QQ_j)=2c(\QQ_j,-\QQ_j,\QQ_j,-\QQ_j)\\
&&\tilde{c}_4=c(\QQ_j,-\QQ_j,\QQ_{j+1},-\QQ_{j+1})=c(\QQ_j,-\QQ_j,\QQ_{j+2},-\QQ_{j+2}).
\end{eqnarray}
One should note that the values of the above equations for both $\tilde{c}_{3,2}$ and $\tilde{c}_4$ are independent of $\QQ_j$ and thus we treat $\tilde{c}_{3,2}$ and $\tilde{c}_4$ as constant parameters. 

Combining all the terms in Eqs.\eqref{eq_SM:F41andF_42} and \eqref{eq_SM:F43} leads to
\begin{eqnarray}\label{eq_SM:F4order_GL}
\mathcal{F}^{(4)}=&&\tilde{c}_0\left[\sum\limits_{\QQ_j\in\mathcal{Q}_6}|P_{\QQ_j}|^4\right]+\tilde{c}_1\left[\sum\limits_{\QQ_j\in\mathcal{Q}_6}|P_{\QQ_j}|^2|P_{\QQ_{j+1}}|^2\right]+\tilde{c}_2\left[\sum\limits_{\QQ_j\in\mathcal{Q}_6}|P_{\QQ_j}|^2|P_{\QQ_{j+2}}|^2\right]\nonumber \\&&+\tilde{c}_3\left[\sum\limits_{\QQ_j\in\mathcal{Q}_6}|P_{\QQ_j}|^2|P_{-\QQ_j}|^2\right]+\tilde{c}_4\sum\limits_{\QQ_j\in\mathcal{Q}_6}\left[P_{\QQ_j}P_{-\QQ_j}(P^*_{\QQ_{j+1}}P^*_{-\QQ_{j+1}}+P^*_{\QQ_{j+2}}P^*_{-\QQ_{j+2}})\right],
\end{eqnarray}
where $\tilde{c}_3=\tilde{c}_{3,1}+\tilde{c}_{3,2}$. Eq.(\ref{eq_SM:F4order_GL}) is equivalent to Eq.(6) in main text with their coefficients related by $c_{0,4}=\tilde{c}_{0,4}$, $c_p=\tilde{c}_p-2\tilde{c}_0$ with $p=1,2$ and $c_3=\tilde{c}_3-\tilde{c}_0$.


Next, we show that Eq.\eqref{eq_SM:F4order_GL} is invariant under the MPG 31$'$ symmetry group that is generated by $C_{3z}$ and $\mathcal{T}$ symmetries. 

{\bf (i) $\tilde{c}_0$ term:} Using the symmetry transformation in Eq.(\ref{eq_SM:C3z_T_Pvector}) for MPG 31$'$, the $\tilde{c}_0$ term transforms as
\begin{eqnarray}\label{eq:Qijsymanalysis1c0}&C_{3z}:\sum\limits_{\QQ_j\in\mathcal{Q}_6}|\tilde{P}_{\QQ_j}|^4=\sum\limits_{\QQ_j\in\mathcal{Q}_6}|P_{\QQ_{j-2}}|^4=\sum\limits_{\QQ_j\in\mathcal{Q}_6}|P_{\QQ_{i}}|^4\nonumber\\
&\mathcal{T}:\sum\limits_{\QQ_j\in\mathcal{Q}_6}|\tilde{P}_{\QQ_j}|^4=\sum\limits_{\QQ_j\in\mathcal{Q}_6}|P^*_{\QQ_{j+3}}|^4=\sum\limits_{\QQ_j\in\mathcal{Q}_6}|P_{\QQ_{j+3}}|^4=\sum\limits_{\QQ_j\in\mathcal{Q}_6}|P_{\QQ_{i}}|^4,\end{eqnarray}
which is invariant under both $C_{3z}$ and $\mathcal{T}$. 


{\bf (ii) $\tilde{c}_{1,2,3}$ terms:} Using the symmetry transformation in Eq.(\ref{eq_SM:C3z_T_Pvector}) for MPG 31$'$, the $\tilde{c}_p$ terms with $p=1,2,3$ transform as
\begin{eqnarray}\label{eq:Qijsymanalysis1c123}&C_{3z}:\sum\limits_{\QQ_j\in\mathcal{Q}_6}|\tilde{P}_{\QQ_j}|^2|\tilde{P}_{\QQ_{j+p}}|^2=\sum\limits_{\QQ_j\in\mathcal{Q}_6}|P_{\QQ_{j-2}}|^2P_{\QQ_{j+p-2}}|^2=\sum\limits_{\QQ_j\in\mathcal{Q}_6}|P_{\QQ_{i}}|^2P_{\QQ_{j+p}}|^2\\
&\mathcal{T}:\sum\limits_{\QQ_j\in\mathcal{Q}_6}|\tilde{P}_{\QQ_j}|^2|\tilde{P}_{\QQ_{j+p}}|^2=\sum\limits_{\QQ_j\in\mathcal{Q}_6}|P_{\QQ_{j+3}}|^2P_{\QQ_{j+p+3}}|^2=\sum\limits_{\QQ_j\in\mathcal{Q}_6}|P_{\QQ_{i}}|^2P_{\QQ_{j+p}}|^2,
\end{eqnarray}
which is invariant under both $C_{3z}$ and $\mathcal{T}$. 

{\bf (iii) $\tilde{c}_{4}$ term:} Using the symmetry transformation in Eq.(\ref{eq_SM:C3z_T_Pvector}) for MPG 31$'$, the $\tilde{c}_{4}$ term transforms as
\begin{eqnarray}\mathcal{C}_{3z}&:&\sum\limits_{\QQ_j\in\mathcal{Q}_6}\tilde{P}_{\QQ_j}\tilde{P}_{\QQ_{j+3}}(\tilde{P}_{\QQ_{j+p}}\tilde{P}_{\QQ_{j+p+3}})^*\nonumber\\&&=\sum\limits_{\QQ_j\in\mathcal{Q}_6}P_{\QQ_{j-2}}P_{\QQ_{j+1}}(P_{\QQ_{j+p-2}}P_{\QQ_{j+p+1}})^*=\sum\limits_{\QQ_j\in\mathcal{Q}_6}P_{\QQ_{i}}P_{\QQ_{j+3}}(P_{\QQ_{j+p}}P_{\QQ_{j+p+3}})^*\\\mathcal{T}&:&\sum\limits_{\QQ_j\in\mathcal{Q}_6}\tilde{P}_{\QQ_j}\tilde{P}_{\QQ_{j+3}}(\tilde{P}_{\QQ_{j+p}}\tilde{P}_{\QQ_{j+p+3}})^*\nonumber\\&&=\sum\limits_{\QQ_j\in\mathcal{Q}_6}(P_{\QQ_{i}}P_{\QQ_{j+3}})^*P_{\QQ_{j+p}}P_{\QQ_{j+p+3}}=\sum\limits_{\QQ_j\in\mathcal{Q}_6}P_{\QQ_{i}}P_{\QQ_{j+3}}(P_{\QQ_{j+p}}P_{\QQ_{j+p+3}})^*
\end{eqnarray}
for both $p=1,2$, from which one can see $\sum\limits_{\QQ_j\in\mathcal{Q}_6}P_{\QQ_j}P_{-\QQ_j}(P^*_{\QQ_{j+p-2}}P^*_{-\QQ_{j+p+1}})$ is invariant under the symmetry operations in the MPG 31$'$, so that the entire $\tilde{c}_4$ term is also invariant under the symmetry operations in the MPG 31$'$.


\subsection{SC Ground states in the $U(6)$ subspace}\label{sec:SCgstates}

In this section, we will first describe our strategy to minimize the fourth-order term $\mathcal{F}^{(4)}$ within the $U(6)$ subspace and then describe our main results of the superconducting ground states. Our strategy to solve the problem includes two steps. In the first step, we will derive the GL equations from the free energy $\mathcal{F}$ in Eq.(\ref{eq:momGL4_0}) and solve the GL equations within a certain solution ansatz subspace, which are expected to give us possible minima of $\mathcal{F}$. In the second step, we will compare the minimal free energies of different solution ans\"atze to identify the ground states $\mathcal{F}$ by combining analytical and numerical approaches.  

We first define $P_{\QQ_j}=\alpha_je^{i\phi_i}$ ($ j\in\mathbb{Z}_6$), and rewrite the vector $\mathcal{P}$ in Eq. \eqref{eq:Pvector} as
\begin{eqnarray}\label{eq_SM:Pvector_form1}
\mathcal{P}=(\alpha_0e^{i\phi_0},\alpha_1 e^{i\phi_1},\alpha_2e^{i\phi_2},\alpha_3e^{i\phi_3},\alpha_4e^{i\phi_4},\alpha_5e^{i\phi_5})^T
\end{eqnarray}
where $\alpha_j$ is the magnitude of the component $P_{\QQ_j}$ which is chosen to be positive and $\phi_i\in[0,2\pi)$ is the phase of $P_{\QQ_j}$. With this form of of $P_{\QQ_j}$, we can rewrite the total free energy $\mathcal{F}=\mathcal{F}^{(2)}+\mathcal{F}^{(4)}$, with $\mathcal{F}^{(2)}$ in Eq.\eqref{eq_SM:freen2simpl} and $\mathcal{F}^{(4)}$ in Eq.(\ref{eq_SM:F4order_GL}) as
\begin{eqnarray}\label{eq_SM:genfreen}
\mathcal{F}[\{\alpha, \phi\};\{c\}]&=&\sum\limits_{j\in\mathbb{Z}_6}\Delta T\alpha_j^2+c_0\Big(\sum\limits_{j\in\mathbb{Z}_6}\alpha_j^2\Big)^2+c_1\sum\limits_{j\in\mathbb{Z}_6}\alpha_j^2\alpha_{j+1}^2+c_2\sum\limits_{j\in\mathbb{Z}_6}\alpha_j^2\alpha_{j+2}^2+c_3\sum\limits_{j\in\mathbb{Z}_6}\alpha_j^2\alpha_{j+3}^2\nonumber\\&+&4c_4\Big\{\alpha_0\alpha_3\alpha_1\alpha_4\cos(\phi_0+\phi_3-\phi_1-\phi_4)+\alpha_1\alpha_4\alpha_2\alpha_5\cos(\phi_1+\phi_4-\phi_2-\phi_5)\nonumber\\&+&\alpha_2\alpha_5\alpha_0\alpha_3\cos(\phi_2+\phi_5-\phi_0-\phi_3)\Big\},
\end{eqnarray}
where $\Delta T=T-T_{c,0}$, $\{\alpha, \phi\}$ labels the variables of the order parameters and $\{c\}$ labels the parameter space. 

The ground state will be the configuration of $\{\alpha_j,\phi_j\big|j\in\mathbb{Z}_6\}$ that minimizes the free energy Eq. \eqref{eq_SM:genfreen}. To identify all the minima of the free energy, we first consider the GL equation,
\begin{eqnarray}\label{eq:genGLeq}
\frac{\partial\mathcal{F}}{\partial\alpha_j}=0,\quad \frac{\partial\mathcal{F}}{\partial\phi_j}=0; \quad \forall j\in\mathbb{Z}_6. 
\end{eqnarray} 

Using the free energy $\mathcal{F}$ in Eq.(\ref{eq_SM:genfreen}), we find a set of GL equations, given by 
\begin{eqnarray}
&\alpha_0\Delta T+\alpha_0\Bigg\{2c_0\Delta_0^2+c_1(\alpha_1^2+\alpha_5^2)+c_2(\alpha_2^2+\alpha_4^2)+2c_3\alpha_3^2\Bigg\}+4c_4\Bigg\{\alpha_3\alpha_1\alpha_4\cos(\theta_1)+\alpha_2\alpha_5\alpha_3\cos(\theta_3)\Bigg\}=0 \label{eq:GLeqphaseqs1}\\
&\alpha_1\Delta T+\alpha_1\Bigg\{2c_0\Delta_0^2+c_1(\alpha_2^2+\alpha_0^2)+c_2(\alpha_3^2+\alpha_5^2)+2c_3\alpha_4^2\Bigg\}+4c_4\Bigg\{\alpha_4\alpha_2\alpha_5\cos(\theta_2)+\alpha_3\alpha_0\alpha_4\cos(\theta_1)\Bigg\}=0\label{eq:GLeqphaseqs2}\\
&\alpha_2\Delta T+\alpha_2\Bigg\{2c_0\Delta_0^2+c_1(\alpha_3^2+\alpha_1^2)+c_2(\alpha_4^2+\alpha_0^2)+2c_3\alpha_5^2\Bigg\}+4c_4\Bigg\{\alpha_5\alpha_3\alpha_0\cos(\theta_3)+\alpha_4\alpha_1\alpha_5\cos(\theta_2)\Bigg\}=0\label{eq:GLeqphaseqs3}\\
&\alpha_3\Delta T+\alpha_3\Bigg\{2c_0\Delta_0^2 +c_1(\alpha_4^2+\alpha_2^2)+c_2(\alpha_5^2+\alpha_1^2)+2c_3\alpha_0^2\Bigg\}+4c_4\Bigg\{\alpha_0\alpha_4\alpha_1\cos(\theta_1)+\alpha_5\alpha_2\alpha_0\cos(\theta_3)\Bigg\}=0\label{eq:GLeqphaseqs4}\\
&\alpha_4\Delta T+\alpha_4\Bigg\{2_0\Delta_0^2 +c_1(\alpha_5^2+\alpha_3^2)+c_2(\alpha_0^2+\alpha_2^2)+2c_3\alpha_1^2\Bigg\}+4c_4\Bigg\{\alpha_1\alpha_5\alpha_2\cos(\theta_2)+\alpha_0\alpha_3\alpha_1\cos(\theta_1)\Bigg\}=0\label{eq:GLeqphaseqs5}\\
&\alpha_5\Delta T+\alpha_5\Bigg\{2c_0\Delta_0^2 +c_1(\alpha_0^2+\alpha_4^2)+c_2(\alpha_1^2+\alpha_3^2)+2c_3\alpha_2^2\Bigg\}+4c_4\Bigg\{\alpha_2\alpha_0\alpha_3\cos(\theta_3)+\alpha_1\alpha_4\alpha_2\cos(\theta_2)\Bigg\}=0\label{eq:GLeqphaseqs6}\\
&c_4\alpha_0\Big\{\alpha_3\alpha_1\alpha_4\sin(\theta_1)+\alpha_3\alpha_2\alpha_5\sin(\theta_3)\}=0\label{eq:GLeqphaseqs7}\\
&c_4\alpha_1\Big\{\alpha_4\alpha_2\alpha_5\sin(\theta_2)+\alpha_4\alpha_3\alpha_0\sin(\theta_1)\}=0\label{eq:GLeqphaseqs8}\\
&c_4\alpha_2\Big\{\alpha_5\alpha_3\alpha_0\sin(\theta_3)+\alpha_5\alpha_4\alpha_1\sin(\theta_2)\}=0\label{eq:GLeqphaseqs9}\\
&c_4\alpha_3\Big\{\alpha_0\alpha_4\alpha_1\sin(\theta_1)+\alpha_0\alpha_5\alpha_2\sin(\theta_3)\}=0\label{eq:GLeqphaseqs10}\\
&c_4\alpha_4\Big\{\alpha_1\alpha_5\alpha_2\sin(\theta_2)+\alpha_1\alpha_0\alpha_3\sin(\theta_1)\}=0\label{eq:GLeqphaseqs11}\\
&c_4\alpha_5\Big\{\alpha_2\alpha_0\alpha_3\sin(\theta_3)+\alpha_2\alpha_1\alpha_4\sin(\theta_2)\}=0\label{eq:GLeqphaseqs12}
\end{eqnarray}
where 
\begin{eqnarray}\label{eq_SM:order_parameter_amplitude_Delta0}
\Delta_0^2 =\sum\limits_{j\in\mathbb{Z}_6}\alpha_j^2
\end{eqnarray}
gives the amplitude of the order parameter vector $\mathcal{P}$ and the $\theta$ angles are defined by
\begin{eqnarray}
    \theta_1=\phi_0+\phi_3-\phi_1-\phi_4;\quad \theta_2=\phi_1+\phi_4-\phi_2-\phi_5;\quad \theta_3=-\theta_1-\theta_2. 
\end{eqnarray}
Among this set of nonlinear GL equations, Eqs.(\ref{eq:GLeqphaseqs1})-(\ref{eq:GLeqphaseqs6}) describe the extremes of free energy with respect to the amplitude of each component $\alpha_j$, while Eqs.(\ref{eq:GLeqphaseqs7})-(\ref{eq:GLeqphaseqs12}) describe the extremes with respect to the phase $\phi_i$. Since only the $c_4$ term in the free energy (\ref{eq_SM:genfreen}) depends on the phase $\phi$, this phase dependence is only relevant for non-zero $c_4$. In the discussion below, we will only focus on the $c_4=0$ situations, so the phase $\phi$ dependence is removed from the free energy $\mathcal{F}$ in Eq.(\ref{eq_SM:genfreen}). Correspondingly, the GL equations are simplified as 
\begin{eqnarray}
&\alpha_0\Delta T+\alpha_0\Bigg\{2c_0\Delta_0^2+c_1(\alpha_1^2+\alpha_5^2)+c_2(\alpha_2^2+\alpha_4^2)+2c_3\alpha_3^2\Bigg\}=0 \label{eq:GLeqphaseqs1a}\\
&\alpha_1\Delta T+\alpha_1\Bigg\{2c_0\Delta_0^2+c_1(\alpha_2^2+\alpha_0^2)+c_2(\alpha_3^2+\alpha_5^2)+2c_3\alpha_4^2\Bigg\}=0\label{eq:GLeqphaseqs2a}\\
&\alpha_2\Delta T+\alpha_2\Bigg\{2c_0\Delta_0^2+c_1(\alpha_3^2+\alpha_1^2)+c_2(\alpha_4^2+\alpha_0^2)+2c_3\alpha_5^2\Bigg\}=0\label{eq:GLeqphaseqs3a}\\
&\alpha_3\Delta T+\alpha_3\Bigg\{2c_0\Delta_0^2 +c_1(\alpha_4^2+\alpha_2^2)+c_2(\alpha_5^2+\alpha_1^2)+2c_3\alpha_0^2\Bigg\}=0\label{eq:GLeqphaseqs4a}\\
&\alpha_4\Delta T+\alpha_4\Bigg\{2c_0\Delta_0^2 +c_1(\alpha_5^2+\alpha_3^2)+c_2(\alpha_0^2+\alpha_2^2)+2c_3\alpha_1^2\Bigg\}=0\label{eq:GLeqphaseqs5a}\\
&\alpha_5\Delta T+\alpha_5\Bigg\{2c_0\Delta_0^2 +c_1(\alpha_0^2+\alpha_4^2)+c_2(\alpha_1^2+\alpha_3^2)+2c_3\alpha_2^2\Bigg\}=0\label{eq:GLeqphaseqs6a}. 
\end{eqnarray}

This set of GL equations can be solved and all the extremes (including minima, maxima and saddle points) of the free energy $\mathcal{F}$ can be obtained. We further substitute the extreme solutions back into Eq.(\ref{eq_SM:genfreen}) to find the minimal free energy $\mathcal{F}$ of different solutions and identify the superconducting ground state, by comparing these minimal free energies.

It is instructive to rewrite the order parameter components $\alpha_j$'s as
\begin{eqnarray}
    \alpha_j=\Delta_0 a_j, 
\end{eqnarray}
where $\Delta_0$ is the amplitude given in Eq.(\ref{eq_SM:order_parameter_amplitude_Delta0}) while $a_j$ describes the direction of the vector $\mathcal{P}$ and satisfies
\begin{eqnarray}
\sum\limits_{j\in\mathbb{Z}_6}a_{j}^2=\sum\limits_{j\in\mathbb{Z}_6}\Big(\frac{\alpha_{j}}{\Delta_0}\Big)^2=1. 
\label{eq_SM:constraints1}
\end{eqnarray} 
Consequently, the vector $\mathcal{P}$ in Eq. \eqref{eq_SM:Pvector_form1} can be written as
\begin{eqnarray}\label{eq_SM:Pvector_form2}
\mathcal{P}=\Delta_0 (a_0 e^{i\phi_0}, a_1 e^{i\phi_1},a_2 e^{i\phi_2},a_3 e^{i\phi_3},a_4 e^{i\phi_4},a_5 e^{i\phi_5}).
\end{eqnarray}

Substituting Eq.(\ref{eq_SM:Pvector_form2}) into the free energy $\mathcal{F}$ in Eq.(\ref{eq_SM:genfreen}) leads to
\begin{eqnarray}\label{eq_SM:freeenergy_1}
\mathcal{F}&=&\Delta T\Delta_0^2+f(\{a,\phi\};\{c\})\Delta_0^4
\end{eqnarray}
where
\begin{eqnarray}\label{eq:optfunc}
f(\{a,\phi\};\{c\})&=&c_0+c_1\sum\limits_{j\in\mathbb{Z}_6}a_{j}^2a_{j+1}^2+c_2\sum\limits_{j\in\mathbb{Z}_6}a_{j}^2a_{j+2}^2+c_3\sum\limits_{j\in\mathbb{Z}_6}a_{j}^2a_{j+3}^2
\end{eqnarray}
with $c_4=0$. 

If we treat $\Delta_0$ in Eq.(\ref{eq_SM:freeenergy_1}) as a variable, this free energy includes the $\Delta_0^2$ and $\Delta_0^4$ terms and can be minimized with respect to $\Delta_0^2$ as
\begin{equation}\label{eq:lowfreenphase}
\mathcal{F}_{min}=-\frac{\Delta T^2}{4 f(\{a,\phi\};\{c\})}
\end{equation}
at 
\begin{eqnarray}
    \Delta_0=\sqrt{-\frac{\Delta T}{2 f(\{a,\phi\};\{c\})}}, 
\end{eqnarray}
when $f(\{a,\phi\};\{c\})$ is positive and $\Delta T<0$ is required for a non-zero order parameter. The smallest value of $f(\{a,\phi\};\{c\})$ under the constraint (\ref{eq_SM:constraints1}) will minimize the free energy $\mathcal{F}_{min}$. Of course, this is equivalent to solving the GL equations (\ref{eq:GLeqphaseqs1a})-(\ref{eq:GLeqphaseqs6a}), so we substitute all the extreme solutions from Eq.(\ref{eq:GLeqphaseqs1a})-(\ref{eq:GLeqphaseqs6a}) into the free energy (\ref{eq:lowfreenphase}) to search for the superconducting ground state. Finally, we also perform the numerical minimization of the free energy $\mathcal{F}$ in Eq.\eqref{eq_SM:genfreen} to verify the stability of these superconducting ground states. In this work, we only focus on the case with $c_4=0$. More general cases for non-zero $c_4$ and even higher order terms have been previously studied in Ref. \cite{agterberg2011}. 

\begin{figure}[h!]
    \centering
    \includegraphics[width=0.6\linewidth]{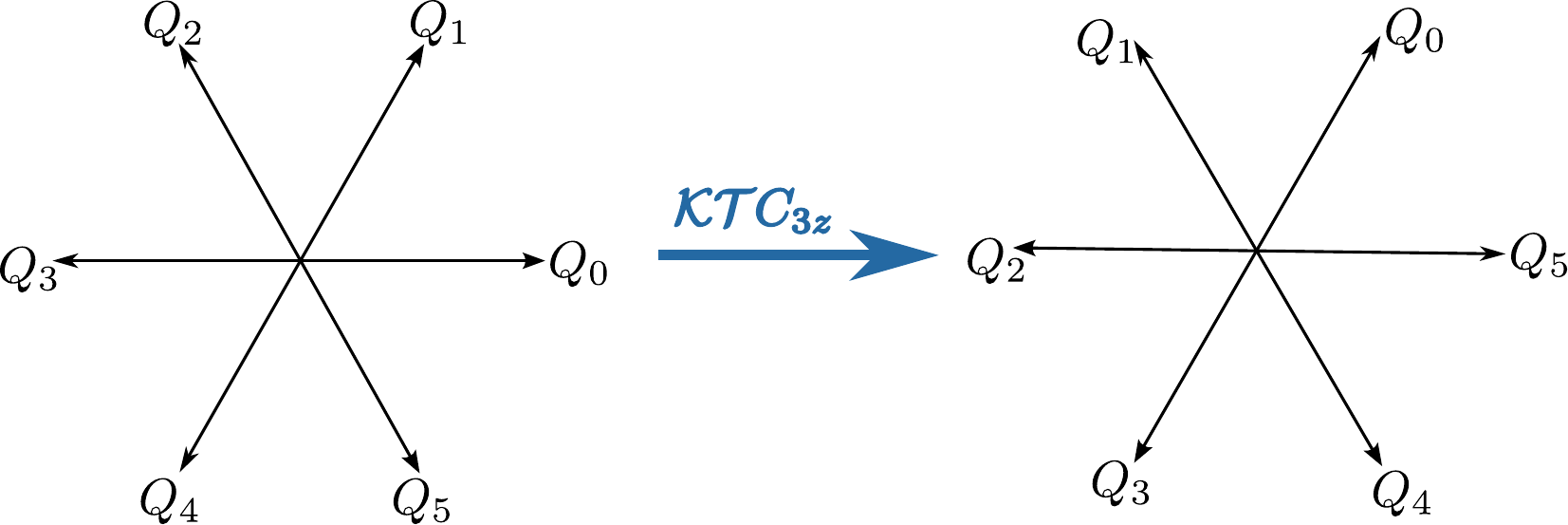}
    \caption{The transformation $\mathcal{KT}C_{3z}$ that leads to cyclic permutations $\mathcal{C}_1$\label{fig:KTC3z}, the emergent symmetry of the free energy.}
    \label{fig:placeholder}
\end{figure}

Before proceeding further, we want to comment on symmetry property of the state vector $\mathcal{P}$ to keep the free energy $\mathcal{F}$ invariant. Since the free energy is real, the action of conjugation ($\mathcal{K}$) on $P_{\QQ_j}$, namely
\begin{eqnarray}\mathcal{K}:\tilde{P}_{\QQ_j}=P^*_{\QQ_j},\end{eqnarray} 
keeps the free energy in Eq.\eqref{eq_SM:F4order_GL} invariant. 
Combining the conjugation $\mathcal{K}$ with the $C_{3z}$ and $\mathcal{T}$ symmetries in the MPG 31$'$, we find the symmetry
\begin{eqnarray}\mathcal{C}_1=\mathcal{KT}C_{3z}:\tilde{P}_{\QQ_j}=P_{-\QQ_{j-2}}=P_{\QQ_{j+1}}\end{eqnarray}
for the free energy Eq.\eqref{eq_SM:F4order_GL}. We can repeat the symmetry transformation of $\mathcal{KT}C_{3z}$, so that all the cyclic permutations of $\mathcal{P}$ in Eq.\eqref{eq_SM:Pvector_form1}
\begin{eqnarray}\label{eq:cyclicsym}
    \mathcal{C}_k: P_{\QQ_j} \to \tilde{P}_{\QQ_j} = P_{\QQ_{j+k}} \rightarrow \alpha_j\to\tilde{\alpha_j}=\alpha_{j+k} \quad \phi_j\to\tilde{\phi_j}=\phi_{j+k}, \quad \forall \quad j,k\in\mathbb{Z}_6,
\end{eqnarray}
are the symmetry operators for the free energy in Eq.\eqref{eq_SM:F4order_GL}. Any two states that are related by the $\mathcal{C}_j$ symmetry operator are degenerate, and this degeneracy can be removed for the superconducting ground state due to spontaneous symmetry breaking. This cyclic permutation ($\mathcal{C}_1$) is schematically depicted in Fig.\ref{fig:KTC3z}. 

Our aim below is to solve the set of GL equations (\ref{eq:GLeqphaseqs1a})-(\ref{eq:GLeqphaseqs6a}) and identify the superconducting ground state described by the order parameter vector $\mathcal{P}$ that minimizes the free energy in Eq.(\ref{eq_SM:freeenergy_1}).
We assume $c_0 \gg |c_{1, 2, 3, 4}|$ and $c_0 > 0$, so that the function $f$ in Eq.(\ref{eq_SM:freeenergy_1}) is always positive, which is required for the existence of minima of $\mathcal{F}$. 

Below we will consider different ans\"atze for the order parameter vector $\mathcal{P}$, labeled by $\mathcal{P}_\lambda$, where $\lambda$ denotes different ans\"atze. We notice that Eqs.(\ref{eq:GLeqphaseqs1a})-(\ref{eq:GLeqphaseqs6a}) contain the common factor $\alpha_j$ with $j=0, 1, ..., 5$, so if any $\alpha_j = 0$, the corresponding GL equation will be naturally satisfied. This motivates us to enumerate the solution ans\"atze with different numbers of non-zero components $\alpha_j$. Below, the {\bf phase 1} has only one non-zero component with the ansatz form $\mathcal{P}_1=\Delta_0 (a_0e^{i\phi_0},0,0,0,0,0)^T$. Due to the symmetry of cyclic permutation, we would expect all other ans\"atze with one non-zero component are equivalent to $\mathcal{P}_1$. For the phases with two non-zero components, there are three inequivalent ans\"atze, namely $\mathcal{P}_{2a}=e^{i\theta}\Delta_0(a_0e^{i\phi_1},a_1e^{-i\phi_1},0,0,0,0)^T$ for {\bf phase 2a}, 
$\mathcal{P}_{2b}=e^{i\theta}\Delta_0(a_0e^{i\phi_1},0,a_2e^{-i\phi_2},0,0,0)^T$ for {\bf phase 2b}
and $\mathcal{P}_{2c}=e^{i\theta}\Delta_0(a_0e^{i\phi_1},0,0,a_3e^{-i\phi_1},0,0)^T$ for {\bf phase 2c}. There are four inequivalent ans\"atze for the phases with three non-zero components of $\mathcal{P}$, which can be listed as $\mathcal{P}_{3a}=e^{i\theta}\Delta_0(a_0e^{i\phi_1},a_1e^{-i(\phi_1+\phi_2)},a_2e^{-i\phi_2},0,0,0)^T$ for phase {\bf phase 3a}, $\mathcal{P}_{3b}=e^{i\theta}\Delta_0(a_0e^{i\phi_1},a_1e^{-i(\phi_1+\phi_2)},0,a_3e^{-i\phi_2},0,0)^T$ for phase {\bf phase 3b}, $\mathcal{P}_{3c}=e^{i\theta}\Delta_0(a_0e^{i\phi_1},a_1e^{-i(\phi_1+\phi_2)},0,0,a_4e^{-i\phi_2},0)^T$ for phase {\bf phase 3c} and $\mathcal{P}_{3d}=e^{i\theta}\Delta_0(a_0e^{i\phi_1},0,a_2e^{-i(\phi_1+\phi_2)},0,a_4e^{-i\phi_2},0)^T$ for phase {\bf phase 3d}. For four non-zero component ans\"atze, we have three candidates, namely,
$\mathcal{P}_{4a}=e^{i\theta}\Delta_0(a_0e^{i\phi_1},a_1e^{-i(\phi_1+\phi_2)},a_2e^{-i\phi_2},a_3e^{-i\phi_3},0,0)^T$ for {\bf phase 4a}, $\mathcal{P}_{4b}=e^{i\theta}\Delta_0(a_0e^{i\phi_1},a_1e^{-i(\phi_1+\phi_2)},a_2e^{-i\phi_2},0,a_4e^{-i\phi_3},0)^T$ for {\bf phase 4b} and $\mathcal{P}_{4c}=e^{i\theta}\Delta_0(a_0e^{i\phi_1},a_1e^{-i(\phi_1+\phi_2)},0,a_3e^{-i\phi_2},a_4e^{-i\phi_3},0)^T$ for {\bf phase 4c}. For five-component phase, we have one possible ansatz $\mathcal{P}_5=e^{i\theta}\Delta_0(a_0e^{i\phi_1},a_1e^{-i(\phi_1+\phi_2)},a_2e^{-i\phi_2},a_3e^{-i\phi_3},a_4e^{-i\phi_4},0)^T$ called {\bf phase 5}. For six-component, the general {\bf phase 6} ansatz will be $\mathcal{P}_6=e^{i\theta}\Delta_0(a_0e^{i\phi_1},a_1e^{-i(\phi_1+\phi_2)},a_2e^{-i\phi_2},a_3e^{-i\phi_3},a_4e^{-i\phi_4},a_5e^{-i\phi_5})^T$.

Below we will calculate the minimal free energy for each phase and then compare them to determine the superconducting ground states in the $\{c\}$ parameter space. 

\textit{\textbf{Phase 1:}}
The order parameter vector of the ground state ansatz for this phase is written as
\begin{eqnarray}\label{eq:ansatz1}
\mathcal{P}_1=\Delta_0(a_0e^{i\phi_0},0,0,0,0,0)^T
\end{eqnarray}
with $a_0, \Delta_0 \neq 0$. Thus, the GL equations in Eqs.\eqref{eq:GLeqphaseqs2a}-\eqref{eq:GLeqphaseqs6a} are all naturally satisfied, and Eq.\eqref{eq:GLeqphaseqs1a} reads
\begin{eqnarray}
\Delta T+2c_0\Delta_0^2=0,
\end{eqnarray}
in which the common factors $a_0$ and $\Delta_0$ have been dropped since they are non-zero. The normalization condition in Eq.\eqref{eq_SM:constraints1} requires $a_0=1$, and when $\Delta T<0$ (the temperature $T$ is below $T_{c,0}$), we have
\begin{eqnarray}\label{eq:phase1sol}
    \Delta_0 = \sqrt{-\Delta T/2c_0}. 
\end{eqnarray}
Thus, we find the solution $\{\Delta_0, a_0\} = \{\sqrt{-\Delta T/c_0}, 1\}$ within the order parameter vector described by $\mathcal{P}_1$. For this solution, the function $f$ in Eq.\eqref{eq:optfunc} is given by
\begin{eqnarray}
f_1[\{c\}]=c_0,
\end{eqnarray}
and is larger than 0, so the solution indeed gives a free energy minimum,

\begin{eqnarray}\label{eq:phase1_minimal_energy}\mathcal{F}_{min,1}=-\frac{\Delta T^2}{4c_0}.\end{eqnarray}

\textit{\textbf{Phase 2a:}} The ground state ansatz is given by 
\begin{eqnarray}\label{eq:ansatz2a}
\mathcal{P}_{2a}=e^{i\theta}\Delta_0(a_0e^{i\phi_1},a_1e^{-i\phi_1},0,0,0,0)^T,
\end{eqnarray}
with $\Delta_0, a_0, a_1 \neq 0$. For this ansatz, the GL equations \eqref{eq:GLeqphaseqs3a}-\eqref{eq:GLeqphaseqs6a} are all naturally satisfied, and Eqs.\eqref{eq:GLeqphaseqs1a} and \eqref{eq:GLeqphaseqs2a} read
\begin{eqnarray}
&&\Delta T+\Delta_0^2\Big[2c_0+c_1a_1^2\Big]=0\label{eq_SM:GLeqn_Phase2a1}\\
&&\Delta T+\Delta_0^2\Big[2c_0+c_1a_0^2\Big]=0.\label{eq_SM:GLeqn_Phase2a2}
\end{eqnarray}
Together with the normalization condition in Eq.\eqref{eq_SM:constraints1}, Eqs.\eqref{eq_SM:GLeqn_Phase2a1} and \eqref{eq_SM:GLeqn_Phase2a2} yield the solution $\{\Delta_0, a_0, a_1\}$ given by
\begin{eqnarray}\label{eq:phase2asol}
    a_0=a_1=1/\sqrt{2}; \quad \Delta_0 = \sqrt{-\frac{\Delta T}{2c_0+c_1/2}}.
\end{eqnarray} The corresponding function $f$ in Eq.\eqref{eq:optfunc} is given by
\begin{eqnarray}
f_{2a}[\{c\}]=c_0 + c_1/4,
\end{eqnarray}
and the minimal free energy is given by
\begin{eqnarray}\label{eq:minfreen2a}
    \mathcal{F}_{min,2a}= -\frac{\Delta T^2}{4c_0+c_1}.
\end{eqnarray}

\textit{\textbf{Phase 2b:}} The ground state ansatz is given by 
\begin{eqnarray}\label{eq:ansatz2b}
\mathcal{P}_{2b}=e^{i\theta}\Delta_0(a_0e^{i\phi_1},0,a_2e^{-i\phi_1},0,0,0)^T,
\end{eqnarray}
with $\Delta_0, a_0, a_2 \neq 0$. For this ansatz, the GL equations \eqref{eq:GLeqphaseqs1a}-\eqref{eq:GLeqphaseqs6a} are all naturally satisfied, and Eqs.\eqref{eq:GLeqphaseqs1a} and \eqref{eq:GLeqphaseqs3a} read
\begin{eqnarray}
&&\Delta T+\Delta_0^2\Big[2c_0+c_2a_2^2\Big]=0\label{eq_SM:GLeqn_Phase2b1}\\
&&\Delta T+\Delta_0^2\Big[2c_0+c_2a_0^2\Big]=0.\label{eq_SM:GLeqn_Phase2b2}
\end{eqnarray}
Together with the normalization condition in Eq.\eqref{eq_SM:constraints1}, Eqs.\eqref{eq_SM:GLeqn_Phase2b1} and \eqref{eq_SM:GLeqn_Phase2b2} yield the solution $\{\Delta_0, a_0, a_2\}$ given by
\begin{eqnarray}\label{eq:phase2bsol}
    a_0=a_2=1/\sqrt{2}; \quad \Delta_0 = \sqrt{-\frac{\Delta T}{2c_0+c_2/2}}.
\end{eqnarray}

The corresponding function $f$ in Eq.\eqref{eq:optfunc} is given by
\begin{eqnarray}
f_{2b}[\{c\}]=c_0 + c_2/4,
\end{eqnarray}
and the minimal free energy is given by
\begin{eqnarray}\label{eq:minfreen2b}
    \mathcal{F}_{min,2b}= -\frac{\Delta T^2}{4c_0+c_2}.
\end{eqnarray}

\textit{\textbf{Phase 2c:}} The ground state ansatz is given by 
\begin{eqnarray}\label{eq:ansatz2c}
\mathcal{P}_{2c}=e^{i\theta}\Delta_0(a_0e^{i\phi_1},0,0,a_3e^{-i\phi_1},0,0)^T,
\end{eqnarray}
with $\Delta_0, a_0, a_3 \neq 0$. For this ansatz, the GL equations \eqref{eq:GLeqphaseqs1a}-\eqref{eq:GLeqphaseqs6a} are all naturally satisfied, and Eqs.\eqref{eq:GLeqphaseqs1a} and \eqref{eq:GLeqphaseqs4a} read
\begin{eqnarray}
&&\Delta T+\Delta_0^2\Big[2c_0+2c_3a_3^2\Big]=0\label{eq_SM:GLeqn_Phase2c1}\\
&&\Delta T+\Delta_0^2\Big[2c_0+2c_3a_0^2\Big]=0.\label{eq_SM:GLeqn_Phase2c2}
\end{eqnarray}
Together with the normalization condition in Eq.\eqref{eq_SM:constraints1}, Eqs.\eqref{eq_SM:GLeqn_Phase2c1} and \eqref{eq_SM:GLeqn_Phase2c2} yield the solution $\{\Delta_0, a_0, a_3\}$ given by
\begin{eqnarray}\label{eq:phase2csol}
    a_0=a_3=1/\sqrt{2}; \quad \Delta_0 = \sqrt{-\frac{\Delta T}{2c_0+c_3}}.
\end{eqnarray} The corresponding function $f$ in Eq.\eqref{eq:optfunc} is given by
\begin{eqnarray}
f_{2c}[\{c\}]=c_0 + c_3/2,
\end{eqnarray}
and the minimal free energy is given by
\begin{eqnarray}\label{eq:minfreen2c}
    \mathcal{F}_{min,2c}= \frac{\Delta T^2}{4c_0+2c_3}.
\end{eqnarray}

\textit{\textbf{Phase 3a:}} The ground state ansatz is given by 
\begin{eqnarray}\label{eq:ansatz3a}
\mathcal{P}_{3a}=e^{i\theta}\Delta_0(a_0e^{i\phi_1},a_1e^{-i(\phi_1+\phi_2)},a_2e^{i\phi_2},0,0,0)^T,
\end{eqnarray}
with $\Delta_0, a_0, a_1, a_2 \neq 0$. For this ansatz, the GL equations \eqref{eq:GLeqphaseqs1a}-\eqref{eq:GLeqphaseqs6a} are all naturally satisfied, and Eqs.\eqref{eq:GLeqphaseqs1a}-\eqref{eq:GLeqphaseqs3a} read
\begin{eqnarray}
&&\Delta T+\Delta_0^2\Big[2c_0+c_1a_1^2+c_2a_2^2\Big]=0\label{eq_SM:GLeqn_Phase3a1}\\
&&\Delta T+\Delta_0^2\Big[2c_0+c_1(a_0^2+a_2^2)\Big]=0\label{eq_SM:GLeqn_Phase3a2}\\
&&\Delta T+\Delta_0^2\Big[2c_0+c_1a_1^2+c_2a_0^2\Big]=0.\label{eq_SM:GLeqn_Phase3a3}
\end{eqnarray}
Together with the normalization condition in Eq.\eqref{eq_SM:constraints1}, Eqs.\eqref{eq_SM:GLeqn_Phase3a1}-\eqref{eq_SM:GLeqn_Phase3a3} yield the solution $\{\Delta_0, a_0, a_1, a_2\}$ given by
\begin{eqnarray}\label{eq:phase3asol}
    a_0=a_2=\frac{a_1}{\sqrt{2-c_2/c_1}}=\sqrt{\frac{c_1}{4c_1-c_2}}; \quad \Delta_0 = \sqrt{-\frac{\Delta T}{2c_0+\frac{2c_1^2}{4c_1-c_2}}}.
\end{eqnarray}The corresponding function $f$ in Eq.\eqref{eq:optfunc} is given by
\begin{eqnarray}
f_{3a}[\{c\}]=c_0 + \frac{c_1^2}{4c_1-c_2},
\end{eqnarray}
and the minimal free energy is given by
\begin{eqnarray}\label{eq:minfreen3a}
    \mathcal{F}_{min,3a}= -\frac{\Delta T^2}{4c_0+\frac{4c_1^2}{4c_1-c_2}}.
\end{eqnarray}

\textit{\textbf{Phase 3b:}} The ground state ansatz is given by 
\begin{eqnarray}\label{eq:ansatz3b}
\mathcal{P}_{3b}=e^{i\theta}\Delta_0(a_0e^{i\phi_1},a_1e^{-i(\phi_1+\phi_2)},0,a_3e^{i\phi_2},0,0)^T,
\end{eqnarray}
with $\Delta_0, a_0, a_1, a_3 \neq 0$. For this ansatz, the GL equations \eqref{eq:GLeqphaseqs1a}-\eqref{eq:GLeqphaseqs6a} are all naturally satisfied, and Eqs.\eqref{eq:GLeqphaseqs1a},\eqref{eq:GLeqphaseqs2a} and \eqref{eq:GLeqphaseqs4a} read
\begin{eqnarray}
&&\Delta T+\Delta_0^2\Big[2c_0+c_1a_1^2+2c_3a_3^2\Big]=0\label{eq_SM:GLeqn_Phase3b1}\\
&&\Delta T+\Delta_0^2\Big[2c_0+c_1a_0^2+c_2a_3^2\Big]=0\label{eq_SM:GLeqn_Phase3b2}\\
&&\Delta T+\Delta_0^2\Big[2c_0+c_2a_1^2+2c_3a_0^2\Big]=0.\label{eq_SM:GLeqn_Phase3b3}
\end{eqnarray}
Together with the normalization condition in Eq.\eqref{eq_SM:constraints1}, Eqs.\eqref{eq_SM:GLeqn_Phase3b1}, \eqref{eq_SM:GLeqn_Phase3b2} and \eqref{eq_SM:GLeqn_Phase3b3} yield the solution $\{\Delta_0, a_0, a_3\}$ given by
\begin{eqnarray}\label{eq:phase3bsol}
    &&\frac{a_0}{\sqrt{c_2(c_2-c_1-2c_3)}}=\frac{a_1}{\sqrt{-2c_3(c_1+c_2-2c_3)}}=\frac{a_3}{\sqrt{c_1(c_1-c_2-2c_3)}}=\frac{1}{\sqrt{c_1^2+(c_2-2c_3)^2-2c_1(c_2+2c_3)}};\nonumber\\&&\Delta_0 = \sqrt{-\frac{\Delta T}{2c_0- \frac{4c_1c_2c_3}{c_1^2+(c_2-2c_3)^2-2c_1(c_2+2c_3)}}}.
\end{eqnarray} The corresponding function $f$ in Eq.\eqref{eq:optfunc} is given by
\begin{eqnarray}
f_{3b}[\{c\}]=c_0 - \frac{2c_1c_2c_3}{c_1^2+(c_2-2c_3)^2-2c_1(c_2+2c_3)},
\end{eqnarray}
and the minimal free energy is given by
\begin{eqnarray}\label{eq:minfreen3b}
    \mathcal{F}_{min,3b}= \frac{\Delta T^2}{4c_0- \frac{8c_1c_2c_3}{c_1^2+(c_2-2c_3)^2-2c_1(c_2+2c_3)}}.
\end{eqnarray}

\textit{\textbf{Phase 3c:}} The ground state ansatz is given by 
\begin{eqnarray}\label{eq:ansatz3c}
\mathcal{P}_{3c}=e^{i\theta}\Delta_0(a_0e^{i\phi_1},a_1e^{-i(\phi_1+\phi_2)},0,0,a_4e^{i\phi_2},0)^T,
\end{eqnarray}
with $\Delta_0, a_0, a_1, a_4 \neq 0$. For this ansatz, the GL equations \eqref{eq:GLeqphaseqs1a}-\eqref{eq:GLeqphaseqs6a} are all naturally satisfied, and Eqs.\eqref{eq:GLeqphaseqs1a}, \eqref{eq:GLeqphaseqs3a} read
\begin{eqnarray}
&&\Delta T+\Delta_0^2\Big[2c_0+c_1a_1^2+c_2a_4^2\Big]=0\label{eq_SM:GLeqn_Phase3c1}\\
&&\Delta T+\Delta_0^2\Big[2c_0+c_1a_0^2+2c_3a_4^2\Big]=0\label{eq_SM:GLeqn_Phase3c2}\\
&&\Delta T+\Delta_0^2\Big[2c_0+c_2a_0^2+2c_3a_1^2\Big]=0.\label{eq_SM:GLeqn_Phase3c3}
\end{eqnarray}
Together with the normalization condition in Eq.\eqref{eq_SM:constraints1}, Eqs.\eqref{eq_SM:GLeqn_Phase3c1}- \eqref{eq_SM:GLeqn_Phase3c3} yield the solution $\{\Delta_0, a_0,a_1,a_4\}$ given by
\begin{eqnarray}\label{eq:phase3csol}
    &&\frac{a_0}{\sqrt{-2c_3(c_2+c_1-2c_3)}}=\frac{a_1}{\sqrt{c_2(c_1-c_2+2c_3)}}=\frac{a_4}{\sqrt{c_1(c_1-c_2-2c_3)}}=\frac{1}{\sqrt{c_1^2+(c_2-2c_3)^2-2c_1(c_2+2c_3)}};\nonumber\\&&\Delta_0 = \sqrt{-\frac{\Delta T}{2c_0- \frac{4c_1c_2c_3}{c_1^2+(c_2-2c_3)^2-2c_1(c_2+2c_3)}}}.
\end{eqnarray}
The corresponding function $f$ in Eq.\eqref{eq:optfunc} is given by
\begin{eqnarray}
f_{3c}[\{c\}]=c_0 - \frac{2c_1c_2c_3}{c_1^2+(c_2-2c_3)^2-2c_1(c_2+2c_3)},
\end{eqnarray}
and the minimal free energy is given by
\begin{eqnarray}\label{eq:minfreen3c}
    \mathcal{F}_{min,3c}= -\frac{\Delta T^2}{4c_0- \frac{8c_1c_2c_3}{c_1^2+(c_2-2c_3)^2-2c_1(c_2+2c_3)}}.
\end{eqnarray}

\textit{\textbf{Phase 3d:}} The ground state ansatz is given by 
\begin{eqnarray}\label{eq:ansatz3d}
\mathcal{P}_{3d}=e^{i\theta}\Delta_0(a_0e^{i\phi_1},0,a_2e^{-i(\phi_1+\phi_2)},0,a_4e^{i\phi_2},0)^T,
\end{eqnarray}
with $\Delta_0, a_0, a_2, a_4 \neq 0$. For this ansatz, the GL equations \eqref{eq:GLeqphaseqs1a}-\eqref{eq:GLeqphaseqs6a} are all naturally satisfied, and Eqs.\eqref{eq:GLeqphaseqs1a}, \eqref{eq:GLeqphaseqs3a} and \eqref{eq:GLeqphaseqs5a} read
\begin{eqnarray}
&&\Delta T+\Delta_0^2\Big[2c_0+c_2(a_2^2+a_4^2)\Big]=0\label{eq_SM:GLeqn_Phase3d1}\\
&&\Delta T+\Delta_0^2\Big[2c_0+c_2(a_4^2+a_0^2)\Big]=0\label{eq_SM:GLeqn_Phase3d2}\\
&&\Delta T+\Delta_0^2\Big[2c_0+c_2(a_0^2+a_2^2)\Big]=0.\label{eq_SM:GLeqn_Phase3d3}
\end{eqnarray}
Together with the normalization condition in Eq.\eqref{eq_SM:constraints1}, Eqs.\eqref{eq_SM:GLeqn_Phase3d1}- \eqref{eq_SM:GLeqn_Phase3d3} yield the solution $\{\Delta_0, a_0, a_2,a_4\}$ given by
\begin{eqnarray}\label{eq:phase3dsol}
    &&a_0=a_2=a_4=\frac{1}{\sqrt{3}};\quad\Delta_0 = \sqrt{-\frac{\Delta T}{2c_0+\frac{2c_2}{3}}}.
\end{eqnarray}
The corresponding function $f$ in Eq.\eqref{eq:optfunc} is given by
\begin{eqnarray}
f_{3d}[\{c\}]=c_0 + \frac{c_2}{3},
\end{eqnarray}
and the minimal free energy is given by
\begin{eqnarray}\label{eq:minfreen3d}
    \mathcal{F}_{min,3d}= -\frac{\Delta T^2}{4c_0+ \frac{4c_2}{3}}.
\end{eqnarray}

\textit{\textbf{Phase 4a:}} The ground state ansatz is given by 
\begin{eqnarray}\label{eq:ansatz4a}
\mathcal{P}_{4a}=e^{i\theta}\Delta_0(a_0e^{i\phi_1},a_1e^{-i(\phi_1+\phi_2)},a_2e^{i\phi_2},a_3e^{-i\phi_3},0,0)^T,
\end{eqnarray}
with $\Delta_0, a_0, a_1, a_4 \neq 0$. For this ansatz, the GL equations \eqref{eq:GLeqphaseqs1a}-\eqref{eq:GLeqphaseqs6a} are all naturally satisfied, and Eqs.\eqref{eq:GLeqphaseqs1a}-\eqref{eq:GLeqphaseqs4a} read
\begin{eqnarray}
&&\Delta T+\Delta_0^2\Big[2c_0+c_1a_1^2+c_2a_2^2+2c_3a_3^2\Big]=0\label{eq_SM:GLeqn_Phase4a1}\\
&&\Delta T+\Delta_0^2\Big[2c_0+c_1(a_0^2+a_1^2)+c_2a_3^2\Big]=0\label{eq_SM:GLeqn_Phase4a2}\\
&&\Delta T+\Delta_0^2\Big[2c_0+c_1(a_1^2+a_3^2)+c_2a_0^2\Big]=0\label{eq_SM:GLeqn_Phase4a3}\\
&&\Delta T+\Delta_0^2\Big[2c_0+c_2a_1^2+c_1a_2^2+2c_3a_0^2\Big]=0.\label{eq_SM:GLeqn_Phase4a4}
\end{eqnarray}
Together with the normalization condition in Eq.\eqref{eq_SM:constraints1}, Eqs.\eqref{eq_SM:GLeqn_Phase4a1}-\eqref{eq_SM:GLeqn_Phase4a4} yield the solution $\{\Delta_0, a_0,a_1, a_2,a_3\}$ given by
\begin{eqnarray}\label{eq:phase4asol}
    &&\frac{a_0}{\sqrt{c_2}}=\frac{a_1}{\sqrt{c_1+c_2-2c_3}}=\frac{a_2}{\sqrt{c_1+c_2-2c_3}}=\frac{a_3}{\sqrt{c_2}}=\frac{1}{\sqrt{2c_1+4c_2-4c_3}};\nonumber\\&&\Delta_0 = \sqrt{-\frac{\Delta T}{2c_0 + \frac{1}{2}\Big(c_1+\frac{c_2^2}{c_1+2c_2-2c_3}\Big)}}.
\end{eqnarray} The corresponding function $f$ in Eq.\eqref{eq:optfunc} is given by
\begin{eqnarray}
f_{4a}[\{c\}]=c_0 + \frac{1}{4}\Big(c_1+\frac{c_2^2}{c_1+2c_2-2c_3}\Big),
\end{eqnarray}
and the minimal free energy is given by
\begin{eqnarray}\label{eq:minfreen4a}
    \mathcal{F}_{min,4a}= -\frac{\Delta T^2}{4c_0 + \Big(c_1+\frac{c_2^2}{c_1+2c_2-2c_3}\Big)}.
\end{eqnarray}


\textit{\textbf{Phase 4b:}} The ground state ansatz is given by 
\begin{eqnarray}\label{eq:ansatz4b}
\mathcal{P}_{4b}=e^{i\theta}\Delta_0(a_0e^{i\phi_1},a_1e^{-i(\phi_1+\phi_2)},a_2e^{-i\phi_2},0,a_4e^{-i\phi_3},0)^T,
\end{eqnarray}
with $\Delta_0, a_0, a_1, a_2, a_4 \neq 0$. For this ansatz, the GL equations \eqref{eq:GLeqphaseqs1a}-\eqref{eq:GLeqphaseqs6a} are all naturally satisfied, and Eqs.\eqref{eq:GLeqphaseqs1a},\eqref{eq:GLeqphaseqs2a},\eqref{eq:GLeqphaseqs3a} and \eqref{eq:GLeqphaseqs5a} read
\begin{eqnarray}
&&\Delta T+\Delta_0^2\Big[2c_0+c_1a_1^2+c_2(a_2^2+a_4^2)\Big]=0\label{eq_SM:GLeqn_Phase4b1}\\
&&\Delta T+\Delta_0^2\Big[2c_0+c_1(a_0^2+a_2^2)+2c_3a_4^2\Big]=0\label{eq_SM:GLeqn_Phase4b2}\\
&&\Delta T+\Delta_0^2\Big[2c_0+c_1a_1^2+c_2(a_0^2+a_4^2)\Big]=0\label{eq_SM:GLeqn_Phase4b3}\\
&&\Delta T+\Delta_0^2\Big[2c_0+c_2(a_0^2+a_2^2)+2c_3a_1^2\Big]=0.\label{eq_SM:GLeqn_Phase4b4}
\end{eqnarray}
Together with the normalization condition in Eq.\eqref{eq_SM:constraints1}, Eqs.\eqref{eq_SM:GLeqn_Phase4b1}-\eqref{eq_SM:GLeqn_Phase4b4} yield the solution $\{\Delta_0, a_0, a_2,a_3\}$ given by
\begin{eqnarray}\label{eq:phase4bsol}
    &&\frac{a_0}{\sqrt{-(c_1+c_2-2c_3)c_3}}=\frac{a_1}{\sqrt{c_2(-c_1+c_2-c_3)}}=\frac{a_2}{\sqrt{-c_3(c_1+c_2-2c_3)}}=\frac{a_4}{\sqrt{c_1(c_2+2c_3)-c_1^2-c_2c_3}}\nonumber\\&=&\frac{1}{\sqrt{(c_1-c_2)^2-2(2c_1+c_2)c_3+4c_3^2}};\quad\Delta_0 = \sqrt{-\frac{\Delta T}{2c_0 + 2\frac{c_2c_3(c_3-2c_1)}{(c_1-c_2)^2-2(2c_1+c_2)c_3+4c_3^2}}}.
\end{eqnarray} The corresponding function $f$ in Eq.\eqref{eq:optfunc} is given by
\begin{eqnarray}
f_{4b}[\{c\}]=c_0 + \frac{c_2c_3(c_3-2c_1)}{(c_1-c_2)^2-2(2c_1+c_2)c_3+4c_3^2},
\end{eqnarray}
and the minimal free energy is given by
\begin{eqnarray}\label{eq:minfreen4b}
    \mathcal{F}_{min,4b}= -\frac{\Delta T^2}{4c_0 + \frac{4c_2c_3(c_3-2c_1)}{(c_1-c_2)^2-2(2c_1+c_2)c_3+4c_3^2}}.
\end{eqnarray}

\textit{\textbf{Phase 4c:}} The ground state ansatz is given by 
\begin{eqnarray}\label{eq:ansatz4c}
\mathcal{P}_{4c}=e^{i\theta}\Delta_0(a_0e^{i\phi_1},a_1e^{-i(\phi_1+\phi_2)},0,a_3e^{i\phi_2},a_4e^{-i\phi_3},0)^T,
\end{eqnarray}
with $\Delta_0, a_0, a_1, a_3, a_4 \neq 0$. For this ansatz, the GL equations \eqref{eq:GLeqphaseqs1a}-\eqref{eq:GLeqphaseqs6a} are all naturally satisfied, and Eqs.\eqref{eq:GLeqphaseqs1a},\eqref{eq:GLeqphaseqs2a},\eqref{eq:GLeqphaseqs4a} and \eqref{eq:GLeqphaseqs5a} read
\begin{eqnarray}
&&\Delta T+\Delta_0^2\Big[2c_0+c_1a_1^2+2c_3a_3^2+c_2a_4^2\Big]=0\label{eq_SM:GLeqn_Phase4c1}\\
&&\Delta T+\Delta_0^2\Big[2c_0+c_1a_0^2+2c_3a_4^2+c_2a_3^2\Big]=0\label{eq_SM:GLeqn_Phase4c2}\\
&&\Delta T+\Delta_0^2\Big[2c_0+c_1a_4^2+2c_3a_0^2+c_2a_1^2\Big]=0\label{eq_SM:GLeqn_Phase4c3}\\
&&\Delta T+\Delta_0^2\Big[2c_0+c_1a_3^2+2c_3a_1^2+c_2a_0^2\Big]=0.\label{eq_SM:GLeqn_Phase4c4}
\end{eqnarray}
Together with the normalization condition in Eq.\eqref{eq_SM:constraints1}, Eqs.\eqref{eq_SM:GLeqn_Phase4c1}- \eqref{eq_SM:GLeqn_Phase4c4} yield the solution $\{\Delta_0, a_0, a_2,a_3\}$ given by
\begin{eqnarray}\label{eq:phase4csol}
    &&a_0=a_1=a_3=a_4=\frac{1}{2};\nonumber\\&&\Delta_0 = \sqrt{-\frac{\Delta T}{2c_0+\frac{c_1+c_2+2c_3}{4}}}.
\end{eqnarray}
The corresponding function $f$ in Eq.\eqref{eq:optfunc} is given by
\begin{eqnarray}
f_{4c}[\{c\}]=c_0+\frac{c_1+c_2+2c_3}{8},
\end{eqnarray}
and the minimal free energy is given by
\begin{eqnarray}\label{eq:minfreen4c}
    \mathcal{F}_{min,4c}= -\frac{\Delta T^2}{4c_0 +\frac{c_1+c_2+2c_3}{2}}.
\end{eqnarray}

\textit{\textbf{Phase 5:}} The ground state ansatz is given by 
\begin{eqnarray}\label{eq:ansatz5}
\mathcal{P}_5=e^{i\theta}\Delta_0(a_0e^{i\phi_1},a_1e^{-i(\phi_1+\phi_2)},a_2e^{-i\phi_2},a_3e^{-i\phi_3},a_4e^{-i\phi_4},0)^T,
\end{eqnarray}
with $\Delta_0, a_0, a_1, a_2, a_3, a_4 \neq 0$. For this ansatz, the GL equations \eqref{eq:GLeqphaseqs1a}-\eqref{eq:GLeqphaseqs6a} are all naturally satisfied, and Eqs.\eqref{eq:GLeqphaseqs1a}-\eqref{eq:GLeqphaseqs5a} read
\begin{eqnarray}
&&\Delta T+\Delta_0^2\Big[2c_0+c_1a_1^2+2c_3a_3^2+c_2(a_2^2+a_4^2)\Big]=0\label{eq_SM:GLeqn_Phase5a1}\\
&&\Delta T+\Delta_0^2\Big[2c_0+c_1(a_0^2+a_2^2)+2c_3a_4^2+c_2a_3^2\Big]=0\label{eq_SM:GLeqn_Phase5a2}\\
&&\Delta T+\Delta_0^2\Big[2c_0+c_1(a_1^2+a_3^2)+c_2(a_0^2+a_4^2)\Big]=0\label{eq_SM:GLeqn_Phase5a3}\\
&&\Delta T+\Delta_0^2\Big[2c_0+c_1(a_2^2+a_4^2)+2c_3a_0^2+c_2a_1^2\Big]=0\label{eq_SM:GLeqn_Phase5a4}\\
&&\Delta T+\Delta_0^2\Big[2c_0+c_1a_3^2+2c_3a_1^2+c_2(a_0^2+a_2^2)\Big]=0.\label{eq_SM:GLeqn_Phase5a5}
\end{eqnarray}
Together with the normalization condition in Eq.\eqref{eq_SM:constraints1}, Eqs.\eqref{eq_SM:GLeqn_Phase5a1}-\eqref{eq_SM:GLeqn_Phase5a5}yield the solution $\{\Delta_0, a_0, a_2,a_3\}$ given by
\begin{eqnarray}\label{eq:phase5sol}
    &&\frac{a_0}{\sqrt{c_1^2+c_2^2-2c_1(c_2+c_3)}}=\frac{a_1}{\sqrt{2c_2(-c_1+c_2-c_3)}}=\frac{a_2}{\sqrt{c_1^2-(c_2-2c_3)^2}}\nonumber\\&=&\frac{a_3}{\sqrt{2c_2(-c_1+c_2-c_3)}}=\frac{a_4}{\sqrt{c_1^2+c_2^2-2c_1(c_2+c_3)}}\nonumber\\&=&\frac{1}{\sqrt{c_1^2+7c_2^2-8c_2c_3+4c_3^2-4c_1(2c_2+c_3)}};\quad\Delta_0 = \sqrt{-\frac{\Delta T}{2c_0 + 2\frac{c_2\{c_2^2-c_1(c_1+4c_3)\}}{c_1^2+7c_2^2-8c_2c_3+4c_3^2-4c_1(2c_2+c_3)}}}.
\end{eqnarray} The corresponding function $f$ in Eq.\eqref{eq:optfunc} is given by
\begin{eqnarray}
f_5[\{c\}]=c_0 + \frac{c_2\{c_2^2-c_1(c_1+4c_3)\}}{c_1^2+7c_2^2-8c_2c_3+4c_3^2-4c_1(2c_2+c_3)},
\end{eqnarray}
and the minimal free energy is given by
\begin{eqnarray}\label{eq:minfreen5}
    \mathcal{F}_{min,5}= -\frac{\Delta T^2}{4c_0 + \frac{4c_2\{c_2^2-c_1(c_1+4c_3)\}}{c_1^2+7c_2^2-8c_2c_3+4c_3^2-4c_1(2c_2+c_3)}}.
\end{eqnarray}

\textit{\textbf{Phase 6:}} The ground state ansatz is given by 
\begin{eqnarray}\label{eq:ansatz6}
\mathcal{P}_6=e^{i\theta}\Delta_0(a_0e^{i\phi_1},a_1e^{-i(\phi_1+\phi_2)},a_2e^{-i\phi_2},a_3e^{i\phi_3},a_4e^{i\phi_4},a_5e^{-i\phi_5})^T,
\end{eqnarray}
with $\Delta_0, a_0, a_1, a_4 \neq 0$. For this ansatz, the GL equations \eqref{eq:GLeqphaseqs1a}-\eqref{eq:GLeqphaseqs6a} are all naturally satisfied,
\begin{eqnarray}
&&\Delta T+\Delta_0^2\Big[2c_0+c_1(a_1^2+a_5^2)+2c_3a_3^2+c_2(a_2^2+a_4^2)\Big]=0\label{eq_SM:GLeqn_Phase6a1}\\
&&\Delta T+\Delta_0^2\Big[2c_0+c_1(a_0^2+a_2^2)+2c_3a_4^2+c_2(a_5^2+a_3^2)\Big]=0\label{eq_SM:GLeqn_Phase6a2}\\
&&\Delta T+\Delta_0^2\Big[2c_0+c_1(a_1^2+a_3^2)+c_2(a_0^2+a_4^2)+2c_3a_5^2\Big]=0\label{eq_SM:GLeqn_Phase6a3}\\
&&\Delta T+\Delta_0^2\Big[2c_0+c_1(a_2^2+a_4^2)+2c_3a_0^2+c_2(a_1^2+a_5^2)\Big]=0\label{eq_SM:GLeqn_Phase6a4}\\
&&\Delta T+\Delta_0^2\Big[2c_0+c_1(a_3^2+a_5^2)+2c_3a_1^2+c_2(a_0^2+a_2^2)\Big]=0\label{eq_SM:GLeqn_Phase6a5}\\
&&\Delta_0^2\Big[2c_0+c_1(a_0^2+a_4^2)+2c_3a_2^2+c_2(a_1^2+a_3^2)\Big]=0.\label{eq_SM:GLeqn_Phase6a6}
\end{eqnarray}
Together with the normalization condition in Eq.\eqref{eq_SM:constraints1}, Eqs.\eqref{eq_SM:GLeqn_Phase6a1}-\eqref{eq_SM:GLeqn_Phase6a6} yield the solution $\{\Delta_0, a_0, a_2,a_3,a_4,a_5\}$ given by
\begin{eqnarray}\label{eq:phase6sol}
    &&a_0=a_1=a_2=a_3=a_4=a_5=\frac{1}{\sqrt{6}};\quad\Delta_0 = \sqrt{-\frac{\Delta T}{2c_0 + \frac{c_1+c_2+c_3}{3}}}.
\end{eqnarray} The corresponding function $f$ in Eq.\eqref{eq:optfunc} is given by
\begin{eqnarray}
f_6[\{c\}]=c_0 + \frac{c_1+c_2+c_3}{6},
\end{eqnarray}
and the minimal free energy is given by
\begin{eqnarray}\label{eq:minfreen6}
    \mathcal{F}_{min,6}= -\frac{\Delta T^2}{4c_0 + \frac{2(c_1+c_2+c_3)}{3}}.
\end{eqnarray}

\begin{table}[h]
\begin{center}
\resizebox{\textwidth}{!}{
\begin{tabular}{| c | c | c |} 
\hline
Phase & Ground state vector $\mathcal{P}$
& $\mathcal{F}_{min}$\\[3pt]
\hline
\hline
1&$\boldsymbol{e^{i\theta}{\alpha_0}(1,0,0,0,0,0)^T}$&$\frac{\Delta T^2}{4c_0}$\\[5pt]
\hline
2a&$\boldsymbol{e^{i\theta}{\alpha_0}(e^{i\phi_1},e^{-i\phi_1},0,0,0,0)^T}$&$-\frac{\Delta T^2}{4c_0+c_1}$\\[5pt]
\hline
{2b}&{$e^{i\theta}{\alpha_0}(e^{i\phi_1},0,e^{-i\phi_1},0,0,0)^T$}&$-\frac{\Delta T^2}{4c_0+c_2}$\\[5pt]
\hline
2c&$\boldsymbol{e^{i\theta}{\alpha_0}(e^{i\phi_1},0,0,e^{-i\phi_1},0,0)^T}$&$\frac{\Delta T^2}{4c_0+2c_3}$\\[5pt]
\hline
3a&$\boldsymbol{e^{i\theta}\alpha_0(\frac{\cos{\epsilon}}{\sqrt{2}}e^{i\phi_1},e^{-i(\phi_1+\phi_2)}\sin{\epsilon},\frac{\cos{\epsilon}}{\sqrt{2}}e^{-i\phi_1},0,0,0)^T}$&$-\frac{\Delta T^2}{4c_0+4c_1^2/(4c_1-c_2)}$\\[8pt]
\hline
{3b}&{$e^{i\theta}(\alpha_0e^{i\phi_1},\alpha_1e^{-i(\phi_1+\phi_2)},0,\alpha_3e^{i\phi_2},0,0)^T$}&$\frac{\Delta T^2}{4c_0-{8c_1c_2c_3}/(c_1^2+(c_2-2c_3)^2-2c_1(c_2+2c_3))}$\\[5pt]
\hline
{3c}&{$e^{i\theta}(\alpha_0e^{i\phi_1},\alpha_1e^{-i(\phi_1+\phi_2)},0,0,\alpha_3e^{i\phi_2},0)^T$}&$-\frac{\Delta T^2}{4c_0- {8c_1c_2c_3}/(c_1^2+(c_2-2c_3)^2-2c_1(c_2+2c_3))}$\\[5pt]
\hline
3d&$\boldsymbol{e^{i\theta}\alpha_0(e^{i\phi_1},0,e^{i\phi_2},0,e^{-i(\phi_1+\phi_2
)},0)^T}$&$-\frac{\Delta T^2}{4c_0+ \frac{4c_2}{3}}$\\[5pt]
\hline
4a&$\boldsymbol{e^{i\theta}(\alpha_0e^{i\phi_1},\alpha_1e^{-i(\phi_1+\phi_2)},\alpha_1e^{i\phi_2},\alpha_0e^{-i\phi_3},0,0)^T}$&$-\frac{\Delta T^2}{4c_0 + c_1+{c_2^2}/(c_1+2c_2-2c_3)}$\\[5pt]
\hline
4b&$\boldsymbol{e^{i\theta}(\alpha_0e^{i\phi_1},\alpha_1e^{-i(\phi_1+\phi_2)},\alpha_0e^{i\phi_2},0,\alpha_4ie^{-i\phi_3},0)^T}$&$-\frac{\Delta T^2}{4c_0 + {4c_2c_3(c_3-2c_1)}/\{(c_1-c_2)^2-2(2c_1+c_2)c_3+4c_3^2\}}$\\[5pt]
\hline
{4c}&{$e^{i\theta}{\alpha_0}(e^{i\phi_1},e^{-i(\phi_1+\phi_2)},0,e^{i\phi_2},e^{i\phi_4},0)^T$}&$-\frac{\Delta T^2}{4c_0 +\frac{c_1+c_2+2c_3}{2}}$\\[5pt]
\hline
5&$\boldsymbol{e^{i\theta}(\alpha_0e^{i\phi_1},\alpha_1e^{-i(\phi_1+\phi_2)},\alpha_2e^{-i\phi_2},\alpha_1e^{-i\phi_3},\alpha_0e^{-i\phi_4},0)^T}$&$-\frac{\Delta T^2}{4c_0 + \frac{4c_2\{c_2^2-c_1(c_1+4c_3)\}}{c_1^2+7c_2^2-8c_2c_3+4c_3^2-4c_1(2c_2+c_3)}}$\\[10pt]
\hline
6&$\boldsymbol{e^{i\theta}\alpha_0(e^{i\phi_1},e^{-i(\phi_1+\phi_2)},e^{-i\phi_2},e^{i\phi_3},e^{i\phi_4},e^{-i\phi_5})^T}$&$-\frac{\Delta T^2}{4c_0 + \frac{2(c_1+c_2+c_3)}{3}}$\\[5pt]
\hline
\end{tabular}}
\end{center}
\caption{General form of the possible ground states from solving the total free energy for $c_4=0$. $\alpha_j=\Delta_0a_j$, where $\Delta_0$ is the global and $a_j$ is the relative amplitude of the components in $\mathcal{P}$ for each phase. Details on these magnitudes are discussed in Sec.\ref{sec:SCgstates}. Ground states highlighted are the ones that exist in some parameter regions of fourth order GL parameters $\{c\}$. The other states never form a global minima. The last column shows the minimal free energy $\mathcal{F}_{min}$ for each possible ground state $\mathcal{P}$. \label{tab:gstates}}
\end{table}

Table III summarized all phases possible as local minima of of the free energy. We have derived the minimal free energy $\mathcal{F}_{min}$ for a certain ansatz of SC order parameter, which is a function of the second order and fourth order GL parameters, $\Delta T$ and $\{c\}$, respectively. To determine the ground state for a particular choice of $\Delta T$ and $\{c\})$, we need to compare the minimal free energy $\mathcal{F}_{min,\lambda}$ for different ans\"atze (we below label different ans\"atze by the index $\lambda$) and identify the ansatz with the lowest value of $\mathcal{F}_{min,\lambda}$. Note that $\Delta T$ is a constant in the $U(6)$ subspace, the minimal free energy requires the conditions
\begin{eqnarray}\label{eq_SM:paramcons}
&&\bigcap\limits_{\substack{\lambda'\in\mathcal{Y}\\\lambda'\neq\lambda}}\Big\{\mathcal{F}_{min,\lambda}<\mathcal{F}_{min,\lambda'}\Big\} \Rightarrow\bigcap\limits_{\substack{\lambda'\in\mathcal{Y}\\\lambda'\neq\lambda}}\Big\{-\frac{(\Delta T)^2}{f_{\lambda}[\{c\}]}<-\frac{(\Delta T)^2}{f_{\lambda'}[\{c\}]}\Big\}
\Rightarrow\bigcap\limits_{\substack{\lambda'\in\mathcal{Y}\\\lambda'\neq\lambda}}\Big\{f_{\lambda}[\{c\}]<f_{\lambda'}[\{c\}]\Big\},\end{eqnarray}
where the $\mathcal{Y}$ is the set of the all the ansatz labels,
\begin{eqnarray}\mathcal{Y}=\{1,2a,2b,2c,3a,3b,3c,3d,4a,4b,4c,5,6\}.
\end{eqnarray}
These sets of inequalities in Eq.(\ref{eq_SM:paramcons}) give the domains in $\{c\}$ parameter space for a particular ansatz to be the ground state with lower energy than other ans\"atze. We below always make the assumption $c_0\gg|c_{1,2,3}|$, which is necessary to avoid the unphysical situation that the free energy is not lower bounded and negative infinite free energy appears in some regions of the GL parameter space due to zero $f[\{c\}]$ according to Eqs.\eqref{eq:optfunc} and \eqref{eq:lowfreenphase}. We numerically compare the free energy of different ans\"atze for the conditions in Eqs.\eqref{eq_SM:paramcons} and find that except $\{2b,3b,3c,4c\}$, all the other ans\"atze can exist in some regions of GL parameter $\{c\}$. 

These finite SC phases can spontaneously break crystal symmetry and time reversal. By examining the transformation property of these SC phases under the symmetry operator $\mathcal{S}$ in MPG31$'$ (See Eqs.(\ref{eq_SM:C3z_T_Pvector})), one can compare the order parameter after the symmetry transformation, denoted as $\tilde{\mathcal{P}}$, with $\mathcal{P}$ before the symmetry transformation, and if one finds 
\begin{eqnarray}\label{eq:SSBcond}\tilde{\mathcal{P}}\neq e^{i\alpha}\mathcal{P}
\end{eqnarray}
for a certain $\alpha\in[0,2\pi]$, the symmetry $\mathcal{S}$ is spontaneously broken. The physical meaning and symmetry properties of the possible SC ground states are summarized as the following. 

{\bf Phase 1} with the order parameter in Eqs.\eqref{eq:ansatz1},\eqref{eq:phase1sol} and the minimal free energy form in Eq.(\ref{eq:phase1_minimal_energy}) represents a single-$\QQ$ Fulde-Ferrell(FF) state \cite{FForiginal}. From the transformation in Eqs.(\ref{eq_SM:C3z_T_Pvector}), we find 
\begin{eqnarray}
&&C_{3z}:\tilde{\mathcal{P}}_1 = U_{C_{3z}}\mathcal{P}_1=\Delta_0(0,0,a_0e^{i\phi_0},0,0,0)^T,\\&&\mathcal{T}:\tilde{\mathcal{P}}_1 =U_{\mathcal{T}}\mathcal{K}\mathcal{P}_{1}= \Delta_0(0,0,0,a_0e^{-i\phi_0},0,0)^T
\end{eqnarray}
for the $C_{3z}$ and $\mathcal{T}$ symmetries, respectively, both of which are different from $\mathcal{P}_1$ itself. Thus, we conclude that both the $C_{3z}$ and $\mathcal{T}$ symmetries are spontaneously broken. 

{\bf Phase 2a} with the order parameter in Eqs.\eqref{eq:ansatz2a},\eqref{eq:phase2asol} and the minimal free energy in Eq.\eqref{eq:minfreen2a} represents a double-$\QQ$ FF state \cite{agterberg2009a}. From the transformation form in Eqs.(\ref{eq_SM:C3z_T_Pvector}), we find
\begin{eqnarray}&&C_{3z}:\tilde{\mathcal{P}}_{2a}=U_{C_{3z}}\mathcal{P}_{2a}=\Delta_0a_0e^{i\theta}(0,0,0,0,e^{i\phi_1},e^{-i\phi_1})^T,\\&&\mathcal{T}:\tilde{\mathcal{P}}_{2a}=U_{\mathcal{T}}\mathcal{K}\mathcal{P}_{2a}=\Delta_0a_0e^{-i\theta}(0,0,e^{i\phi_1},e^{-i\phi_1},0,0)^T.\end{eqnarray}

As one can see, both symmetries are broken according to Eq.\eqref{eq:SSBcond}, making the system breaking $C_{3z}$ and $\mathcal{T}$ spontaneously.

{\bf Phase 2c} with the order parameter in Eqs.\eqref{eq:ansatz2c},\eqref{eq:phase2csol} and minimum free energy \eqref{eq:minfreen2c} represents a single-$\QQ$ Larkin-Ovchinikov (LO) state \cite{LOstate} or a stripe phase \cite{agterberg2020b}. From Eq.\eqref{eq:ansatz2c}, we see that it can break $C_{3z}$ symmetry spontaneously, but not the TR symmetry, as
\begin{eqnarray}&&C_{3z}:\tilde{\mathcal{P}}_{2c}=U_{C_{3z}}\mathcal{P}_{2c}=\Delta_0a_0e^{i\theta}(0,0,e^{-i\phi_1},0,0,e^{i\phi_1})^T,\\&&\mathcal{T}:\tilde{\mathcal{P}}_{2c}=U_{\mathcal{T}}\mathcal{K}\mathcal{P}_{2c}=\Delta_0a_0e^{-i\theta}(e^{i\phi_1},0,0,e^{-i\phi_1},0,0)^T=e^{i2\theta}\mathcal{P}_{2c}. 
\end{eqnarray}

{\bf Phase 3a} with the order parameter in Eqs.\eqref{eq:ansatz3a},\eqref{eq:phase3asol} and minimum free energy \eqref{eq:minfreen3a} represents a three-$\QQ$ FF state. From Eq.\eqref{eq:ansatz3a}, we see that it can break both $\mathcal{T},C_{3z}$ symmetries spontaneously,
\begin{eqnarray}&&C_{3z}:\tilde{\mathcal{P}}_{3a}=U_{C_{3z}}\mathcal{P}_{3a}=\Delta_0e^{i\theta}(a_0e^{i\phi_2},0,0,0,a_0e^{i\phi_1},a_1e^{-i(\phi_1+\phi_2)})^T ,\\&&\mathcal{T}:\tilde{\mathcal{P}}_{3a}=U_{\mathcal{T}}\mathcal{K}\mathcal{P}_{3a}=\Delta_0e^{-i\theta}(0,0,0,a_0e^{-i\phi_1},a_1e^{i(\phi_1+\phi_2)},a_0e^{-i\phi_2})^T.\end{eqnarray}

{\bf Phase 3d} with the order parameter in Eqs.\eqref{eq:ansatz3d},\eqref{eq:phase3dsol} and minimum free energy \eqref{eq:minfreen3d} represents the vortex-antivortex phase, which corresponds to formation of vortex-antivortex hexagonal lattice\cite{agterberg2011,liangfuval}. From Eq.\eqref{eq:ansatz3d}, the transformation rules yield
\begin{eqnarray}
&& C_{3z}:\tilde{\mathcal{P}}_{3d} =U_{C_{3z}}\mathcal{P}_{3d}=\Delta_0a_0(e^{-i(\phi_1+\phi_2)},0,e^{i\phi_2},0,e^{i\phi_1},0)^T, \\
&& \mathcal{T}:\tilde{\mathcal{P}}_{3d}=U_{\mathcal{T}}\mathcal{K}\mathcal{P}_{3d}=e^{-i\theta}\Delta_0a_0(0,e^{-i\phi_2},0,e^{-i\phi_1},0,e^{i(\phi_1+\phi_2)})^T. \end{eqnarray}

From the above equations, one can see that the {\bf phase 3d} breaks TR symmetry spontaneously. For the $C_{3z}$ symmetry, it is more tricky. At the first sight, it seems the $C_{3z}$ symmetry only exists when both the phase factors $\phi_1$ and $\phi_2$ are zero. We may define 
\begin{eqnarray}
\mathcal{P}_{3d,0}=e^{i\theta}\Delta_0(a_0,0,a_0,0,a_0,0)
\end{eqnarray}
with $\phi_1 = \phi_2 =0$ so that $C_{3z}:\tilde{\mathcal{P}}_{3d, 0} = \mathcal{P}_{3d,0}$.  
Next we will show that $\mathcal{P}_{3d}$ with any values of $\phi_1$ and $\phi_2$ can be connected to $\mathcal{P}_{3d,0}$ with a lattice translation. To see that, we consider the real space wavefunction of the {\bf phase 3d} in Eq.\eqref{eq:realeigvec} for $\mathcal{P}_{3d}$ and $\mathcal{P}_{3d,0}$, which are given by 
\begin{eqnarray}
    \Psi^{(3d)}_\alpha(\rr)=\frac{1}{V}\sum\limits_{\QQ_j\in\mathcal{Q}_6} P^{(3d)}_{\QQ_j}\Psi_{c,\alpha}^{(0)}(\QQ_j)e^{i\QQ_j\cdot\rr}
\end{eqnarray}
and 
\begin{eqnarray}
    \Psi^{(3d,0)}_\alpha(\rr)=\frac{1}{V}\sum\limits_{\QQ_j\in\mathcal{Q}_6} P^{(3d,0)}_{\QQ_j}\Psi_{c,\alpha}^{(0)}(\QQ_j)e^{i\QQ_j\cdot\rr}
\end{eqnarray}
where $P^{(3d)}_{\QQ_j}$ and $P^{(3d,0)}_{\QQ_j}$ are the $j$th components of $\mathcal{P}_{3d}$ and $\mathcal{P}_{3d,0}$, respectively. Explicitly, we have 
\begin{eqnarray}\Psi^{(3d)}_{\alpha}(\rr)&=&\frac{\Delta_0a_0e^{i\theta}}{V}\Big[\Psi_{0,\alpha}(\QQ_0)e^{i(\QQ_0\cdot\rr+\phi_1)}+\Psi_{0,\alpha}(\QQ_2)e^{i(\QQ_2\cdot\rr-\phi_1-\phi_2)}+\Psi_{0,\alpha}(\QQ_4)e^{i(\QQ_4\cdot\rr+\phi_2)}\Big]\nonumber\\&=&\frac{\Delta_0a_0e^{i\theta}}{V}\Big[\Psi_{0,\alpha}(\QQ_0)e^{ik_0(x+\frac{\phi_1}{k_0})}+\Psi_{0,\alpha}(\QQ_2)e^{ik_0\{-\frac{1}{2}(x+\frac{\phi_1}{k_0})+\frac{\sqrt{3}}{2}(y-\frac{\phi_1}{\sqrt{3}k_0}-\frac{2\phi_2}{\sqrt{3}k_0})\}}\nonumber\\&+&\Psi_{0,\alpha}(\QQ_4)e^{ik_0\{-\frac{1}{2}(x+\frac{\phi_1}{k_0})-\frac{\sqrt{3}}{2}(y-\frac{\phi_1}{\sqrt{3}k_0}-\frac{2\phi_2}{\sqrt{3}k_0})\}}\Big]\nonumber\\&=&\frac{\Delta_0a_0e^{i\theta}}{V}\Big[\Psi_{0,\alpha}(\QQ_0)e^{i(\QQ_0\cdot(\rr+\dd(\phi_{1},\phi_{2})))}+\Psi_{0,\alpha}(\QQ_2)e^{i(\QQ_2\cdot(\rr+\dd(\phi_{1},\phi_{2})))}+\Psi_{0,\alpha}(\QQ_4)e^{i(\QQ_4\cdot(\rr+\dd(\phi_{1},\phi_{2})))}\Big]\nonumber\\&=&\Psi^{(3d,0)}_{\alpha}(\rr+\dd(\phi_{1},\phi_{2}))\label{eq:3dGoldstone}\end{eqnarray}
where $\QQ_j$ is given in Eq.\eqref{eq:sixpointsset}, $\alpha={0,\pm1}$ labels three components of the order parameters and 
\begin{eqnarray}\dd(\phi_{1},\phi_{2})=\frac{\phi_1}{k_0}\hat{e}_x-\left(\frac{\phi_1}{2}+\phi_2\right)\hat{e}_y\end{eqnarray}
is a vector representing the translation of the vortex-antivortex lattice. Thus, $\Psi^{(3d)}_{\alpha}(\rr)$ can be viewed as the Goldstone mode of $\Psi^{(3d,0)}_{\alpha}(\rr)$ with a translation of $\dd(\phi_{1},\phi_{2})$\cite{k8sb-rqxf} characterized by two $U(1)$ phases $(\phi_1,\phi_2)$ between different components of $\mathcal{P}_{3d}$. Since $\Psi^{(3d,0)}_{\alpha}(\rr)$ has $C_{3z}$ symmetry, we would expect $\Psi^{(3d)}_{\alpha}(\rr)$ is also $C_{3z}$ invariant, but its rotation center is shifted by the vector $\dd(\phi_{1},\phi_{2})$. 
Mathematically, we have
\begin{eqnarray}\Psi^{(3d,0)}_{\alpha}(\rr)=\Psi^{(3d)}_{\alpha}(\rr-\dd(\phi_1,\phi_2)),\end{eqnarray}
which implies
\begin{eqnarray}\label{eq:trans3d1}t[-\dd(\phi_1,\phi_2)]:\quad \tilde{\mathcal{P}}_{3d}=U_t[-\dd(\phi_1,\phi_2)] \mathcal{P}_{3d} = \Delta_0a_0e^{i\theta}(1,0,1,0,1,0) = \mathcal{P}_{3d,0}\end{eqnarray}
according to Eq.\eqref{eq:realeigvec}. 
As $\mathcal{P}_{3d,0}$ is invariant under $C_{3z}$, we must have
\begin{eqnarray}t[\dd(\phi_1,\phi_2)]C_{3z}t[-\dd(\phi_1,\phi_2)]:\quad \tilde{\mathcal{P}}_{3d}=U_t[\dd(\phi_1,\phi_2)]U_{C_{3z}}\mathcal{P}_{3d,0}=\Delta_0a_0e^{i\theta}(e^{i\phi_1},0,e^{-i(\phi_1+\phi_2)},0,e^{i\phi_2},0),
\end{eqnarray}
which indicates that $\mathcal{P}_{3d}$ is invariant under the three-fold rotation symmetry with the rotation center at $\dd(\phi_1,\phi_2)$, as described by $t[\dd(\phi_1,\phi_2)]C_{3z}t[-\dd(\phi_1,\phi_2)]$. 


{\bf Phase 4a} with the order parameter in Eqs.\eqref{eq:ansatz4a},\eqref{eq:phase4asol} and minimum free energy \eqref{eq:minfreen4a} represents a superposition state between the two-$\QQ$ FF state and a single-$\QQ$ LO/PDW state. From Eq.\eqref{eq:ansatz4a}, we find 
\begin{eqnarray}&&C_{3z}:\tilde{\mathcal{P}}_{4a}=U_{C_{3z}}\mathcal{P}_{4a}=\Delta_0e^{i\theta}(a_1e^{i\phi_2},a_0e^{-i\phi_3},0,0,a_0e^{i\phi_1},a_1e^{-i(\phi_1+\phi_2)})^T,\\&&\mathcal{T}:\tilde{\mathcal{P}}_{4a}=U_{\mathcal{T}}\mathcal{K}\mathcal{P}_{4a}=\Delta_0e^{-i\theta}(a_0e^{i\phi_3},0,0,a_0^{-i\phi_1},a_1e^{i(\phi_1+\phi_2)},a_1^{-i\phi_2})^T,\end{eqnarray}
and thus {\bf Phase 4a} spontaneously break both $\mathcal{T}$ and $C_{3z}$ symmetries,

{\bf Phase 4b} with the order parameter in Eqs.\eqref{eq:ansatz4a},\eqref{eq:phase4bsol} and minimum free energy \eqref{eq:minfreen4b}, represent a four-$\QQ$ FF state. From Eq.\eqref{eq:ansatz4b}, we find
\begin{eqnarray}&&C_{3z}:\tilde{\mathcal{P}}_{4b}=U_{C_{3z}}\mathcal{P}_{4b}=\Delta_0e^{i\theta}(a_0e^{-i\phi_2},0,a_4e^{-i\phi_3},0,a_0e^{i\phi_1},a_1e^{-i(\phi_1+\phi_2)})^T,\\&&\mathcal{T}:\tilde{\mathcal{P}}_{4b}=U_{\mathcal{T}}\mathcal{K}\mathcal{P}_{4b}=\Delta_0e^{-i\theta}(0,a_4e^{i\phi_3},0,a_0e^{-i\phi_1},a_1e^{i(\phi_1+\phi_2)},a_0e^{i\phi_2})^T,\end{eqnarray}
so that {\bf Phase 4b} spontaneously break both $\mathcal{T}$ and $C_{3z}$ symmetries. 

{\bf Phase 5} with the order parameter in Eqs.\eqref{eq:ansatz5},\eqref{eq:phase5sol} and minimum free energy \eqref{eq:minfreen5}, represent a superposition of the two-$\QQ$ FF  and the three-$\QQ$ FF states. From Eq.\eqref{eq:ansatz5}, we find
\begin{eqnarray}&&C_{3z}:\tilde{\mathcal{P}}_{5}=U_{C_{3z}}\mathcal{P}_{5}=e^{i\theta}\Delta_0(a_2e^{-i\phi_2},a_1e^{-i\phi_3},a_0e^{-i\phi_4},0,a_0e^{i\phi_1},a_1e^{-i(\phi_1+\phi_2)})^T,\\&&\mathcal{T}:\tilde{\mathcal{P}}_{5}=U_{\mathcal{T}}\mathcal{K}\mathcal{P}_{5}=\Delta_0e^{-i\theta}(a_1e^{-i\phi_3},a_0e^{-i\phi_4},0,a_0e^{-i\phi_1},a_1e^{i(\phi_1+\phi_2)},a_2e^{i\phi_2})^T,\end{eqnarray}
so that {\bf Phase 5} spontaneously break both $\mathcal{T}$ and $C_{3z}$ symmetries. 

For {\bf Phase 6}, we find all six components in $\mathcal{P}$ are non-zero and equal, according to Eq.(\ref{eq:phase6sol}). Consequently, the presence and absence of $\mathcal{T}$ and $C_{3z}$ of {\bf Phase 6} rely on the relative phases between different components, which requires to take into account the phase dependent terms, such as the $c_4$ term. This phase has been previously explored in Ref.\cite{agterberg2011}.

From the above discussion, we find that a variety of superconducting ground states can spontaneously break either $C_{3z}$ or $\mathcal{T}$ symmetry. In particular, the {\bf Phase 3d} (vortex-antivortex lattice phase) breaks time-reversal ($\mathcal{T}$) but preserves three-fold rotation ($t[\dd(\phi_1,\phi_2)]C_{3z}t[-\dd(\phi_1,\phi_2)]$), while the {\bf Phase 2c} (stripe phase) breaks three-fold rotation but preserves time reversal. The {\bf phases 1, 2a, 3a, 4a, 4b, 5} all break both three-fold rotation and time reversal. 

The real space distribution of the order parameters, as well as their different components ($A_{1g}$ or $E$ components), are depicted in Fig.\ref{fig:S4} for {\bf phases 2a, 3a, 4a and 4b}, in Fig. \ref{fig:S1} for {\bf phase 3d} and in Fig. \ref{fig:S2} for {\bf phase 2c}. The {\bf phase 1} is the single-$\QQ$ FF phase with only phase modulation, while its amplitude is uniform, and thus we do not plot its amplitude. We find the {\bf phase 2a, 2c} and {\bf 4a} are stripe phases. {\bf Phase 2a}, as shown in Fig.\ref{fig:S4}(a-c), has a sinusoidal modulation in the direction of $\QQ_0-\QQ_1$, where $\QQ_j$ is defined in Eq.\eqref{eq:sixpointsset}. Likewise, {\bf phase 2c} has stripe modulation along $\QQ_0-\QQ_3$, as seen in Fig.\ref{fig:S2}(a-c). {\bf Phase 4a} is a combination of two stripe modulations of different periodicity, both along the $\QQ_0$ direction, as seen in Fig.\ref{fig:S4}(g-i), which can be viewed as a mixture of two stripe phases {\bf 2a and 2c}. {\bf Phase 3a} forms a planar oblique lattice with the longer diagonal oriented along $\QQ_1$, as seen in Fig.\ref{fig:S4}(d-f). {\bf Phase 4b \text{and} 3d} form hexagonal lattices, as seen in Fig.\ref{fig:S4}(j-l) and Fig.\ref{fig:S1}(a-c), respectively.

\subsection{Example: phase transition between {\bf Phases 3d} and {\bf 2c}}
As an example, we will focus on two representative phases, the {\bf Phase 3d} and {\bf Phase 2c}, in which the former breaks $\mathcal{T}$ and the latter breaks $C_{3z}$. It turns out that these two phases can be controlled by a single tuning parameter in the phase diagram. Below we will first figure out the existence regions in the parameter space for these two phases.  

For {\bf Phase 2c}, the constraints in Eq.\eqref{eq_SM:paramcons} can be explicitly written down. When $c_1\leq 0$, {\bf Phase 2c} turns out to be the ground state in the following parameter regimes, $c_2<4 c_1\cap c_3<\frac{1}{2} (c_1+2 c_2)$ or $4 c_1<c_2\leq \frac{43 c_1}{14}-\frac{1}{14} \sqrt{57} \sqrt{c_1^2}\cap c_3<\frac{2 c_1^2}{4 c_1-c_2}$ or $\frac{43 c_1}{14}-\frac{1}{14} \sqrt{57} \sqrt{c_1^2}<c_2\leq 0\cap c_3<\frac{1}{2} (c_1+2 c_2)-\frac{1}{2} \sqrt{3} \sqrt{4 c_1 c_2-c_2^2}$ or $c_2>0\cap c_3<\frac{c_1}{2}$. When $c_1>0$, {\bf Phase 2c} can also be the ground state in the following parameter regimes, $c_2\leq -\frac{1}{2} (3 c_1)\cap c_3<\frac{1}{2} (c_1+2 c_2)$ or $-\frac{1}{2} (3 c_1)<c_2\leq 0\cap c_3<\frac{2 c_2}{3}$ or $0<c_2\leq \frac{c_1}{7}\cap c_3<\frac{1}{4} (c_1+3 c_2)-\frac{1}{4} \sqrt{c_1^2+22 c_1 c_2-7 c_2^2}$ or $\frac{c_1}{7}<c_2<c_1\cap \Big(c_3<\frac{1}{4} (c_1+3 c_2)-\frac{1}{4} \sqrt{c_1^2+22 c_1 c_2-7 c_2^2}\cup \frac{1}{2} (c_1+2 c_2)-\frac{1}{2} \sqrt{3} \sqrt{4 c_1 c_2-c_2^2}<c_3<0\Big)$ or $c_1\leq c_2<4 c_1\cap c_3<0$ or $c_2>4 c_1\cap c_3<\frac{2 c_1^2}{4 c_1-c_2}$. 

Similarly, we can also explicitly simplify the constraints in Eq.\eqref{eq_SM:paramcons} 
for {\bf Phase 3d}. For {\bf phase 3d}, when $c_1<0$, this phase exists when $c_2<\frac{69 c_1}{2}-\frac{9}{2} \sqrt{57} \sqrt{c_1^2}$ and $\frac{3 c_1^2+2 c_1 c_2-5 c_2^2}{6 c_1-8 c_2}<c_3<\frac{c_1+c_2}{2}-\sqrt{c_1 c_2}$ or $ \frac{1}{4} (2 c_2-c_1)-\frac{1}{4} \sqrt{4 c_1 c_2-3 c_1^2}<c_3<\frac{1}{4} \sqrt{4 c_1 c_2-3 c_1^2}+\frac{1}{4} (2 c_2-c_1)$ or $ c_3>\frac{c_1+c_2}{2}+\sqrt{c_1 c_2}$. It also expands through $\frac{69 c_1}{2}-\frac{9}{2} \sqrt{57} \sqrt{c_1^2}\leq c_2<4 c_1$ and $\frac{1}{4} (2 c_2-c_1)-\frac{1}{4} \sqrt{4 c_1 c_2-3 c_1^2}<c_3<\frac{1}{4} \sqrt{4 c_1 c_2-3 c_1^2}+\frac{1}{4} (2 c_2-c_1)$ or $c_3>\frac{c_1+c_2}{2}+\sqrt{c_1 c_2}$. The last zone of parameters is $3 c_1<c_2<c_1\cap c_3>\frac{c_1+c_2}{2}+\sqrt{c_1 c_2}$ for $c_1<0$. When $c_1=0$, we find the parameter regions are defined by $c_2<0$ with $\frac{5 c_2}{8}<c_3<\frac{c_2}{2}$ or $ c_3>\frac{c_2}{2}$. When $c_1>0$ we find two regions of parameters where the phase exists. The first one is $c_2\leq -\frac{1}{2} (3 c_1)\cap c_3>\frac{3 c_1^2+2 c_1 c_2-5 c_2^2}{6 c_1-8 c_2}$. The second one is $\frac{1}{2} (3 c_1)<c_2<0$ when $\frac{2 c_2}{3}<c_3<\frac{1}{2} (c_1+2 c_2)$ or $ c_3>\frac{3 c_1^2+2 c_1 c_2-5 c_2^2}{6 c_1-8 c_2}$.
The general conditions for the above parameter regions are complicated. We find when $c_1+c_2=0$ is satisfied, the above parameter regions can be greatly simplified. In this case, the {\bf Phase 2c} exists in two parameter regions, (i) $c_2<0$ and $c_3<\frac{2c_2}{3}$ or (ii) $c_2>0$ and $c_3<-\frac{c_2}{2}$, while the existence of {\bf Phase 3d} always requires $c_2<0$ and furthermore, it demands $\frac{2c_2}{3}<c_3<\frac{c_2}{2}$ or $c_3>\frac{2c_2}{7}$. 
Thus, under this simplification, we can consider the phase diagram for {\bf Phase 2c} and {\bf Phase 3d} in the $c_2$-$c_3$ parameter space when $c_1+c_2=0$ in Fig 1 of main text. Particularly, we find the phase boundary between {\bf Phase 2c} and {\bf Phase 3d} exists at $c_3=\frac{2c_2}{3}$ for $c_2<0$ in Fig 1 of main text.

\subsection{Vortex-antivortex lattice and effective angular momentum}
\begin{figure}
    \centering
    \includegraphics[width=\linewidth]{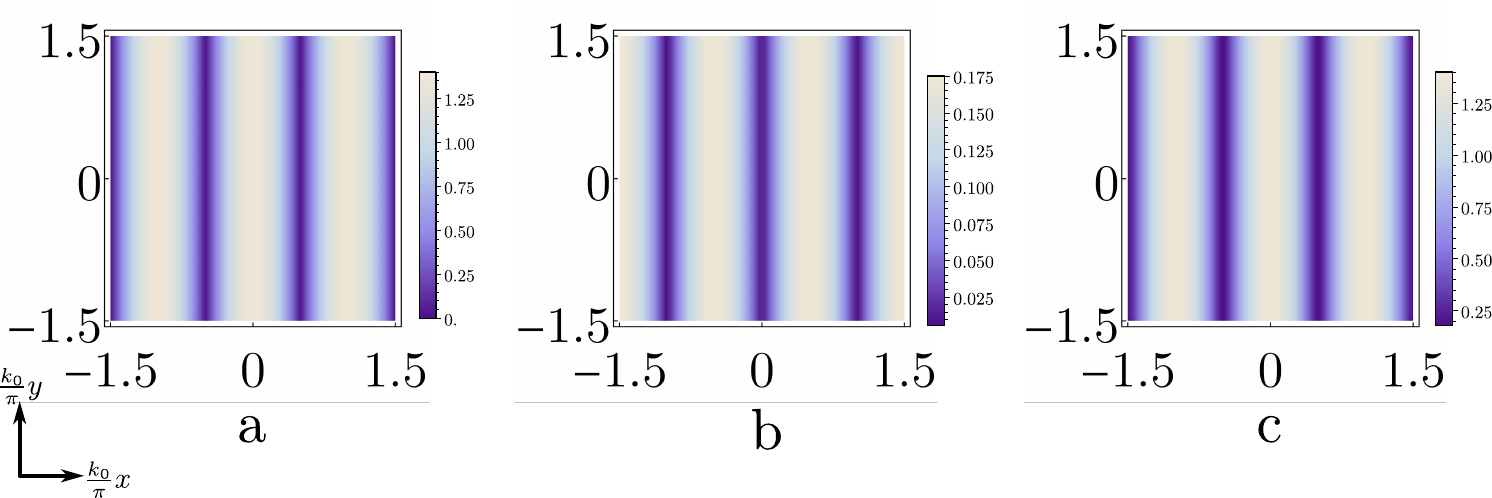}
    \caption{\label{fig:S2}Spatial distribution of different order parameter $\Psi(\rr)$ in striped {\bf phase 2c}. All axes represent the $x,y$ coordinates in scales of $\frac{\pi}{k_0}$. (a) Real-space distribution of the A-component $|\eta_{A}(\rr)|$(b)Real-space distribution of the E components $|\eta_{E}(\rr)|$, defined in Eq.\eqref{eq:Eampli}(c)The total amplitude of the order parameter, $|\Psi(\rr)|$, defined in Eq.\eqref{eq:totampli}. All numerical parameters are same as Fig 1. in main-text.}
\end{figure}

\begin{figure}
    \centering
    \includegraphics[width=\linewidth]{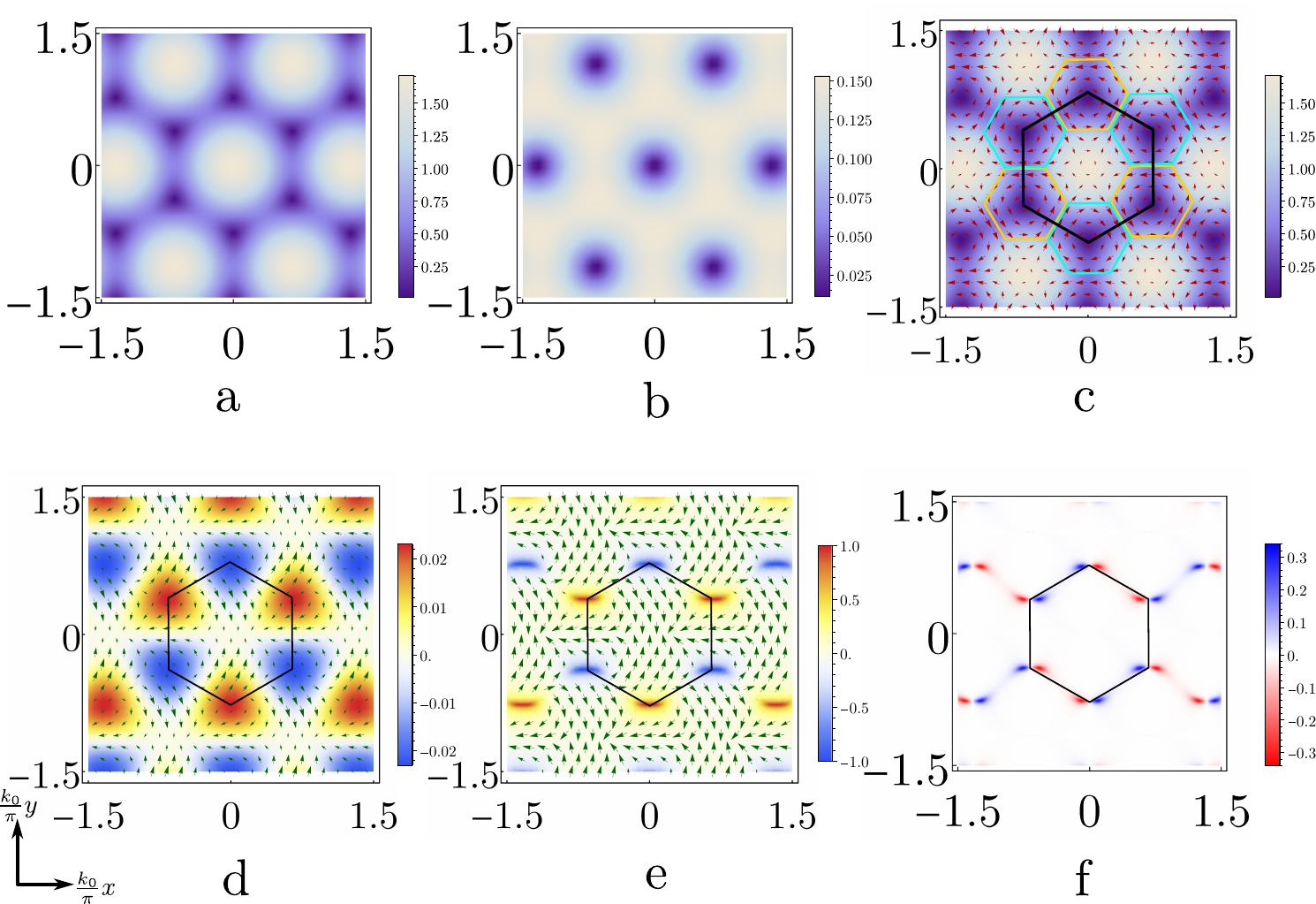}
    \caption{The spatial distribution and angular momentum properties of the order parameter $\Psi(\rr)$ in vortex-antivortex {\bf phase 3d}. All axes represent the $x,y$ coordinates in scales of $\frac{\pi}{k_0}$. $\phi_1=\phi_2=0$ for all settings, since these Goldstone modes only act in translation. All black hexagons represent the unit cells appearing in the corresponding quantity distributions. (a) Spatial distribution of $|\eta_A(\rr)|$ (b) Spatial distribution of $|\eta_E(\rr)$ in Eq.\eqref{eq:Eampli} (c)Spatial distribution of overall amplitude $|\Psi(\rr)|$. Arrows represent local current vector $\frac{1}{2e\gamma}\jj_{S}(\rr)$ in Eq.\eqref{eq:localcurrent}, it forms a hexagonal vortex-antivortex lattice. The bigger hexagon represents the unit cell and the yellow and cyan colored hexagonal plaquettes represent the domain of the vortices and anti-vortices, respectively. (d)Local distribution of intrinsic angular momenta $\LL(\rr)$. Arrows represent the $L_x(\rr),L_y(\rr)$ components and the color-bar represents $L_z(\rr)$ (e)Local distribution of normalized intrinsic angular momenta $\nn(\rr)$. Arrows represent the $n_x(\rr),n_y(\rr)$ components and the color-bar represents $n_z(\rr)$ (f) Distribution of the local winding number $\nn(\rr)\cdot (\partial_x\nn(\rr)\times\partial_\nn(\rr))$. For all plots, the same parameters as Fig. 1(c) in maintext is used.}\label{fig:S1}
\end{figure}

\begin{figure}
    \centering
    \includegraphics[width=\linewidth]{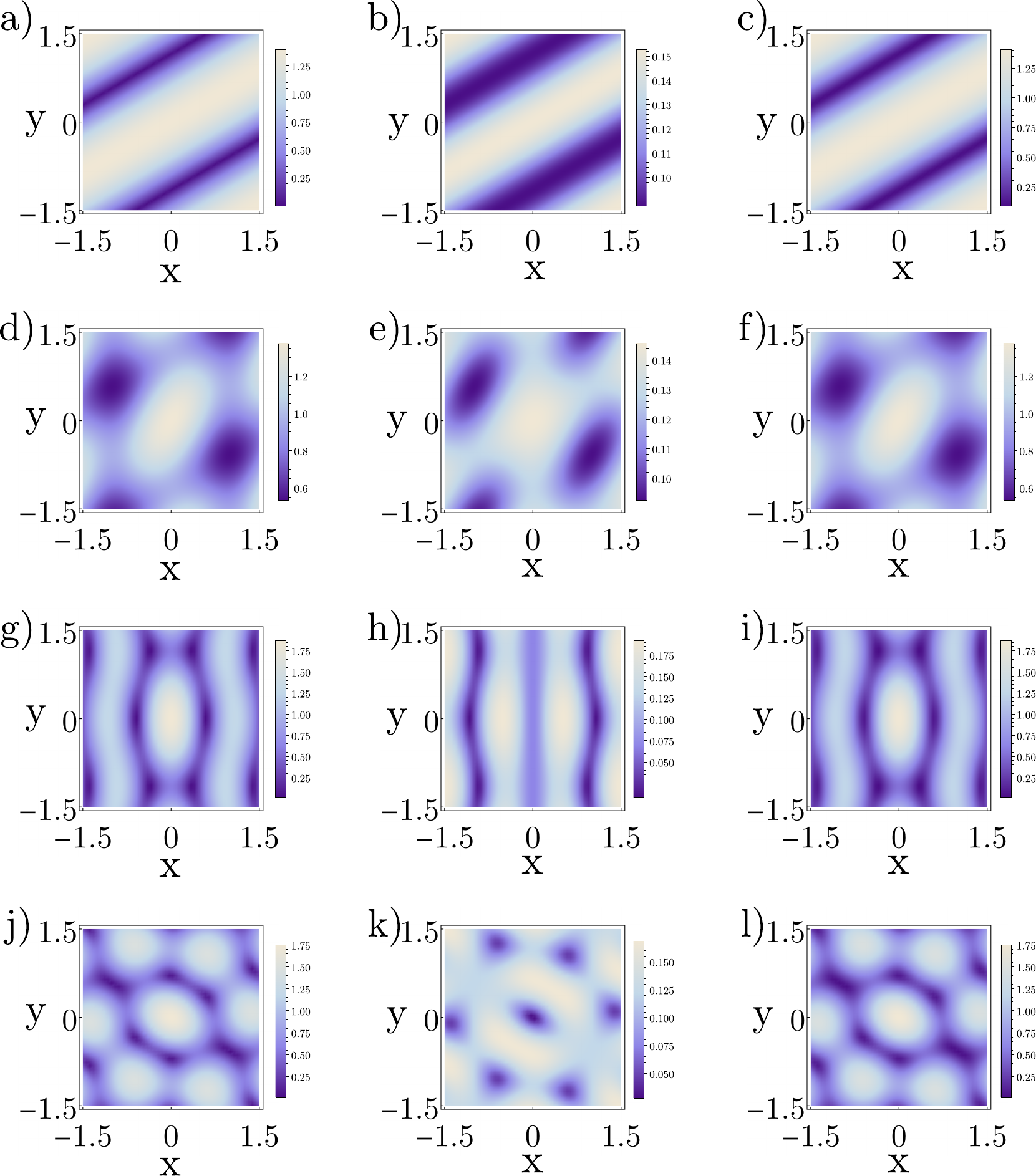}
    \caption{Spatial distribution of different order parameter components for different phases. First, second and third columns represent $\eta_A(\rr)|,|\eta_E(\rr)|$ and $|\Psi(\rr)|$ for these phases. Figs.(a)-(c) show the order parameter amplitudes for {\bf phase 2a} with the ground state $\frac{e^{i\theta}}{\sqrt{2}}\{1,1,0,0,0,0\}$. Figs.(d)-(f) show the order parameter amplitudes for {\bf phase 3a} with the ground state $\frac{e^{i\theta}}{\sqrt{2}}\{\sqrt{1-u^2},\sqrt{2}u,\sqrt{1-u^2},0,0,0\}\}$ with $u=0.3$. Figs.(g)-(i) show the order parameter amplitudes for {\bf phase 4a} with the ground state $\frac{e^{i\theta}}{\sqrt{10}}\{2,1,1,2,0,0\}$. Figs (j)-(l) show the order parameter amplitudes for {\bf phase 4b} with the ground state $\frac{2e^{i\theta}}{5}\{1,1/2,1,0,2,0\}$. All ground state component values are chosen for illustrating the features in order parameter and are not chosen specific to any particular domain of GL parameters. All the in-plane $\{x,y\}$ coordinates are in units of $\pi/k_0$, $k_0$ being the optimal momenta.\label{fig:S4}}
\end{figure}

The real space distribution of order parameter for {\bf Phase 3d} has been shown in Fig.\ref{fig:S1}(c) and discussed in Sec.\ref{sec:SCgstates}. Next we discuss the $A$ and $E$ components of the order parameter for {\bf phase 3d} separately. The amplitude of $A$ order parameter $|\eta_A(\rr)|$ in {\bf phase 3d}, as shown in \ref{fig:S1}(a), is one order larger than the amplitude of $E$ order parameter $|\eta_E(\rr)|$, as shown in Fig.\ref{fig:S1}b, which is defined as
\begin{eqnarray}\label{eq:Eampli}|\eta_{E}(\rr)|=\sqrt{|\eta_{E+}(\rr)|^2+|\eta_{E-}(\rr)|^2}.\end{eqnarray} In Fig.\ref{fig:S1}c, the colors represent the overall amplitude,
\begin{eqnarray}\label{eq:totampli}|\Psi(\rr)|=\sqrt{\eta_{A}(\rr)|^2+|\eta_{E+}(\rr)|^2+|\eta_{E-}(\rr)|^2},\end{eqnarray}
which primarily follows the spatial distribution of $\eta_{A}(\rr)$ in Fig.\ref{fig:S1}a, further testifying the dominance of $\eta_{A}(\rr)$ over the $\eta_E(\rr)$ amplitude. This dominance is evident for all other phases as well, as shown in Figs.\ref{fig:S2} and \ref{fig:S4}.

We examine the supercurrent distribution of {\bf Phase 3d} in Fig.\ref{fig:S1}c, to extract its vortex-antivortex structure. The local supercurrent can be defined from total free energy in Eq.\eqref{eq_SM:GLfreeenergy_1}, defined by 
\begin{eqnarray}\label{eq:genericjs}
\jj_s(\rr)=-2e\nabla_{\bf A}\{\mathcal{F}^{(2)}(\rr)+\mathcal{F}^{(4)}(\rr)\}|_{\bf A=0},
\end{eqnarray}
where $\mathcal{F}^{(2)}(\rr)$ is defined as the inverse Fourier transform of Eq.\eqref{eq:freen2app} and $\mathcal{F}^{(4)}(\rr)$ is defined in Eq.\eqref{eq:momGL4_0}. The vector field ${\bf A}$ is gauge coupled to the momentum as as $\kk\to\kk-2e{\bf A}$ in \eqref{eq:freen2app}, which gives rise to
\begin{eqnarray}\label{eq:localcurrent}\jj_{s}(\rr)=2e(\gamma_A+\gamma_E)\Im\{\Psi^*(\rr)\nabla\Psi(\rr)\}. 
\end{eqnarray}
The supercurrent distribution is shown by the red arrows in Fig.\ref{fig:S1}c for {\bf phase 3d}. The unit cell of the finite momentum order parameter is shown by the black hexagon in Fig.\ref{fig:S1}c, from which one can see that the winding of the supercurrent (red arrows) that forms vortex is around six corners of the hexagon. For the cyan and yellow small hexagons in Fig.\ref{fig:S1}c, the supercurrent is winding in the opposite directions, clockwise and counter-clockwise, leading to the formations of vortices and antivortices, respectively. This alternating vortex-antivortex structure is known as the vortex-antivortex lattice \cite{liangfuval,agterberg2011}.

Since the s-wave $A$ component with zero intrinsic angular momentum and the $E$ components with chiral $p\pm ip$ structure has $\pm1$ intrinsic angular momenta, we can define the effective angular momentum as
\begin{eqnarray}&&\LL(\rr)=\Psi^{\dagger}(\rr){\bf m}\Psi(\rr),\end{eqnarray}
where ${\bf m}$ is the vector of angular momentum component matrices,
\begin{eqnarray}&&m_x=\frac{1}{\sqrt{2}}\left(\begin{matrix}0&1&0\\1&0&1\\0&1&0\end{matrix}\right);m_y=\frac{1}{\sqrt{2}}\left(\begin{matrix}0&-i&0\\i&0&-i\\0&i&0\end{matrix}\right);m_z=\left(\begin{matrix}1&0&0\\0&0&0\\0&0&-1\end{matrix}\right).\end{eqnarray}

Using the form of the order parameter $\Psi(\rr)$, we find
\begin{eqnarray}\label{eq:Lcompr}
&&L_x(\rr)=\sqrt{2}\Re\Big[\eta_A^*(\rr)\{\eta_{E+}(\rr)+\eta_{E-}(\rr)\}\Big]\nonumber\\
&&L_y(\rr)=-\sqrt{2}\Im\Big[\eta_A^*(\rr)\{\eta_{E+}(\rr)-\eta_{E-}(\rr)\}\Big]\nonumber\\
&&L_z(\rr)=|\eta_{E+}(\rr)|^2-|\eta_{E-}(\rr)|^2.
\end{eqnarray}
The expression can be further expanded using the components of the order parameter in main text Eq. \eqref{eq:realeigvec} 
\begin{eqnarray}\eta_{E,\pm}(\rr)=\sum\limits_{j\in\mathbb{Z}_6}P_{\QQ_j}\eta_{E,\pm}^{(0)}(\QQ_j)e^{i\QQ_j\cdot\rr}.\end{eqnarray}
With the eigenvector $\eta_{E,\pm}^{(0)}(\QQ_j)$ in Eq. \eqref{eq_SM:GL2eigstate}, we find
\begin{eqnarray}
L_x(\rr)&&=\frac{2\sqrt{2}}{N(k_0)}\{T_c(k_0)-T_{c,E}-\gamma_Ek_0^2\}\sum\limits_{i,j\in\mathbb{Z}_6}\Re(\zeta'_1\QQ_{j+})\Re\Big\{P_{\QQ_j}P^*_{\QQ_k}e^{i(\QQ_j-\QQ_k)\cdot\rr}\Big\}\nonumber\\
L_y(\rr)&&=-\frac{2\sqrt{2}}{N(k_0)}\{T_c(k_0)-T_{c,E}-\gamma_Ek_0^2\}\sum\limits_{i,j\in\mathbb{Z}_6}\Im(\zeta'_1\QQ_{j+})\Im\Big\{P_{\QQ_j}P^*_{\QQ_k}e^{i(\QQ_j-\QQ_k)\cdot\rr}\Big\}\nonumber\\
L_z(\rr)&&=-\frac{2}{N(k_0)}\sum\limits_{i,j\in\mathbb{Z}_6}|\zeta_1'|^2\Im(\QQ_{j+}\QQ_{j-})\Im\Big(P_{\QQ_j}P_{\QQ_k}^*e^{i(\QQ_j-\QQ_k)\cdot\rr}\Big). \end{eqnarray}
We emphasize this angular momentum $\LL$ is an effective one since the order parameter is a combination of s-wave and p-wave pairing. A non-zero value of $L_z(\rr)$ can demonstrate spontaneous time reversal symmetry breaking. From Eq.(\ref{eq:Lcompr}), one can see that $L_z(\rr)$ describes the difference between the amplitudes $|\eta_{E+}(\rr)|$ and $|\eta_{E-}(\rr)|$. From \eqref{eq:antiunirealsteps}, time reversal operation transforms $\eta_{E,\mp}(\rr)$ as 
\begin{eqnarray}
\mathcal{T}: \quad \tilde{\eta}_{E,\pm}(\rr)=\eta^*_{E,\mp}(\rr).
\end{eqnarray} 
Non-zero $L_z(\rr)$ means different amplitudes between the $\eta_{E,+}(\rr)$ and $\eta_{E,-}(\rr)$ components and thus $\mathcal{T}$ must be broken. Fig.\ref{fig:S1}(d) reveals the distribution of $\LL(\rr)$ in the real space. The color represents the local $L_z(\rr)$ component, while the arrow length and direction represent the magnitude and direction of the in-plane components $(L_x(\rr),L_y(\rr))$, respectively. The distribution follows hexagonal lattices, as colored in black hexagon in Fig.\ref{fig:S1}(d). Each adjacent corners of the unit cell has opposite $L_z(\rr)$ and in-plane component windings. We further normalize $\LL$, defined by the unit-vector $\nn(\rr)=\frac{\LL(\rr)}{|\LL(\rr)|}$, which is shown in Fig.\ref{fig:S1}e. The color shows $n_z(\rr)$ while the arrow represents the in-plane vector $(n_x(\rr),n_y(\rr))$ and the black hexagon denotes the unit cell. The winding structure of the $\nn(\rr)$ field exhibits a hexagonal lattice of meron-antimeron pairs \cite{wang2020polar}. To quantify these features, we plot the local winding density $\mathcal{W}(\rr)$, defined as \begin{eqnarray}\mathcal{W}(\rr)=\frac{1}{4\pi}\nn(\rr)\cdot\{\partial_x\nn(\rr)\times\partial_y\nn(\rr)\},\end{eqnarray} in Fig.\ref{fig:S1}(f), which characterizes the meron-antimeron configuration. Non-zero values of $\mathcal{W}(\rr)$ reveal a dipole structure at each corner of hexagon in Fig.\ref{fig:S1}(f), suggesting that a hexagonal lattice of meron-antimeron dipoles is formed for the unit vector $\nn(\rr)$ field. Thus, the spontaneous breaking of time-reversal symmetry in the vortex-antivortex phase can give rise to meron-antimeron-pair lattice of the normalized unit vector $\nn(\rr)$ for the {\bf phase 3d}.

\section{Supercurrent and SC Diode Effect}~\label{app:3}
In this section, we will calculate the supercurrent and study the SC diode effect for this pair-mixing system. We first apply the generic expression of supercurrent in Eq.\eqref{eq:genericjs} to our particular ground states in the momentum space. We can characterize the motion of the Cooper pairs in the supercurrent with superfluid momentum \cite{tinkham2004} $\qq_s=\frac{\hbar}{2(\gamma_A+\gamma_E)}\vv_s$, where $\vv_s$ is the superfluid velocity and $\gamma_A+\gamma_E$ gives the effective mass of Cooper pairs. The superfluid velocity should be included in the kinetic energy terms of the GL free energy when a supercurrent is driven through a bulk superconductor, and thus we should change the free energy in Eq.\eqref{eq_SM:genfreen} to \begin{eqnarray}\mathcal{F}\equiv \mathcal{F}(\{\QQ_j\},\{P_{\QQ_j}\})\to\mathcal{F}(\{\QQ_j+\qq_s\},\{P_{\QQ_j+\qq_s}\}).\end{eqnarray} 
The corresponding supercurrent is given by\cite{fuliang2022} 
\begin{eqnarray}\label{eq:js1}\JJ_s(\qq_s)&=&-\frac{d\mathcal{F}(\{\QQ_j+\qq_s-2e\mathbf{A}\},\{\alpha,\phi\})}{d \mathbf{A}}\Bigg|_{\mathbf{A}=0}\nonumber\\&=&-\Bigg[\sum\limits_{\QQ_j\in\mathcal{Q}_6}\frac{\partial\mathcal{F}[\{\QQ_j+\qq_s\},\{\alpha,\phi\}]}{\partial \alpha_j}\frac{d\alpha_j}{d \mathbf{A}}+\frac{\partial\mathcal{F}[\{\QQ_j+\qq_s-2e\mathbf{A}\},\{\alpha,\phi\}]}{\partial{\bf \mathbf{A}}}\Bigg]\Bigg|_{\mathbf{A}=0}\nonumber\\&=&-\sum\limits_{\QQ_j\in\mathcal{Q}_6}\frac{\partial\mathcal{F}_{min}[\{\QQ_j+\qq_s-2e\mathbf{A}\}]}{\partial{\bf A}}\Bigg|_{\mathbf{A}=0}\nonumber\\&=&2e\sum\limits_{\QQ_j\in\mathcal{Q}_6}\partial_{\qq_s}\mathcal{F}_{min}[\{\QQ_j+\qq_s\}].\end{eqnarray}
In the third equality, we used the fact that the $\alpha_j$ component for the ground state should satisfy the GL equation $\frac{\partial\mathcal{F}}{\partial \alpha_j}=0$ so that the free energy is minimized, denoted as $\mathcal{F}_{min}$. 
With Eq.\eqref{eq_SM:genfreen}, the expression for $\mathcal{F}_{min}[\{\QQ_j+\qq_s\}]$ can be further expanded as
\begin{eqnarray}\label{eq_SM:free_energy_min_1}
\mathcal{F}_{min}[\{\QQ_j+\qq_s\}]&=&\sum\limits_{j\in\mathbb{Z}_6}\Delta T(\QQ_j+\qq_s)\tilde{\alpha}_i^2+\Big[c_0\Big(\sum\limits_{j\in\mathbb{Z}_6}\tilde{\alpha}_i^2\Big)^2+c_1\sum\limits_{j\in\mathbb{Z}_6}\tilde{\alpha}_{i}^2\tilde{\alpha}_{i+1}^2+c_2\sum\limits_{j\in\mathbb{Z}_6}\tilde{\alpha}_{i}^2\tilde{\alpha}_{i+2}^2+c_3\sum\limits_{j\in\mathbb{Z}_6}\tilde{\alpha}_{i}^2\tilde{\alpha}_{i+3}^2\Big],
\end{eqnarray}
where $\tilde{\alpha}_i=\Delta_0a_j$ with $\{\Delta_0,a\}$ to be different solutions discussed in Sec.\ref{sec:SCgstates}. 

In Eq.(\ref{eq_SM:free_energy_min_1}), the momentum shift $\qq_s$ only appears in the second order GL coefficient \begin{eqnarray}\label{eq:SDEtemp}
\Delta T(\QQ_j+\qq_s)\approx T-T_c^{(0)}(\kk=\QQ_j+\qq_s),
\end{eqnarray} 
with the critical temperature eigenvalue $T_c^{(0)}(\kk)$ given in Eq.\eqref{eq:Tsolnapp}. Here we consider the isotropic approximation and neglect all anisotropic corrections, which is a good approximation when $|\kk|<\sqrt{\frac{T_A}{\gamma'_A}}$ so that $T_c^{(0)}(\kk)\gg\Delta T_c^{(1)}(\kk),\Delta T_c^{(2)}(\kk)$. 

Although the GL coefficient $\Delta T(\QQ_j+\qq_s)$ is isotropic in Eq.\eqref{eq:SDEtemp}, the spontaneous symmetry breaking in the SC ground state $\mathcal{P}$, namely $\tilde{\alpha}_i$ in Eq.(\ref{eq_SM:free_energy_min_1}), can still induce the anisotropy in critical current along different directions. Using Eqs.(\ref{eq:js1}), (\ref{eq_SM:free_energy_min_1}) and (\ref{eq:SDEtemp}), we find \begin{eqnarray}\label{eq:genJsexp}
    |\JJ_s(\qq_s)| &&= \sum\limits_{j\in\mathbb{Z}_6}4\Big[T_{c,A}+T_{c,E}-|\QQ_j+\qq_s|^2(\gamma_A'+\gamma_E')+\sqrt{\Big\{T_{c,A}-T_{c,E}+|\QQ_j+\qq_s|^2(\gamma_E'-\gamma_A')\Big\}^2}\Big]\nonumber\\&&\times\Big[|\QQ_j+\qq_s|(\gamma_A'+\gamma_E')-\frac{\{T_{c,A}-T_{c,E}+|\QQ_j+\qq_s|^2(\gamma_E'-\gamma_A')\}2|\QQ_j+\qq_s|+8|\QQ_j+\qq_s||\zeta_1'|^2}{\sqrt{\Big\{T_{c,A}-T_{c,E}+|\QQ_j+\qq_s|^2(\gamma_E'-\gamma_A')\Big\}^2}}\Big]\nonumber\\&&\times\Big\{-\frac{1}{2f_{\lambda}[\{c\}]}a_{\lambda,j}^2+\frac{1}{4(f_{\lambda}[\{c\}])^2}\Big(c_0+c_1a_{\lambda,j}^2a_{\lambda,j+1}^2+c_2a_{\lambda,j}^2a_{\lambda,j+2}^2+c_3a_{\lambda,j}^2a_{\lambda,j+3}^2\Big)\Big\},
\end{eqnarray}
where $\lambda$ labels the particular ground state we are interested in, $f_{\lambda}[\{c\}]$ is the corresponding optimizing functions and $\{a_{\lambda,j\in\mathbb{Z}_6}\}$ are the components of order parameter defined in \ref{sec:SCgstates}. 

The critical current is defined by maximizing the magnitude of supercurrent $\JJ_s(\qq_s)$, which can be written as 
\begin{eqnarray}\label{eq:defnJc}
J_c(\theta)=\max\limits_{q_s}|\JJ_s(\qq_s)|, \end{eqnarray}
for a certain momentum angle $\theta$ with
$\qq_s=q_s (\cos(\theta),\sin(\theta))$. We further define the normalized critical current,
\begin{eqnarray}\label{eq:critical_current_anisotropy}
    \mathcal{J}(\theta) = \frac{J_c(\theta)}{J_c(0)}, 
\end{eqnarray}
to characterize the anisotropy of critical current. The SDE efficiency $\eta$ is then defined by 
\begin{eqnarray}
    \eta=\frac{J_c(0)-J_c(\pi)}{J_c(0)+J_c(\pi)}=\frac{1-\mathcal{J}(\pi)}{1+\mathcal{J}(\pi)}. 
\end{eqnarray}

\begin{figure}
    \centering
    \includegraphics[width=0.8\linewidth]{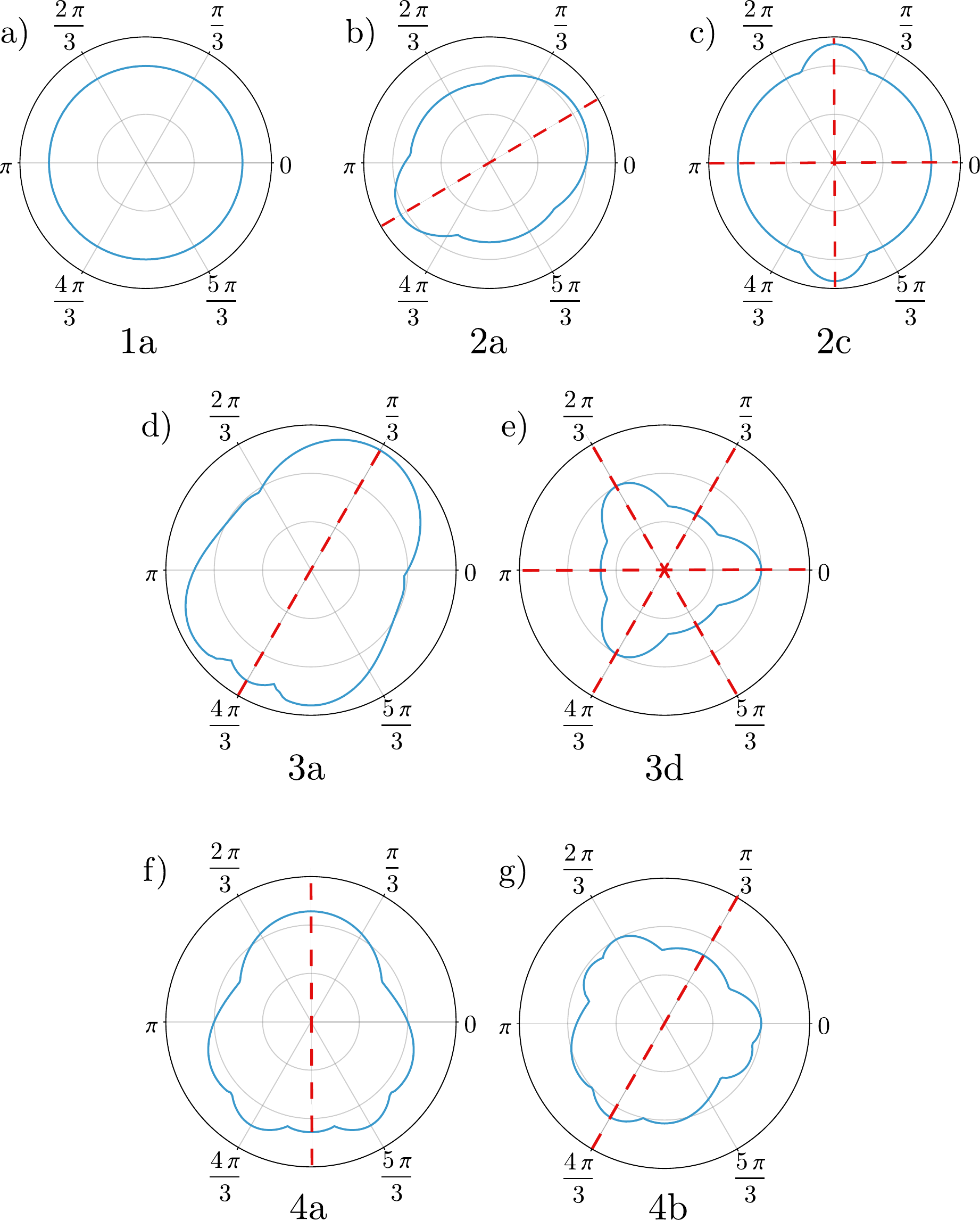}
    \caption{Normalized critical current $\mathcal{J}(\theta)$ as a function of $\theta$ for different Superconductor phases. Each panel is labeled by the ansatze index $\lambda\in\mathcal{Y}$. The red dashed lines represent the additional mirror symmetry axes of $\mathcal{J}(\theta)$.}
    \label{fig:Jcansatze}
\end{figure}

We numerically evaluate $|\JJ_s(\qq_s)|$ in Eq.(\ref{eq:genJsexp}) and then maximize it with respect to $q_s$ to compute the critical current $J_c(\theta)$ in Eq.(\ref{eq:defnJc}), from which we can extract the critical current anisotropy $\mathcal{J}(\theta)$ via Eq.(\ref{eq:critical_current_anisotropy}). The critical current anisotropy $\mathcal{J}(\theta)$ as a function of $\theta$ is shown in Fig.\ref{fig:Jcansatze} (a-g) for the {\bf phases 1, 2a, 2c, 3a, 3d, 4a} and {\bf 4b}, respectively. Among these phases, we find $\mathcal{J}(\theta)$ remains isotropic for {\bf phase 1}, but reveals anisotropy for all other phases ({\bf phases 2a, 2c, 3a, 3d, 4a, 4b}). {\bf Phase 2c} reveals a two-fold rotation symmetry. Thus, the SDE does not appear for the {\bf phases 1} and {\bf 2c}, but exists for other phases. Among the phases with SDE, only {\bf phase 3d} preserves three-fold rotation symmetry, while the other phases, including {\bf phases 2a, 3a, 4a, 4b}, also break three-fold rotation. In the main text, we have discussed the normalized critical current $\mathcal{J}(\theta)$ and the SDE efficiency $\eta$ for the {\bf phase 3d}. Below we will address two questions about the features shown in Fig.\ref{fig:Jcansatze}: (1) why does {\bf phase 1a} show isotropic $\mathcal{J}(\theta)$, even though it breaks both $C_{3z}$ and $\mathcal{T}$? (2) Why does $\mathcal{J}(\theta)$ show additional mirror symmetry with the mirror line shown by red dashed lines in Fig.\ref{fig:Jcansatze}, even though there is no mirror symmetry in the original free energy? 

(1) The minimum free energy of the {\bf phase 1a} is given by 
\begin{eqnarray}\mathcal{F}_{min,1}(\qq_s)=-\frac{{\Delta T(\QQ_0+\qq_s)}^2}{4c_0},\end{eqnarray}
from which we can derive the supercurrent from Eq.\eqref{eq:genJsexp}
\begin{eqnarray}
|\JJ_s(\qq_s)|&=&\frac{1}{c_0}\Big[T_{c,A}+T_{c,E}-|\QQ_0+\qq_s|^2(\gamma_A'+\gamma_E')+\sqrt{\Big\{T_{c,A}-T_{c,E}+|\QQ_0+\qq_s|^2(\gamma_E'-\gamma_A')\Big\}^2}\Big]\nonumber\\&&\times\Big[|\QQ_0+\qq_s|(\gamma_A'+\gamma_E')-\frac{\{T_{c,A}-T_{c,E}+|\QQ_0+\qq_s|^2(\gamma_E'-\gamma_A')\}2|\QQ_0+\qq_s|+8|\QQ_0+\qq_s||\zeta_1'|^2}{\sqrt{\Big\{T_{c,A}-T_{c,E}+|\QQ_0+\qq_s|^2(\gamma_E'-\gamma_A')\Big\}^2}}\Big]\label{eq:jsphase1}.\end{eqnarray}
The function $|\JJ_s(\qq_s-\QQ_0)|$ is isotropic across different directions $\theta$ of $\qq_s-\QQ_0$. Hence, one can say that the critical current for this particular shifted current is independent of $\theta$, i.e.-
\begin{eqnarray}\tilde{J}_c=\max\limits_{\qq_s-\QQ_0}|\JJ_s(\qq_s-\QQ_0)|.\end{eqnarray}
Here, we implicitly assume that the maximum occurs at some $|\qq_s|>|\QQ_0|$. Following the definition of $J_c(\theta)$ in Eq.\eqref{eq:defnJc}, we can see that
\begin{eqnarray}\label{eq:Jciso1}J_c(\theta)=\max\limits_{q_s}|\JJ_s(\qq_s)|=\max\limits_{|\qq_s-\QQ_0|}|\JJ_s(\qq_s-\QQ_0)|=\tilde{J}_c\end{eqnarray}
Thus, one see that $J_c(\theta)$, and subsequently $\mathcal{J}(\theta)$ is isotropic for {\bf phase 1} in Fig.\ref{fig:Jcansatze}a. For other phases, multiple $\QQ_j$ components contributing to the $\mathcal{F}_{min}(\qq_s)$ so that the above argument is no longer valid and the critical current anisotropy generally exists.

(2) Next we will discuss the emergent mirror symmetry by the red dashed lines shown in Fig.\ref{fig:Jcansatze}. Since the orientation of the mirror axis is different for different phases, we first give a general definition of the mirror symmetry. The mirror symmetry $\mathcal{M}^{||}_{\theta_{||}}$ is defined for a mirror plane perpendicular to our 2D system containing the axis parallel to an arbitrary azimuthal direction $\theta_{||}$ in the 2D plane. This acts on any general pairing momentum $\qq$ as
\begin{eqnarray}\label{eq:defnMth}\mathcal{M}^{||}_{\theta_{||}}:\tilde{\qq}=R_z[\theta_{||}-\theta_{\qq}]\qq=q\{\cos(2\theta_{||}-\theta_{\qq}),\sin(2\theta_{||}-\theta_{\qq})\};\quad\qq=q\{\cos(\theta_{\qq}),\sin(\theta_{\qq})\}, \end{eqnarray}
where $R_z(\phi)$ denotes the operation of rotation by angle $\phi$ along the z-axis. Let us take the example of {\bf phase 2a} to demonstrate how it has this emergent symmetry for $\theta_{||}=\frac{\pi}{6}$. First we write down the minimum energy for this phase,

\begin{eqnarray}\mathcal{F}_{min,2a}(\tilde{\qq}_s)=\frac{(\Delta T(\QQ_0+\tilde{\qq}_s))^2+(\Delta T(\QQ_1+\tilde{\qq}_s))^2}{8c_0+2c_1}.\end{eqnarray}
We see the momentum dependence in this minimum free energy manifests in $\Delta T(\QQ_{j}+\qq_s)$ for $j=0,1$. Following from Eq.\eqref{eq:SDEtemp} and referring to the definition of $T_{c}^{(0)}(\QQ_{j}+\qq_s)$ in Eq.\eqref{eq_SM:Tc0kk}, we see that $\Delta T(\QQ_{j}+\qq_s)$ is a polynomial function of quadratic terms $(\QQ_{j}+\qq_s)^2$. Expanding this expression gives
\begin{eqnarray}
(\QQ_j+\qq_s)^2=k_0^2+q_s^2+2k_0q_s\cos\left(\frac{j\pi}{3}-\theta_{\qq_s}\right); \quad \qq_s=q_s\{\cos(\theta_{\qq_s}),\sin(\theta_{\qq_s})\} \quad \forall j\in\mathbb{Z}_6.
\end{eqnarray}
With $\mathcal{M}_{\pi/6}^{||}$, $\mathcal{F}_{min,2a}(\qq_s)$ becomes $\mathcal{F}_{min,2a}(\tilde{\qq}_s)$, the transformation of this energy will be dictated by the transformation of $\qq_s$ in $(\QQ_{j}+\qq_s)^2$ for $j=0,1$. From Eq.\eqref{eq:defnMth}, for $\theta_{||}=\frac{\pi}{6}$, the transformations are 
\begin{eqnarray}&&\mathcal{M}^{||}_{\pi/6}:(\QQ_0+\tilde{\qq}_s)^2=k_0^2+q_s^2+2k_0q_s(\pi/3-\theta_{\qq_s})=(\QQ_1+\qq_s)^2\Rightarrow \Delta T(\QQ_0+\tilde{q}_s)=\Delta T(\QQ_1+\tilde{q}_s)\nonumber\\&&\mathcal{M}^{||}_{\pi/6}:(\QQ_1+\tilde{\qq}_s)^2=k_0^2+q_s^2+2k_0q_s(\theta_{\qq_s})=(\QQ_0+\qq_s)^2\Rightarrow \Delta T(\QQ_1+\tilde{q}_s)=\Delta T(\QQ_0+\tilde{q}_s),\end{eqnarray}
and thus, we find
\begin{eqnarray}\label{eq:fmin_2a_SDE}\mathcal{M}_{\pi/6}^{||}:&&\mathcal{F}_{min,2a}(\tilde{\qq}_s)=\frac{(\Delta T(\QQ_0+\tilde{\qq}_s))^2+(\Delta T(\QQ_1+\tilde{\qq}_s))^2}{8c_0+2c_1}\\&&=\frac{(\Delta T(\QQ_1+\qq_s))^2+(\Delta T(\QQ_0+\qq_s))^2}{8c_0+2c_1}=\mathcal{F}_{min,2a}(\qq_s). \end{eqnarray}
Thus, we prove that the minimum free energy of {\bf phase 2a} in Eq.\eqref{eq:fmin_2a_SDE} remains invariant under this emergent mirror symmetry, which does not exist in our original free energy in Eq.\eqref{eq_SM:genfreen}. 
From Eq.(\ref{eq:js1}), this $\mathcal{M}_{\pi/6}^{||}$ symmetry of $\mathcal{F}_{min,2a}(\qq_s)$ gets inherited by the corresponding supercurrent $\JJ_{s}(\qq_s)$ and consequently, shows up in $\mathcal{J}(\theta)$ shown in Fig.\ref{fig:Jcansatze}(b) for {\bf phase 2a}. Under similar arguments, {\bf Phase 2c} will also show mirror symmetry $\mathcal{M}^{||}_{0,\frac{\pi}{2}}$ in Fig.\ref{fig:Jcansatze}(c), {\bf Phase 3a} and {\bf Phase 4b}
have the mirror $\mathcal{M}^{||}_{\frac{\pi}{3}}$ in Fig.\ref{fig:Jcansatze}(d) and (g), respectively, {\bf Phase 3d} shows three mirror axes, i.e.-$\mathcal{M}^{||}_{0,\frac{\pi}{3},2\frac{\pi}{3}}$ in Fig.\ref{fig:Jcansatze}(e), and {\bf Phase 4a} will have $\mathcal{M}^{||}_{\frac{\pi}{2}}$ symmetry in Fig.\ref{fig:Jcansatze}(f). These different critical current patterns of $\mathcal{J}(\theta)$ will allow us to distinguish these different SC phases.

\end{document}